\documentclass[12pt]{article}
\usepackage[utf8]{inputenc}

\usepackage{amsfonts,amsthm,amsmath,amssymb,upgreek,bm}
\usepackage[hidelinks]{hyperref}
\usepackage[paper=letterpaper,margin=.85in]{geometry}
\usepackage{graphicx}
\usepackage{units}
\usepackage{float}
\usepackage[export]{adjustbox}
\usepackage{bbold}
\usepackage{xcolor}
\usepackage[shortlabels]{enumitem}
\usepackage{dsfont}
\usepackage[mathscr]{euscript}
\usepackage[normalem]{ulem}
\usepackage{physics}
\usepackage{tikz-cd}
\usepackage[autostyle]{csquotes}
\usepackage{babel}

\newcommand{\be}{\begin{equation}}
\newcommand{\ee}{\end{equation}}
\renewcommand{\tilde}{\widetilde}

\numberwithin{equation}{section}

\def\tr{\text{Tr}}
\def\Tr{\text{Tr}}

\usepackage{diagbox} 
\usepackage{multirow} 
\usepackage{xcolor}
\definecolor{dgreen}{rgb}{0.0, 0.5, 0.0}
\definecolor{dred}{rgb}{0.5, 0.0 , 0.0}
\usepackage{slashed}
 
\usepackage[compress]{cite}

\usepackage{braket}
\newcommand{\normord}[1]{:\mathrel{#1}:}

\usepackage{physics}

	\usepackage{tikz}
	\usepackage{makecell}

\usepackage{amsthm} % Required for theorem-like environments

\newtheorem{lemma}{Lemma}

\def\ie{\begin{equation}\begin{aligned}}
\def\fe{\end{aligned}\end{equation}}

\newtheorem{thm}{Theorem}

\newcommand{\RN}[1]{%
  \textup{\uppercase\expandafter{\romannumeral#1}}%
}

\begin{document}

\nocite{*}

\thispagestyle{empty}

\vspace*{2.5cm}
\begin{center}

{\bf {\LARGE Principle of Diminishing Potentialities in Large $N$ Algebras}}

\begin{center}

\vspace{1cm}

{\bf  Bik Soon Sia$^{1,2, \dagger}$}    \\
  \bigskip \rm

\bigskip 
${}^1$Yau Mathematical Sciences Center (YMSC), Tsinghua University, Beijing, China

${}^2$Department of Mathematical Sciences, Tsinghua University, Beijing, China

\bigskip
\rmfamily ${}^{\dagger}$\textit{Email}: \texttt{b.s.sia1106@gmail.com} 
\rm
  \end{center}

\vspace{2cm}
{\bf Abstract}
\end{center}
\begin{quotation}
\noindent

A connection between the completion of quantum mechanics featuring events  and the theory of emergent spacetime in quantum gravity where von Neumann algebra plays a vital role is established. In thermal equilibrium, it can be  shown that the Principle of Diminishing Potentialities (PDP) holds for the large $N$ algebra of $\mathcal{N}=4$ Super Yang-Mills (SYM) theory with gauge group $SU(N)$  when the temperature is higher than Hawking-Page temperature. Below Hawking-Page transition and for the case of zero temperature, PDP does not hold. Since the centralizer of thermofield double state on the large $N$ algebra of $\mathcal{N}=4$ SYM theory coincides with the center of the large $N$ algebra which is trivial, we extend the large $N$ algebra by performing crossed product by the maximal abelian subgroup $H $ of the compact symmetry group $G$ of the two-sided eternal black hole. In this case, the centralizer of an extension of thermofield double state is non-trivial and it is given by the action of the maximal abelian subgroup $H$ on the Hilbert space $\mathcal{H}_{TFD} \otimes L^{2} (H)$. This centralizer is by itself commutative and it coincides with its own center. This implies that the first actual event that initiate the ``Events-Trees-Histories''  dynamical evolution in this framework is given by the spectral projectors associated to the action of the Cartan subalgebra $\mathfrak{h}$ of the Lie algebra $\mathfrak{g}$ associated to the group $G$ on the Hilbert space $\mathcal{H}_{TFD} \otimes L^{2} (H)$.

\end{quotation}

\setcounter{page}{0}
\setcounter{tocdepth}{2}
\setcounter{footnote}{0}
\newpage

\setcounter{page}{2}

\tableofcontents

\section{Introduction}\label{sec:introduction}

In quantum mechanics, the characterization of an isolated physical system $S$ is given by four postulates:

 \begin{itemize}
     \item Physical observables are represented by self-adjoint operator $a=a^*$ acting on a separable Hilbert space $\mathcal{H}_{S}$. In general, the collection of all physical observables form an algebra $\mathcal{A}_{S}$

     \item The dynamics of the physical observables/ operators are given by Heisenberg evolution 
     \begin{align}
         a(t)= e^{i H (t-t') }a(t') e^{-i H (t-t')}; \quad  t, t' \in \mathbb{R}
     \end{align}
     where $H=H^*$ is the Hamiltonian of the isolated system $S$

     \item The expectation values of physical quantities in states i.e. operator-valued functional $\omega: \mathcal{A_{S}}\rightarrow \mathbb{C}$. Typically, in non-relativistic quantum mechanics, the states are taken to be density matrices / mixed states $\rho _{\omega}$ on $\mathcal{H}_{S} $ which are trace-class operator 
     \begin{align}
         \omega (a)= \Tr (\rho_{\omega} a);  \quad \Tr (\rho_{\omega})=1
     \end{align}
 Pure states are given by minimal (rank $1$) orthogonal projection $P= P^* = P^2$ which corresponds to an unit ray in $\mathcal{H}_{S}$. However, the treatment of states as density matrices is not universally applicable. For instance, the local algebra of observables $\mathcal{A}_{\mathcal{U}}$ associated to the region $\mathcal{U}$ in relativistic quantum field theory are typically classified as Type    $\RN{3}_1$ hyperfinite factor. Such algebras only admits infinite projector. Nevertheless, the algebraic state $\omega:  \mathcal{A}_{\mathcal{U}} \rightarrow \mathbb{C}$ with $\omega (\mathbb{1})=1$ always make sense.  

 \item Collapse/ Reduction of state upon measurement: When a measurement of a physical observable $\hat{a}$ yields an outcome $\xi_{\hat{a}}$ belonging to the spectrum $Spec(\hat{a})$, the pre-measurement state $\omega$ is replaced by $\omega_{\xi_{\hat{a}}}$
 \begin{align}
     \omega_{\xi_{\hat{a}}} (\cdot)= \frac{1}{\omega (\pi_{\xi_{\hat{a}}} )} \omega (\pi_{\xi_{\hat{a}}}\cdot \pi _{\xi_{\hat{a}}} )  \label{reduction}
 \end{align}
 where $\xi_{\hat{a}}$ is one of the eigenvalue of $\hat{a}$  and $\pi_{\xi_{\hat{a}}}$ is  the corresponding spectral projection. The probability of the outcome of the measurement is given by Born's rule
 \begin{align}
     P (\xi_{\hat{a}})= \omega (\pi_{\xi_{\hat{a}}})
 \end{align}
 \end{itemize}

The first three postulates are widely accepted. The controversy arises with the fourth postulate; to this day, no consensus has been reached. Even now, it remains unclear how the rule (\ref{reduction}) is enforced, leaving no compelling reason for its acceptance.

\par Recently, a profound proposal has been put forward, the so-called``Events-Trees-Histories "(ETH) approach pioneered by Fröhlich et al., as a potential completion of quantum mechanics addressing the fourth postulate \cite{frohlich2022time} \cite{frohlich2020brief} \cite{frohlich2015quantum} \cite{frohlich2020relativistic} \cite{Frohlich:2023dgx}. In the proposal, the fundamental mathematical construct comprises a co-filtration ${\mathcal{A}_{\geq t}};{t\in \mathbb{R}}$ of  algebras of observables indexed by a time parameter $t$, where the choice of temporal parameterization will be discussed subsequently.  Each algebra $\mathcal{A}_{\geq t}$ represents the collection of physical observables measurable from time $t$ onward. The key principle governing such a structure is
\begin{align}
 \mathcal{A}_{\geq t}   \subsetneqq \mathcal{A}_{\geq t'}  \quad \textit{for} \: \:  \: t > t'  \label{Principle of diminishing Potentialities}
\end{align}
(\ref{Principle of diminishing Potentialities}) is called the \textbf{principle of diminishing potentialities} which is a statement that the algebra generated by all operators representing possible events occurring at time  $t$
  or later is a strict subset of the algebra generated by operators representing possible events at an earlier time $t'$ where $t'< t$. Furthermore, we may consider a state $\omega$ on the full algebra $\mathcal{A}= \overline{\lor_{t \in \mathbb{R}} \mathcal{A}_{\geq t}}^{\vert \vert \cdot \vert \vert}$  where the closure is taken with respect to the norm of the system's physical Hilbert space. While the state $\omega$ may be pure with respect to the full algebra $\mathcal{A}$, it typically appears mixed when restricted to any proper subalgebra $\mathcal{A}_{\geq t} \subsetneqq \mathcal{A}$. 
 This establishes a framework for how a pure state can evolve into a mixed state, i.e. $\omega_{t}=\omega \vert_{\mathcal{A}_{\geq t}}$ generically becomes mixed. What is notable is that this framework provides a more natural mechanism compared to the standard fourth postulate for explaining how potential events transition to being actualized. One necessary condition for events to happen at time $t$ or later in such a system is that the centralizer $C_{\omega_{t}} (\mathcal{A}_{\geq t})$ of the state $\omega_{t}$ on $\mathcal{A}_{\geq t}$
 \begin{align}
     \mathcal{C}_{\omega_\geq t} (\mathcal{A}_{\geq t}) = \{ X \in \mathcal{A}_{\geq t} \vert \omega_{t} ([X,A])=0; \quad \forall A \in \mathcal{A}_{\geq t} \}
 \end{align}
 needs to be non-trivial. This offers a more plausible justification for state reduction without invoking an instantaneous, disruptive measurement process. The ETH approach to quantum mechanics establishes a fundamental form of temporal irreversibility (distinct from conventional entropic arguments) as an inherent feature of quantum evolution (stochastic branching evolution). As we will give a review in Section 2, this framework not only provides a comprehensive reformulation of quantum theory but also naturally extends to relativistic regimes \cite{frohlich2020relativistic}. These features make the ETH approach particularly significant for understanding quantum dynamics beyond traditional formulations, potentially offering new perspectives on fundamental physics problems such as the black hole information /unitary paradox.

  On the other hand, there has been a growing recent trend in studying the emergence of classical spacetime from the underlying microscopic theory using techniques from algebraic quantum field theory and von Neumann algebras within holography and AdS/CFT. Many of the recent developments stem from the study of subalgebra-subregion duality introduced by Liu and Leutheusser\cite{leutheusser2023emergent} \cite{leutheusser2022subalgebra}. They examined the large $N$ limit of
$\mathcal{N}=4$ SYM theory with gauge group $SU(N)$ in $\mathbb{R}\times S^{3}$, which is conjectured to be dual to the Type IIB superstring theory in $Ads_5 \times S_{5}$. From a novel  perspective, they analyzed the large 
$N$ algebra $\mathcal{A}_{\beta}$ generated by single-trace operators in SYM at some finite temperature $\frac{1}{\beta}$. In fact, the large $N$ structure of $\mathcal{N}=4$ SYM is given by the generalized free field theory. Their construction depends on the preferred canonical thermal state (the Gibbs state) i.e. $\rho_{\beta}= \frac{1}{\Tr e^{- \beta H} }e^{- \beta H}$, which, upon canonical purification, becomes the thermofield double state $\omega_{TFD}$. With a prefered state selected, they perform the GNS (Gelfand-Naimark-Segal) construction to build the GNS Hilbert space $\mathcal{H}_{TFD}$ and the GNS representation of the algebra $\pi_{\omega_{TFD} }(\mathcal{A}_{\beta})''=\mathcal{M}_{\beta}$. When the temperature is below the Hawking Page temperature $\beta > \beta_{HP}$, $\omega_{TFD}$ is conjectured to be dual to two causally disconnected thermal $Ads_{5}$ spacetimes entangled with one other. Here, the algebra $\mathcal{M}_{\beta}$ describing all operations / measurements accessible from a single asymptotic boundary  forms a Type $\RN{1}_{\infty}$ factor.
Conversely, above the Hawking Page temperature $\beta < \beta_{HP}$, $\omega_{TFD}$ is conjectured to be dual to the eternal $Ads$-Schwarzschild black hole, with its two asymptotic boundaries  connected through the bulk geometry. In this case, $\mathcal{M}_{\beta}$ associated with one asymptotic boundary constitutes a Type $\RN{3}_1$ factor. Crucially, the entanglement structure between the two boundary field theories is encoded in the von Neumann algebra type assigned to each asymptotic boundary, with distinct entanglement patterns manifesting as different bulk geometries in the large
$N$ limit.

\par The construction by Liu and Leutheusser of the emergent large $N$ algebra and large $N$ Hilbert space is not merely a formal/straightfoward large $N$ limit of the gauge theory. This is because a significant fraction of operators in the finite $N$ theory do not survive in the large $N$ limit. Specifically, they fail to exhibit well-defined or regular behaviour as $N$ approaches infinity. Notably, above Hawking-Page transition,  all $\mathcal{N}=4$ SYM global charges, including the Hamiltonian $H$ in radial quantization  become ill-defined or singular as $N \rightarrow \infty$. These operators are thereby excluded from the effective large $N$ description. This behaviour echoes with the fact that  modular automorphism of the Type $\RN{3}_{1} $ factor $\mathcal{M}_{\beta}$ associated with the thermofield double state $\omega_{TFD}$ is outer, meaning the modular Hamiltonian $h_{TFD}$ accessed by operations in $\mathcal{A}_{\beta}$ lies outside the algebra $\mathcal{A}_{\beta}$. In this case, the modular Hamiltonian $h_{TFD}$ associated to thermofield double state is
\begin{align}
    h_{TFD}= \beta H_{R} - \beta H_{L} \label{modular Hamiltonian}
\end{align}
where the subscript $R/L$ denotes the operator for right/left asymptotic boundaries.  We will simply denote the algebra $(\mathcal{M}_{\beta})_{R/L}$  as $\mathcal{M}_{R/L}$. Now, we restrict our discussion to the right asymptotic boundary.  Since the right Hamiltonian $H_{R}$ itself does not have a large $N$ limit, there is no equations of motion for the large $N$ generalized free fields (subtracted single trace operators).  The consequence is that time-slice axiom (the algebra of observables associated to a region $V$ can be solely reconstructed from the algebra of observables associated to its subset $U \subset V$ if $U$ contains a Cauchy surface of $V$ ) no longer holds. Thus, it can be the case that the algebra $\mathcal{M}_{R}$ restricted to different Cauchy slices are inequivalent, in the sense that if $\Sigma_1$ and $\Sigma_2$ are different Cauchy slices, then $\mathcal{M}_{R} \vert_{\Sigma_1}$ and $\mathcal{M}_{R} \vert_{\Sigma_2}$ cannot be expressed in terms of one another. Hence, this allows us to consider a proper time band subalgebra $\mathcal{M}_{R} \vert_{I \times S^3} $ where $I= (t',t)$ is a time interval. Nevertheless, the modular Hamiltonian (\ref{modular Hamiltonian}) with respect to the thermofield double state is well defined and it generates a one-parameter modular automorphism of the algebra $\mathcal{M}_{R}$, i.e.

\begin{align}
\Delta_{TFD}^{-iu} A(t, \Vec{x}) \Delta_{TFD}^{iu}= A(t + \beta u, \Vec{x}), \quad \Delta^{-iu}_{TFD} \mathcal{M}_{R} \Delta^{iu}_{TFD}= \mathcal{M}_{R},  \: \: \forall u \in \mathbb{R} \label{time translation formula} 
\end{align}

where $A(t, \Vec{x}) \in \mathcal{M}_{R} \vert_ {\Sigma_t}$ denotes some local operator at the boundary time $t$. However, we cannot interpret this as Heisenberg's equation or equation of motion because the modular Hamiltonian $ h_{TFD}= -\log \Delta_{TFD}$ depends on the entire asymptotic boundary and not just on a single time slice. Moreover, using these properties together with further structural features of the large
$N$ algebras constructed by Liu-Leutheusser, it is an inevitable consequence that  in thermal equilibrium and in the case when the temperature is above the Hawking Page temperature $\beta < \beta_{HP}$, the following relation holds
\begin{align}
    \mathcal{M}_{R} \vert_{\geq t} \subsetneqq \mathcal{M}_{R}\vert _{\geq t'} , \quad \textit{for} \: \: t> t'    \label{PDP in the introduction}
\end{align}
Here, $\mathcal{M}_{R}\vert_{\geq t}= \mathcal{M}_{R} \vert_{[t, \infty) \times S^3}$ where $[t, \infty)$ denotes a time band from time $t$ to the future infinity.
 This is in reminiscence of
the principle of diminishing potentialities (\ref{Principle of diminishing Potentialities}) proposed by  Fröhlich et al   which posits that such algebraic constraints play a fundamental role in governing state dynamics, particularly for processes where events occur which are not described by the Schrödinger equation.  In the case when the temperature is below the Hawking Page temperature $\beta > \beta_{HP}$ and in the case of zero temperature, one can also argue that  (\ref{PDP in the introduction}) does not hold based on the exponential type method in\cite{gesteau2024toward}.

\par The applicability of the principle (\ref{Principle of diminishing Potentialities}) to  the asymptotic boundary algebra becomes especially significant if one can also apply the dynamical evolution in the ETH approach to the preferred state $\omega_{TFD}  $ since $\mathcal{H}_{TFD}$ describes $\mathcal{O}(1)$ perturbation around the two-sided eternal black hole. If this can be implemented, this will imply that the semi-classical black hole can also be treated / modeled as an isolated open system characterized by (\ref{Principle of diminishing Potentialities}) where events   progressively occur as time progresses from the past to future.  The selection of the time parameter in this case is unambiguous, since the fundamental time is truly the asymptotic boundary time $t$ in holography. However, this immediately leads to a fundamental obstruction : the centralizer of the thermal state $\mathcal{C}_{TFD} (\mathcal{M}_{I})$ is trivial. One immediate way to see this is that the commutators of generalized free fields are given by $c$ numbers. This implies that the centralizer $\mathcal{C}_{TFD} (\mathcal{M}_{R})$ coincides with the center of the algebra $\mathcal{M}_{R}$ which is trivial.  This algebraic property makes such implementation impossible, as it precludes the occurrence of events while maintaining exact thermal equilibrium, revealing a tension between equilibrium thermodynamics and dynamical evolution in the ETH framework. 

\par Nevertheless, we  argue that this difficulty can be circumvented by incorporating the additional symmetries of the eternal black hole. Following the work of Liu and Leutheusser, a key development involves introducing a ``renormalized Hamiltonian'' into the right boundary algebra $\mathcal{M}_{R}$
  via the modular crossed product construction developed by Witten et.al \cite{witten2022gravity} \cite{chandrasekaran2023large}. The resulting algebra $\mathcal{N}_{R}= \mathcal{M}_{R} \rtimes \mathbb{R}_{TFD}$  is a Type $\RN{2}_{\infty}$ factor. Here,  $\mathcal{N}_{R}$ is constructed by adjoining $\mathcal{M}_{R}$ by $h_{TFD} + X$ where $h_{TFD}$ is the modular Hamiltonian associated to thermofield double state $TFD$,  and $X$ is a real variable.  As a Type $\RN{2}_{\infty}$ factor, $\mathcal{N}_{R}$ admits a renormalized trace  $\tau$ which is unique up to a multiplicative constant. Consequently, this trace $\tau$ enables the consistent definition of density matrices and entropy differences for the algebra $\mathcal{N}_{R}$. The crossed product algebra $\mathcal{N}_{R}$ acts on the Hilbert space $\mathcal{H}_{TFD} \otimes L^{2} (\mathbb{R})$ where the second factor is the space of square integrable function of $X$. The typically prefered state $\hat{\omega}$ for $\mathcal{N}_{R}$ in Dirac bra-ket notation is given by a tensor product state
  \begin{align}
      \ket{\Psi_{\hat{\omega}}}= \ket{TFD} \otimes f(X)
  \end{align}
where $f(X)$ is a square integrable function of $X$. In the case of modular crossed product algebra $\mathcal{N}_{R}$, its centralizer of the thermofield double state $\mathcal{C}_{TFD,f} (\mathcal{N}_{R})$ is indeed non-trivial. However,  the action of time translation on the Hilbert space $\mathcal{H}_{TFD} \otimes L^{2} (\mathbb{R})$ which is represented as $e^{is (h_{TFD} + X)}$ is inner with respect to the modular crossed product algebra $\mathcal{N}_{R}$. This leads to the following relation
\begin{align}
    \mathcal{N}_{R, \geq t} = \mathcal{N}_{R, \geq t-s} = \mathcal{N}_{R}, \forall s >0
\end{align}
In particular, this means that the Principle of Diminishing Potentialities (\ref{Principle of diminishing Potentialities}) does not hold for the modular crossed product algebra $\mathcal{N}_{R}$. 

\par Alternatively, we consider the crossed product of the right algebra $\mathcal{M}_{R}$ by the maximal abelian subgroup $H$ of a compact group automorphisms $G$. Here, $G$ refers to the symmetry group (with time translation excluded) of two-sided eternal black hole with vanishing angular momentum and charges. We denote the resulting algebra as $\mathcal{Y}^{H}_{R}$. In this case, the action of time translation is outer with respect to the algebra $\mathcal{Y}^{H}_{R}$ and thus the Principle of Diminishing Potentialities continues to hold
\begin{align}
    \mathcal{Y}^{H}_{R, \geq t} \subsetneqq \mathcal{Y}^{H}_{R, \geq t'} \quad \textit{for} \: \: t > t'
\end{align}
    Furthermore, the centralizer of a natural extension of thermofield double state denoted as $\ket{TFD,k}$ ( where $k(h) \in L^{2} (H)$ is a normalized function) on the algebra $\mathcal{Y}^{H}_{R}$ is non-trivial and it is given by
    \begin{align}
        \mathcal{C}_{TFD,k} (\mathcal{Y}^{H}_{R})= \{ W(h) w(h) \vert h \in H \}
    \end{align}
    where $W(H)$ it the action of the maximal abelian symmetry subgroup $H$ on the GNS Hilbert space $\mathcal{H}_{TFD}$ and $w(h)$ is the right action of the group $H$ on $L^{2} (H)$. Since this centralizer is by itself commutative, it coincides with its center $Z_{TFD,k}(\mathcal{Y}^{H}_{R})= \mathcal{C}_{TFD,k} (\mathcal{Y}^{H}_{R})\cap \mathcal{C}_{TFD,k} (\mathcal{Y}^{H}_{R})'$. From here, we can deduce that \textbf{the first actual event  initiating the entire ETH dynamics} is given by the spectral projectors associated to the action of the Cartan subalgebra $\mathfrak{h}$ of the Lie algebra $\mathfrak{g}$ associated to the group $G$ on the Hilbert space $\mathcal{H}_{TFD} \otimes L^{2} (H)$. Therefore,  we find a framework to apply the evolution according to the ETH formalism for the two-sided eternal black hole in asymptotically $Ads_{5}$ spacetime.

    \par Now,  we outline the structure of this paper. In section 2, we summarize the Events-Trees-Histories Approach (ETH) to quantum mechanics -- both in the setting of non-relativistic and relativistic quantum theory. In section 3, we give a comprehensive review of the algebraic approach to $\mathcal{N}=4$ SYM theory with gauge group $SU (N)$ and its implication to holography. In section 4, we show that the Principle of Diminishing Potentialities (PDP) holds for the large $N$ generalized free field algebra  in the case of thermal equilibrium and above Hawking Page transition. Below Hawking Page transition and also for the case of zero temperature, we give arguments that PDP does not hold in large $N$ algebra. We also give a rigorous proof that event does not occur at all in thermal equilibrium. Finally, we extend the large $N$ algebra and carry out procedures described in the previous paragraph to find out the first actual event that initiate the ETH dynamics.

    \par We also note that in a recent work \cite{Leutheusser:2025zvp} by Liu and Leutheusser, the relations between diminishing potentialities (shrinking of  algebras) and complexity/ growth of interior volume of black hole is also carried out. See \cite{Leutheusser:2025zvp} for more details.
\section{Events-Trees-Histories Approach} 
\subsection{Isolated Open System}

\label{Event-Trees-Histories Approach}
 
We start with a physical system $S$. The observables associated with $S$ is represented by some abstract self-adjoint operators  $X= X^{*}$. In relativistic quantum theory, $X$  can be generated by the smeared function of some Hermitian operator- valued distribution  $\phi (\tau, \Vec{x}) $ which depends on the spacetime point $(\tau, \Vec{x})$, i.e.

\begin{align}
    X = \int_{\mathcal{U}} d \Vec{x} d \tau \:  \phi(\tau, \Vec{x}) f(\tau, \Vec{{x}}) 
\end{align}
where $\mathcal{U}$ is an open set of the spacetime; $f(\tau, \Vec{x})$ is a test-function with its support contained in $\mathcal{U}$. We denote the set of all such observables  as $\mathcal{O}_{S}$.  By assumption, $\mathcal{O}_{S}$ consists of  all bounded functions of these abstract self-adjoint operators, i.e. if $F$ is a bounded function, then $F(X) \in \mathcal{O}_{S}$. 
We take the physical system $S$ to be an \textbf{isolated-open system}. As explained in \cite{frohlich2022time}, such a  system $S$ has degrees of freedom that interact only negligibly with those of its complement $S^{C}$ for a period of time much longer than the temporal interval in which the evolution of $S$ is monitored (definition of an isolated system ). This ensures that any interactions between $S$ and $S^{C}$ are too slow to affect the dynamics of $S$ during the experiment. So, for all practical purposes, $S$ can be  effectively treated as it does not interact with $S^{C}$ at all.   Despite isolation,  states in $S \vee S' $ can still be entangled. Moreover, an isolated system is called open if it can emit modes to $S^{C}$ that eventually cannot be detected anymore by any operations in $S$. Yet, these missing modes can still be in a state  entangled with a state in $S$.

\par In the ETH approach of quantum mechanics, the  \textbf{time} $t$ which is described by $\mathbb{R}$ is the fundamental concept.  As we will see, the time $t$ can be inferred by detecting events occurring in an isolated-open system $S$. At every time $t \in \mathbb{R}$, we assume there is a Hilbert space representation of $\mathcal{O}_{S}$ by self-adjoint operators acting on a separable Hilbert space $\mathcal{H}_{S}$
\begin{align}
    \mathcal{O}_{S} \ni Y \mapsto Y(t) \in \mathcal{B} (\mathcal{H}_{S})
\end{align}
where $Y= F(X)$ and $\mathcal{B}(\mathcal{H}_{S})$ is the algebra of  bounded operators of $\mathcal{H}_{S}$. In the case of relativistic quantum theory, the Hilbert space representation for each time $t$ is given by
\begin{align}
  Y=  F[ \int_{\mathcal{U}} d\Vec{x} d \tau \phi (\tau, \Vec{x}) f(\tau, \Vec{x}) ] \mapsto F[\int_{\mathcal{U}} d \Vec{x} d \tau \phi (\tau + t, \vec{x}) f(\tau , \Vec{x})]= Y(t)
\end{align}

We will also assume that the system $S$ is an autonomous system, i.e. the Hamiltonian $H_{S}$ of the system $S$ is time independent
\begin{align}
    H_{S}(t)= H \quad \forall t \in \mathbb{R}
\end{align}
As usual, since $S$ is an isolated system,  the self-adjoint operators acting on $\mathcal{H}_{S}$ representing $Y \in \mathcal{O}_{S}$ at two different times $t_1$ and $t_2$ are related to each other by unitary time evolution generated by the Hamiltonian $H$
\begin{align}
    Y(t_2)= e^{iH(t_2 -t_1)} Y(t_1) e^{-i H (t_2 - t_1)} \label{Heisenberg Picture}
\end{align}
It is important to note that the evolution in Heisenberg picture (\ref{Heisenberg Picture}) is only valid in a closed system $S$ where the degrees of freedom in $S$ and $S^{C}$ do not interact. If there are non-negligible interactions between the degrees of freedom of $S$ and $S^{C}$, the unitary time evolution of observables will break down and the dynamics of the observables require a more general description including effects such as decoherence, dissipation and thermalization.

\par Next, we can introduce the algebra of observables $\mathcal{Y}_{I}$ , $I \subset \mathbb{R}$ which is a bounded  interval of the time axis. The algebra $\mathcal{Y}_{I}$ is generated by arbitrary linear combinations of arbitrary products of bounded operator $Y(t)$  acting on the separable Hilbert space $\mathcal{H}_{S}$ with $t \in I $ and the additional requirement that the identity $\mathbb{1}$ belongs to all  algebras $\mathcal{Y}_{I}$ for any arbitrary interval $I$. Then, we define algebras $\mathcal{A}_{\geq t}$ as follows 
\begin{align}
    \mathcal{A}_{\geq t} = \overline{\bigvee_{I \subset [t, \infty)} \mathcal{Y}_{I}}^{\textit{weak}} \label{filtration of algebra}
\end{align}
where the closure is with respect to the weak operator topology of $\mathcal{H}_{S}$. Such algebras $\mathcal{A}_{\geq t}$ are called algebra of potentialities and they are defined to be  von Neumann algebras. As we will see later, potentialities here mean potential events. An isolated open system $S$ is characterized by this co-filtration of algebra of potentialities $\{ \mathcal{A}_{\geq t} \vert t \in \mathbb{R}   \}$. The principle that governs such structure is the \textbf{ principle of diminishing potentialities} (PDP)
\begin{align}
    \mathcal{A}_{\geq t}  \subsetneqq \mathcal{A}_{\geq t'} ,  \quad \textit{for} \: \: t > t'
 \end{align}
Note that this requirement is non-trivial because the consequence of the definition (\ref{filtration of algebra}) is only the relations $\mathcal{A}_{\geq t} \subseteq \mathcal{A}_{t'}$ for $t > t'$.   PDP is a stronger requirement that the algebra $\mathcal{A}_{\geq t}$ is a strict subalgebra of $\mathcal{A}_{\geq t'}$ for $t>t'$. We also define the full algebra of potentialities $\mathcal{A}$ to be
\begin{align}
    \mathcal{A}= \overline{\bigvee_{t \in \mathbb{R}} \mathcal{A}_{\geq t}}^{\vert \vert \cdot \vert \vert}
\end{align}
where the closure now is with respect to the norm of the Hilbert space $\mathcal{H}_{S}$. Now, we may also introduce a state $\omega$ as an operator-valued distribution on the full algebra $\mathcal{A}$ i.e. $\omega: \mathcal{A} \rightarrow \mathbb{C}$ to extract the information about the system $S$. In non-relativistic quantum mechanics (usually the algebra of observables is characterized by a Type $\RN{1}$ factor), states are taken to be density matrices $\rho_{\omega}$ on $\mathcal{H}_{S}$, i.e. $\rho = \rho^{*} \geq 0$ with $\tr \rho_{\omega}=1$ and $\omega (A)= \Tr(\rho_{\omega}A)$ with $A \in \mathcal{A}$. Certainly the domain of  states can be extended to $\mathcal{B}(\mathcal{H_{S}})$ with no difficulties. Note that for an autonomous system with finitely many degrees of freedom, the algebra $\mathcal{A}_{\geq t}$ coincides with $\mathcal{B}(\mathcal{H}_{S})$ for all time $t \in \mathbb{R}$. In such systems, the PDP does not hold; therefore, they are not used to describe isolated open systems.
 Instead,  realistic physical systems used to characterize isolated open systems must have infinitely many degrees of freedom.   In the Heisenberg picture, only operators acting on $\mathcal{H}_{S}$ evolves in time while states $\omega$ are independent of time. It is often said that Heisenberg picture is equivalent to the 
Schr\"{o}dinger picture where operators / physical quantities are independent of time while states $\rho_{\omega}$ evolves in time according to the Schr\"{o}dinger equation
\begin{align}
    \rho_{\omega } (t)= e^{i H (t-t') }\rho_{\omega} (t') e^{-i H (t-t')}
\end{align}
However, the equivalence between these pictures only hold during the absence of measurements/ observations. In standard quantum mechanics, the assigned rule is that when a measurement takes place, a collapse or reduction of state occurs to interrupt the Schr\"{o}dinger evolution and the state $\omega$ where the measurement is performed is replaced by (supposed the physical quantity being measured is represented by the self-adjoint operator $A$ and the measurement takes place at some time $t$)
\begin{align}
    \tilde{\rho}_{\omega} (t) = \frac{1}{\Tr (\tilde{\rho}_{\omega} \pi_{a} (t))} \pi_{a} (t) \rho_{\omega} \pi_{a} (t)
\end{align}
where $a \in Spec(A)$ is one of the eigenvalue of $A$ which represents the outcome of the measurement  and $\pi_{a} (t)$ is the spectral projection corresponding to the eigenvalue $a$ of the operator $A(t)$ at time $t$. The weak point of this rule, as pointed out by 
Fr\"{o}hlich et al.  in \cite{frohlich2022time} \cite{frohlich2020brief} \cite{frohlich2015quantum} \cite{frohlich2020relativistic} \cite{Frohlich:2023dgx}
and many others is that the notion of measurement is vague and does not correspond to a mathematically and physically well-defined operation in the formalism of quantum mechanics. 

\par In the ETH approach to quantum mechanics, the unitary evolution of operators retains its fundamental role but the evolution of states subject to measurements, featuring events is subjected to a statistical rule. In this spirit,  
the Heisenberg picture provides a more accurate operator-centric framework, but
it does not by itself account for the`` statistical law'' that governs the states dynamics. To properly describe this statistical law in the ETH approach, we first introduce the concept of \textbf{potentialities/ potential events}. A potentiality or potential event associated with the system $S$ that is localized at time $\geq t$ is given by a partition of unity by orthogonal projections on $\mathcal{H}_{S}$
\begin{align}
    \{ \pi_{\xi} \vert \xi \in \mathcal{X} \} \subset  \mathcal{A}_{\geq t}
\end{align}
with properties

\begin{align}
    \pi_{\xi} \pi_{\eta}= \delta_{\xi \eta} \pi_{\xi}; \quad \pi_{\xi}^{*}= \pi_{\xi}= \pi_{\xi}^2; \quad \sum_{\xi \in \mathcal{X}} \pi_{\xi}=1 
\end{align}

where for simplicity the set $\mathcal{X}$ here is assumed to be a countable set, i.e. the potential events are identified with spectral projections of self adjoint operators with discrete spectrum \footnote{More generally potential events can be identified with spectral projections of maximal set of commuting self-adjoint operators with continuous spectrum}. Now, suppose that a state $\omega$ occupies an   isolated open system $S$. The state $\omega$ might be a pure state on the full algebra $\mathcal{A}$. However, the state $\omega_{t}$ at time $t$ 
\begin{align}
    \omega_{t} := \omega \vert_{\mathcal{A}_{\geq t}}
\end{align}

is usually a mixed state due to PDP: $\mathcal{A}_{\geq t} \subsetneqq \mathcal{A}$ and the principle of entanglement in quantum mechanics. As we will see now, the authors of
\cite{frohlich2022time} \cite{frohlich2020brief} \cite{frohlich2020relativistic} used this insight to pave the way for a natural conception of actual events and lays the foundation for a theory of direct or projective measurements/ observations. In the standard formulation of quantum mechanics (Copenhagen interpretation), it is natural to say that a potential event $\{ \pi_{\xi} \vert \xi \in \mathcal{X} \} \in \mathcal{A}_{\geq t}$ actually occurs in a state $\omega$ at time $\geq t$ or \textbf{actualizing at time $\geq t$} if and only if
\begin{align}
    \omega_{t} (A)
= \sum_{\xi \in \mathcal{X}}\omega (\pi_{\xi} A \pi_{\xi}) \quad \forall A \in \mathcal{A}_{\geq t} \label{incoherent superposition}
\end{align}
which means that the state $\omega_{t}$ is an incoherent superposition of  projectors  $\pi_{\xi}$ of a potential event actualizing in time $\geq t$, this means that the right hand side of (\ref{incoherent superposition}) does not have off-diagonal element, i.e. a mixture/ mixed state. However, without knowing the state $\omega_t$ is mixed at the very first place,  it is subtle to know which  set of complete orthogonal projectors i.e. potential event that will be actualized to become an actual event such that (\ref{incoherent superposition})  always hold. We want to find out the criterion that allow us to  decide whether a potential event localized at time $\geq t$ actualizes itself in a state $\omega$  to become an actual event. We therefore need a systematic way to determine which, if any, partition of unity satisfies this condition (\ref{incoherent superposition}). In order to do this, we pass to the \textbf{centralizer of a state $\omega$ on $\mathcal{A}_{\geq t}$} that we denote as $\mathcal{C}_{\omega_{t}} (\mathcal{A}_{\geq t})$, i.e. the subalgebra of $\mathcal{A}_{\geq t}$ that commutes with the state $\omega_{t}$

\begin{align}
    \mathcal{C}_{\omega_{t}} (\mathcal{A}_{\geq t})= \{ X \in \mathcal{A}_{\geq t} \vert \omega_{t} ([X,A])=0; \quad \forall A \in \mathcal{A}_{\geq t}        \}
\end{align}

The motivation is straightforward: if $X \in \mathcal{A}_{\geq t}$ is a self adjoint element of the centralizer $\mathcal{C}_{\omega_t}(\mathcal{A}_{\geq t})$, so that $\omega_t ([X, A])=0$ for all $A$, then its spectral projectors $ \{\pi_{\xi} \vert \xi \in \textit{Spec}(X) \}$ automatically satisfy condition (\ref{incoherent superposition}) i.e,
\begin{align*}
    \omega_t (A) &= \sum_{\xi, \eta \in \textit{Spec}(X)}\omega (\pi_{\xi} A \pi_{\eta})\\
    &= \sum_{\xi, \eta \in \textit{Spec } (X)} \omega (\pi_{\eta} \pi_{\xi}A)= \sum_{\xi, \eta \in \textit{Spec} (X)} \omega (\delta_{\eta \xi} \pi_{\xi}A)\\
    &= \sum_{\xi \in \textit{Spec}(X)} \omega (\pi_{\xi}A)= \sum_{\xi \in \textit{Spec}(X)} \omega (\pi^2 _{\xi}A)\\
    &= \sum _{\xi \in \textit{Spec} (X) }\omega(\pi_{\xi}A \pi_{\xi}) \quad \quad  \quad \forall A \in \mathcal{A}_{\geq t}
\end{align*}

The first line employs the completeness relation in quantum mechanics, introducing two independent partitions of unity. The second line follows from the fact that $\pi_{\xi} A \in \mathcal{A}_{\geq t}$ and $\pi_{\eta} \in \mathcal{C}_{\omega_t} (\mathcal{A}_{\geq t})$. The subsequent steps rely on the properties of projectors. Thus, the non-triviality of this centralizer of the state  $\mathcal{C}_{\omega_{t}} (\mathcal{A}_{\geq t})$ is the natural arena in which to look for partitions of unity that meet the event criterion. Now, if the centralizer $\mathcal{C}_{\omega_t} (\mathcal{A}_{\geq t})$ is non-trivial, we need to further look into its \textbf{center of the centralizer}
$Z_{\omega_{t}}(\mathcal{A}_{\geq t})$ i.e.

\begin{align}
    Z_{\omega_{t}}(\mathcal{A}_{\geq t})= \mathcal{C}_{\omega_{t}} (\mathcal{A}_{\geq t}) \cap \mathcal{C}_{\omega_{t}} (\mathcal{A}_{\geq t})' =\{   Y\in \mathcal{C}_{\omega_{t}} (\mathcal{A}_{\geq t}) \vert [Y,X]=0; \quad \forall X \in \mathcal{C}_{\omega_{t}}(\mathcal{A}_{\geq t})  \}
\end{align}
where $\mathcal{C}_{\omega_t} (\mathcal{A}_{\geq t})' \in \mathcal{B} (\mathcal{H}_{S})$ is the commutant of  $\mathcal{C}_{\omega_t} (\mathcal{A}_{\geq t})$.  
The passage from the centralizer to its center is due to the following  consideration. The centralizer $\mathcal{C}_{\omega_t} (\mathcal{A}_{\geq t})$ may well contain non-commuting self-adjoint elements. Each such element supplies, via its spectral resolution, a family of orthogonal projectors satisfying (\ref{incoherent superposition}), but since the elements do not commute, these families describe mutually incompatible classical decompositions of the system. To eliminate this ambiguity and single out a unique family of projectors, one simply restricts to the center of the centralizer $Z_{\omega_t}(\mathcal{A}_{\ge t})$ because by definition its elements all commute with one another, so they are simultaneously diagonalizable and yield a single, unambiguous frame to describe decoherence. It is this unique family that is identified with the actual event.

\par Therefore, an actual event is setting in at time $\geq t$ with respect to a state $\omega_{t}$ if and only if the center of the centralizer $Z_{\omega_{t}}(\mathcal{A}_{\geq t})$ is nontrivial, i.e. $Z_{\omega_{t}}(\mathcal{A}_{\geq t}) \neq \mathbb{C} \mathbb{1}$ or equivalently $Z_{\omega_{t}} (\mathcal{A}_{\geq t})$ contains  \textbf{at least} two disjoint, orthogonal projections $\pi_{1}, \pi_{2}$, i.e. $\pi_{1} \pi_{2}=0$ and $\pi_{i}=\pi_{i}^{*}=\pi_{i}^2; i=1,2$ with 
\begin{align}
   0< \omega(\pi_{i})<1; \quad i=1,2
\end{align}
Now, given a state $\omega_{t}$ on the algebra $\mathcal{A}_{\geq t}$,  we say that an \textbf{actual event/ actuality} setting in at time $\geq t$  is the partition of unity of orthogonal projection
that generates the center of the centralizer $Z_{\omega_{t}} (\mathcal{A}_{\geq t})$
\begin{align}
    \{ \pi_{\xi} \vert \xi \in \mathcal{X}_{\omega_t}   \} \: \textit{generates}   \:  \: Z_{\omega_{t} (\mathcal{A}_{\geq t})}
\end{align}
here the set $\mathcal{X}_{\omega_t}$ denotes the spectrum of $Z_{\omega_t}(\mathcal{A}_{\geq t})$, i.e. $\mathcal{X}_{\omega_t}=\textit{Spec}(Z_{\omega_t}(\mathcal{A}_{\geq t}))$.  In the ETH approach, we take an active interpretation that if 
$Z_{\omega_{t}} (\mathcal{A}_{\geq t})$ is non-trivial, then the only potential event localized in time $\geq t$ which will \textbf{actually happen} to become an actual event at time $\geq t$ is the one which lies in $Z_{\omega_t}(\mathcal{A}_{\geq t})$.

\par We can further refine the above mechanism to localize the particular time $t$ that an actual event occurs. First, we note that if $\{ \pi_{\xi} \vert \xi \in \mathcal{X}  \} \subset \mathcal{A}_{\geq t}$ is a potential event localized at times $\geq t$, then $\{ e^{is H} \pi_{\xi} e^{-i s H} \vert \xi \in \mathcal{X} \} \in \mathcal{A}_{\geq t+s}$ is localized at times $t+s$. We  say that a potential event $\{  \pi_{\xi} \vert \xi \in \mathcal{X}\} \subset \mathcal{A}_{\geq t}$ might happen at time $t$ if and only if it is also not included in $\mathcal{A}_{\geq t'}$ for all $t' > t$. In the similar way, we say that  an actual event $\{ \pi_{\xi} \vert \xi \in \mathcal{X}_{\omega_t}  \} \subset \mathcal{A}_{\geq t}$ occurs at time $t$ if and only if it is not included in $\mathcal{A}_{\geq t'}$ for all $t' > t$.

\par The final step of the ETH formalism is the \textbf{Collapse Postulate (CP)}. Suppose that $\{ \pi_{\xi} \vert \xi \in \mathcal{X}_{\omega_t}   \} \in Z_{\omega_t}$ is the actual event that occurs at a particular time $t$ and let $\omega_{t}$ be the state in the algebra $\mathcal{A}_{\geq t}$ right before  the time $t$, then the state $\omega_{t + \epsilon}$ on $\mathcal{A}_{\geq t + \epsilon}$ right after the event has occurred with $\epsilon$ tends to zero is given by 
\begin{align}
    \omega_{t+ \epsilon, \xi_{*}} (\cdot)= \frac{1}{\omega_{t+ \epsilon} (\pi_{\xi_*})} \omega_{t+ \epsilon} (\pi_{\xi_*} \cdot \pi_{\xi_*})
\end{align}
where $\xi_*$ is some point in $\mathcal{X}_{\omega_t}$ with $\omega_{t} (\pi_{\xi_*}) >0$. The probability that the system $S$ is in the state $\omega_{t+ \epsilon}$ right after time $t$ with $\epsilon$ tends to zero where the actual event has occured is given by \textbf{Born's rule}
\begin{align}
    P(\xi_{*},t)= \omega_{t} (\pi_{\xi_*})
\end{align}
Therefore, we see that in the ETH approach to quantum mechanics, the evolution of states of an isolated open systen $S$ featuring events is determined by a non-linear stochastic branching process where the branching probabilities of the proccess is given by Born's rule.
The entire ETH formalism is illustrated in Figure $1$.

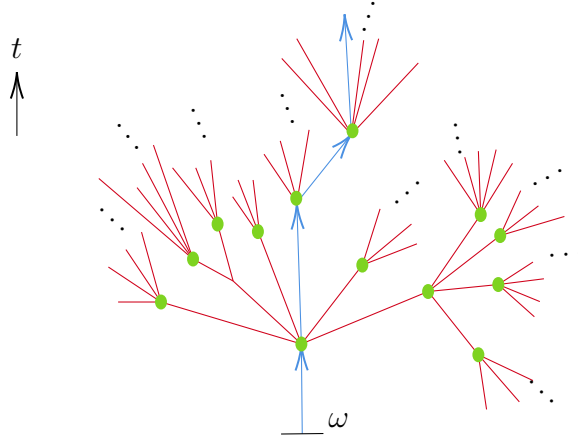
\begin{figure}[!ht]
    \centering    
\tikzset{every picture/.style={line width=0.4pt}} %set default line width to 0.75pt        

\begin{tikzpicture}[x=0.7pt,y=0.7pt,yscale=-1.3,xscale=1.00]
%uncomment if require: \path (0,264); %set diagram left start at 0, and has height of 264

%Straight Lines [id:da5225686456646468] 
\draw    (147.51,109.13) -- (147.51,84.74) ;
\draw [shift={(147.51,82.74)}, rotate = 90] [color={rgb, 255:red, 0; green, 0; blue, 0 }  ][line width=0.75]    (10.93,-3.29) .. controls (6.95,-1.4) and (3.31,-0.3) .. (0,0) .. controls (3.31,0.3) and (6.95,1.4) .. (10.93,3.29)   ;
%Straight Lines [id:da10082775176445835] 
\draw [color={rgb, 255:red, 208; green, 2; blue, 27 }  ,draw opacity=1 ]   (368.62,173.49) -- (300.48,195.19) ;
%Straight Lines [id:da9174506476215468] 
\draw [color={rgb, 255:red, 208; green, 2; blue, 27 }  ,draw opacity=1 ]   (333.25,162.56) -- (300.48,195.19) ;
%Straight Lines [id:da0937810030877273] 
\draw [color={rgb, 255:red, 208; green, 2; blue, 27 }  ,draw opacity=1 ]   (263.79,169.15) -- (300.48,195.19) ;
%Straight Lines [id:da5948082324626548] 
\draw [color={rgb, 255:red, 74; green, 144; blue, 226 }  ,draw opacity=1 ]   (300.95,232.33) -- (300.5,197.19) ;
\draw [shift={(300.48,195.19)}, rotate = 89.27] [color={rgb, 255:red, 74; green, 144; blue, 226 }  ,draw opacity=1 ][line width=0.75]    (10.93,-3.29) .. controls (6.95,-1.4) and (3.31,-0.3) .. (0,0) .. controls (3.31,0.3) and (6.95,1.4) .. (10.93,3.29)   ;
%Straight Lines [id:da527465009495261] 
\draw [color={rgb, 255:red, 74; green, 144; blue, 226 }  ,draw opacity=1 ]   (300.48,195.19) -- (298.1,138.79) ;
\draw [shift={(298.01,136.79)}, rotate = 87.58] [color={rgb, 255:red, 74; green, 144; blue, 226 }  ,draw opacity=1 ][line width=0.75]    (10.93,-3.29) .. controls (6.95,-1.4) and (3.31,-0.3) .. (0,0) .. controls (3.31,0.3) and (6.95,1.4) .. (10.93,3.29)   ;
%Straight Lines [id:da6154558740332191] 
\draw [color={rgb, 255:red, 208; green, 2; blue, 27 }  ,draw opacity=1 ]   (278.19,150.62) -- (300.48,195.19) ;
%Straight Lines [id:da6421143093331907] 
\draw [color={rgb, 255:red, 208; green, 2; blue, 27 }  ,draw opacity=1 ]   (225,177.83) -- (300.48,195.19) ;
%Straight Lines [id:da19452426565818304] 
\draw [color={rgb, 255:red, 208; green, 2; blue, 27 }  ,draw opacity=1 ]   (359.68,145.59) -- (333.25,162.56) ;
%Straight Lines [id:da7689782478730857] 
\draw [color={rgb, 255:red, 208; green, 2; blue, 27 }  ,draw opacity=1 ]   (342.8,139.31) -- (333.25,162.56) ;
%Straight Lines [id:da7590836469273863] 
\draw [color={rgb, 255:red, 208; green, 2; blue, 27 }  ,draw opacity=1 ]   (362.62,153.76) -- (333.25,162.56) ;
%Straight Lines [id:da2722887208335525] 
\draw [color={rgb, 255:red, 208; green, 2; blue, 27 }  ,draw opacity=1 ]   (407.4,150.34) -- (368.62,173.49) ;
%Straight Lines [id:da624648916756513] 
\draw [color={rgb, 255:red, 208; green, 2; blue, 27 }  ,draw opacity=1 ]   (396.92,141.66) -- (368.62,173.49) ;
%Straight Lines [id:da2747713023183821] 
\draw [color={rgb, 255:red, 208; green, 2; blue, 27 }  ,draw opacity=1 ]   (406.36,170.6) -- (368.62,173.49) ;
%Straight Lines [id:da38148817083024733] 
\draw [color={rgb, 255:red, 208; green, 2; blue, 27 }  ,draw opacity=1 ]   (395.66,199.65) -- (368.62,173.49) ;
%Straight Lines [id:da6138189662902622] 
\draw [color={rgb, 255:red, 208; green, 2; blue, 27 }  ,draw opacity=1 ]   (255.43,145.59) -- (263.79,169.15) ;
%Straight Lines [id:da5373933664995805] 
\draw [color={rgb, 255:red, 208; green, 2; blue, 27 }  ,draw opacity=1 ]   (242.22,160.05) -- (263.79,169.15) ;
%Straight Lines [id:da9834520536802548] 
\draw [color={rgb, 255:red, 208; green, 2; blue, 27 }  ,draw opacity=1 ]   (206.13,156.13) -- (225,177.83) ;
%Straight Lines [id:da0510687111687369] 
\draw [color={rgb, 255:red, 208; green, 2; blue, 27 }  ,draw opacity=1 ]   (196.7,163.36) -- (225,177.83) ;
%Straight Lines [id:da5745427764801732] 
\draw [color={rgb, 255:red, 208; green, 2; blue, 27 }  ,draw opacity=1 ]   (214.52,148.9) -- (225,177.83) ;
%Straight Lines [id:da5555832812607933] 
\draw [color={rgb, 255:red, 208; green, 2; blue, 27 }  ,draw opacity=1 ]   (201.94,177.83) -- (225,177.83) ;
%Straight Lines [id:da6657252164301493] 
\draw [color={rgb, 255:red, 208; green, 2; blue, 27 }  ,draw opacity=1 ]   (304.62,106.62) -- (298.01,136.79) ;
%Straight Lines [id:da5040702418172649] 
\draw [color={rgb, 255:red, 74; green, 144; blue, 226 }  ,draw opacity=1 ]   (298.01,136.79) -- (325.94,109.89) ;
\draw [shift={(327.38,108.51)}, rotate = 136.08] [color={rgb, 255:red, 74; green, 144; blue, 226 }  ,draw opacity=1 ][line width=0.75]    (10.93,-3.29) .. controls (6.95,-1.4) and (3.31,-0.3) .. (0,0) .. controls (3.31,0.3) and (6.95,1.4) .. (10.93,3.29)   ;
%Straight Lines [id:da8271139881743225] 
\draw [color={rgb, 255:red, 208; green, 2; blue, 27 }  ,draw opacity=1 ]   (280.39,116.68) -- (298.01,136.79) ;
%Straight Lines [id:da8937342487120986] 
\draw [color={rgb, 255:red, 208; green, 2; blue, 27 }  ,draw opacity=1 ]   (395.66,115.42) -- (396.92,141.66) ;
%Straight Lines [id:da7596222045072833] 
\draw [color={rgb, 255:red, 208; green, 2; blue, 27 }  ,draw opacity=1 ]   (388.32,117.93) -- (396.92,141.66) ;
%Straight Lines [id:da8825213329974066] 
\draw [color={rgb, 255:red, 208; green, 2; blue, 27 }  ,draw opacity=1 ]   (414.01,120.45) -- (396.92,141.66) ;
%Straight Lines [id:da6716731566347722] 
\draw [color={rgb, 255:red, 208; green, 2; blue, 27 }  ,draw opacity=1 ]   (405.2,114.79) -- (396.92,141.66) ;
%Straight Lines [id:da3594351821082057] 
\draw [color={rgb, 255:red, 208; green, 2; blue, 27 }  ,draw opacity=1 ]   (377.3,119.19) -- (396.92,141.66) ;
%Straight Lines [id:da7094594566433506] 
\draw [color={rgb, 255:red, 208; green, 2; blue, 27 }  ,draw opacity=1 ]   (418.42,131.76) -- (407.4,150.34) ;
%Straight Lines [id:da5892687599406228] 
\draw [color={rgb, 255:red, 208; green, 2; blue, 27 }  ,draw opacity=1 ]   (427.96,138.05) -- (407.4,150.34) ;
%Straight Lines [id:da6975328065495097] 
\draw [color={rgb, 255:red, 208; green, 2; blue, 27 }  ,draw opacity=1 ]   (441.91,142.45) -- (407.4,150.34) ;
%Straight Lines [id:da9926087715951009] 
\draw [color={rgb, 255:red, 208; green, 2; blue, 27 }  ,draw opacity=1 ]   (425.02,161.3) -- (406.36,170.6) ;
%Straight Lines [id:da5281003473660824] 
\draw [color={rgb, 255:red, 208; green, 2; blue, 27 }  ,draw opacity=1 ]   (430.9,168.22) -- (406.36,170.6) ;
%Straight Lines [id:da8285048963705055] 
\draw [color={rgb, 255:red, 208; green, 2; blue, 27 }  ,draw opacity=1 ]   (406.36,170.6) -- (428.7,177.65) ;
%Straight Lines [id:da840676930740342] 
\draw [color={rgb, 255:red, 208; green, 2; blue, 27 }  ,draw opacity=1 ]   (406.36,170.6) -- (425.76,183.93) ;
%Straight Lines [id:da2023312965114249] 
\draw [color={rgb, 255:red, 208; green, 2; blue, 27 }  ,draw opacity=1 ]   (395.66,199.65) -- (425.02,210.33) ;
%Straight Lines [id:da7986752549047649] 
\draw [color={rgb, 255:red, 208; green, 2; blue, 27 }  ,draw opacity=1 ]   (395.66,199.65) -- (417.68,219.13) ;
%Straight Lines [id:da3545345285342425] 
\draw [color={rgb, 255:red, 208; green, 2; blue, 27 }  ,draw opacity=1 ]   (395.66,199.65) -- (400.06,222.27) ;
%Straight Lines [id:da3967025275642664] 
\draw [color={rgb, 255:red, 208; green, 2; blue, 27 }  ,draw opacity=1 ]   (243.69,122.33) -- (255.43,145.59) ;
%Straight Lines [id:da27579920115922907] 
\draw [color={rgb, 255:red, 208; green, 2; blue, 27 }  ,draw opacity=1 ]   (231.2,122.33) -- (255.43,145.59) ;
%Straight Lines [id:da9152651429532985] 
\draw [color={rgb, 255:red, 208; green, 2; blue, 27 }  ,draw opacity=1 ]   (251.76,119.19) -- (255.43,145.59) ;
%Straight Lines [id:da18609852773839275] 
\draw [color={rgb, 255:red, 208; green, 2; blue, 27 }  ,draw opacity=1 ]   (266.44,127.36) -- (278.19,150.62) ;
%Straight Lines [id:da47040579546260686] 
\draw [color={rgb, 255:red, 208; green, 2; blue, 27 }  ,draw opacity=1 ]   (274.52,127.36) -- (278.19,150.62) ;
%Straight Lines [id:da9282228443491577] 
\draw [color={rgb, 255:red, 208; green, 2; blue, 27 }  ,draw opacity=1 ]   (263.51,134.28) -- (278.19,150.62) ;
%Straight Lines [id:da35162555401263285] 
\draw [color={rgb, 255:red, 208; green, 2; blue, 27 }  ,draw opacity=1 ]   (215.05,120.45) -- (242.22,160.05) ;
%Straight Lines [id:da6520474514844592] 
\draw [color={rgb, 255:red, 208; green, 2; blue, 27 }  ,draw opacity=1 ]   (212.12,130.51) -- (242.22,160.05) ;
%Straight Lines [id:da1954271898701576] 
\draw [color={rgb, 255:red, 208; green, 2; blue, 27 }  ,draw opacity=1 ]   (191.56,126.73) -- (242.22,160.05) ;
%Straight Lines [id:da418390828885434] 
\draw [color={rgb, 255:red, 208; green, 2; blue, 27 }  ,draw opacity=1 ]   (220.93,112.91) -- (242.22,160.05) ;
%Straight Lines [id:da8175760982268622] 
\draw [color={rgb, 255:red, 74; green, 144; blue, 226 }  ,draw opacity=1 ]   (327.38,108.51) -- (323.86,60.85) ;
\draw [shift={(323.71,58.85)}, rotate = 85.77] [color={rgb, 255:red, 74; green, 144; blue, 226 }  ,draw opacity=1 ][line width=0.75]    (10.93,-3.29) .. controls (6.95,-1.4) and (3.31,-0.3) .. (0,0) .. controls (3.31,0.3) and (6.95,1.4) .. (10.93,3.29)   ;
%Straight Lines [id:da798279798429823] 
\draw [color={rgb, 255:red, 208; green, 2; blue, 27 }  ,draw opacity=1 ]   (298.01,68.91) -- (327.38,108.51) ;
%Straight Lines [id:da6936574992142402] 
\draw [color={rgb, 255:red, 208; green, 2; blue, 27 }  ,draw opacity=1 ]   (342.06,68.91) -- (327.38,108.51) ;
%Straight Lines [id:da5093926626768868] 
\draw [color={rgb, 255:red, 208; green, 2; blue, 27 }  ,draw opacity=1 ]   (333.99,69.54) -- (327.38,108.51) ;
%Straight Lines [id:da015195067913597815] 
\draw [color={rgb, 255:red, 208; green, 2; blue, 27 }  ,draw opacity=1 ]   (363.35,78.96) -- (327.38,108.51) ;
%Straight Lines [id:da9071630810871703] 
\draw [color={rgb, 255:red, 208; green, 2; blue, 27 }  ,draw opacity=1 ]   (289.94,77.08) -- (327.38,108.51) ;
%Straight Lines [id:da8123442242215725] 
\draw [color={rgb, 255:red, 126; green, 211; blue, 33 }  ,draw opacity=1 ]   (396.92,141.66) ;
\draw [shift={(396.92,141.66)}, rotate = 0] [color={rgb, 255:red, 126; green, 211; blue, 33 }  ,draw opacity=1 ][fill={rgb, 255:red, 126; green, 211; blue, 33 }  ,fill opacity=1 ][line width=0.75]      (0, 0) circle [x radius= 2.68, y radius= 2.68]   ;
%Straight Lines [id:da14148249742561092] 
\draw [color={rgb, 255:red, 126; green, 211; blue, 33 }  ,draw opacity=1 ]   (406.36,170.6) ;
\draw [shift={(406.36,170.6)}, rotate = 0] [color={rgb, 255:red, 126; green, 211; blue, 33 }  ,draw opacity=1 ][fill={rgb, 255:red, 126; green, 211; blue, 33 }  ,fill opacity=1 ][line width=0.75]      (0, 0) circle [x radius= 2.68, y radius= 2.68]   ;
%Straight Lines [id:da3265000257755081] 
\draw [color={rgb, 255:red, 126; green, 211; blue, 33 }  ,draw opacity=1 ]   (407.4,150.34) ;
\draw [shift={(407.4,150.34)}, rotate = 0] [color={rgb, 255:red, 126; green, 211; blue, 33 }  ,draw opacity=1 ][fill={rgb, 255:red, 126; green, 211; blue, 33 }  ,fill opacity=1 ][line width=0.75]      (0, 0) circle [x radius= 2.68, y radius= 2.68]   ;
%Straight Lines [id:da8569733860313942] 
\draw [color={rgb, 255:red, 126; green, 211; blue, 33 }  ,draw opacity=1 ]   (327.97,107) ;
\draw [shift={(327.97,107)}, rotate = 0] [color={rgb, 255:red, 126; green, 211; blue, 33 }  ,draw opacity=1 ][fill={rgb, 255:red, 126; green, 211; blue, 33 }  ,fill opacity=1 ][line width=0.75]      (0, 0) circle [x radius= 2.68, y radius= 2.68]   ;
%Straight Lines [id:da5919427159744691] 
\draw [color={rgb, 255:red, 126; green, 211; blue, 33 }  ,draw opacity=1 ]   (368.62,173.49) ;
\draw [shift={(368.62,173.49)}, rotate = 0] [color={rgb, 255:red, 126; green, 211; blue, 33 }  ,draw opacity=1 ][fill={rgb, 255:red, 126; green, 211; blue, 33 }  ,fill opacity=1 ][line width=0.75]      (0, 0) circle [x radius= 2.68, y radius= 2.68]   ;
%Straight Lines [id:da6023766609686475] 
\draw [color={rgb, 255:red, 126; green, 211; blue, 33 }  ,draw opacity=1 ]   (395.66,199.65) ;
\draw [shift={(395.66,199.65)}, rotate = 0] [color={rgb, 255:red, 126; green, 211; blue, 33 }  ,draw opacity=1 ][fill={rgb, 255:red, 126; green, 211; blue, 33 }  ,fill opacity=1 ][line width=0.75]      (0, 0) circle [x radius= 2.68, y radius= 2.68]   ;
%Straight Lines [id:da3167957286491426] 
\draw [color={rgb, 255:red, 126; green, 211; blue, 33 }  ,draw opacity=1 ]   (333.25,162.56) ;
\draw [shift={(333.25,162.56)}, rotate = 0] [color={rgb, 255:red, 126; green, 211; blue, 33 }  ,draw opacity=1 ][fill={rgb, 255:red, 126; green, 211; blue, 33 }  ,fill opacity=1 ][line width=0.75]      (0, 0) circle [x radius= 2.68, y radius= 2.68]   ;
%Straight Lines [id:da04143100115323506] 
\draw [color={rgb, 255:red, 126; green, 211; blue, 33 }  ,draw opacity=1 ]   (277.11,148.7) ;
\draw [shift={(277.11,148.7)}, rotate = 0] [color={rgb, 255:red, 126; green, 211; blue, 33 }  ,draw opacity=1 ][fill={rgb, 255:red, 126; green, 211; blue, 33 }  ,fill opacity=1 ][line width=0.75]      (0, 0) circle [x radius= 2.68, y radius= 2.68]   ;
%Straight Lines [id:da6778281784820415] 
\draw [color={rgb, 255:red, 126; green, 211; blue, 33 }  ,draw opacity=1 ]   (225,177.83) ;
\draw [shift={(225,177.83)}, rotate = 0] [color={rgb, 255:red, 126; green, 211; blue, 33 }  ,draw opacity=1 ][fill={rgb, 255:red, 126; green, 211; blue, 33 }  ,fill opacity=1 ][line width=0.75]      (0, 0) circle [x radius= 2.68, y radius= 2.68]   ;
%Straight Lines [id:da08423867699125032] 
\draw [color={rgb, 255:red, 126; green, 211; blue, 33 }  ,draw opacity=1 ]   (255.43,145.59) ;
\draw [shift={(255.43,145.59)}, rotate = 0] [color={rgb, 255:red, 126; green, 211; blue, 33 }  ,draw opacity=1 ][fill={rgb, 255:red, 126; green, 211; blue, 33 }  ,fill opacity=1 ][line width=0.75]      (0, 0) circle [x radius= 2.68, y radius= 2.68]   ;
%Straight Lines [id:da23536109681050588] 
\draw [color={rgb, 255:red, 126; green, 211; blue, 33 }  ,draw opacity=1 ]   (242.22,160.05) ;
\draw [shift={(242.22,160.05)}, rotate = 0] [color={rgb, 255:red, 126; green, 211; blue, 33 }  ,draw opacity=1 ][fill={rgb, 255:red, 126; green, 211; blue, 33 }  ,fill opacity=1 ][line width=0.75]      (0, 0) circle [x radius= 2.68, y radius= 2.68]   ;
%Straight Lines [id:da024754739626578814] 
\draw    (289.57,232.43) -- (312.33,232.23) ;
%Straight Lines [id:da7596081602491036] 
\draw [color={rgb, 255:red, 126; green, 211; blue, 33 }  ,draw opacity=1 ]   (300.48,195.19) ;
\draw [shift={(300.48,195.19)}, rotate = 0] [color={rgb, 255:red, 126; green, 211; blue, 33 }  ,draw opacity=1 ][fill={rgb, 255:red, 126; green, 211; blue, 33 }  ,fill opacity=1 ][line width=0.75]      (0, 0) circle [x radius= 2.68, y radius= 2.68]   ;
%Straight Lines [id:da14184884700455358] 
\draw [color={rgb, 255:red, 208; green, 2; blue, 27 }  ,draw opacity=1 ]   (289.11,111.18) -- (297.72,134.91) ;
%Straight Lines [id:da5605863110349906] 
\draw [color={rgb, 255:red, 126; green, 211; blue, 33 }  ,draw opacity=1 ]   (297.72,134.91) ;
\draw [shift={(297.72,134.91)}, rotate = 0] [color={rgb, 255:red, 126; green, 211; blue, 33 }  ,draw opacity=1 ][fill={rgb, 255:red, 126; green, 211; blue, 33 }  ,fill opacity=1 ][line width=0.75]      (0, 0) circle [x radius= 2.68, y radius= 2.68]   ;

% Text Node
\draw (143.11,67.78) node [anchor=north west][inner sep=0.75pt]    {$t$};
% Text Node
\draw (313.66,222.61) node [anchor=north west][inner sep=0.75pt]    {$\omega $};
% Text Node
\draw (421.91,128.71) node [anchor=north west][inner sep=0.75pt]  [rotate=-332.58]  {$\dotsc $};
% Text Node
\draw (291.16,89.19) node [anchor=north west][inner sep=0.75pt]  [rotate=-69.69]  {$\dotsc $};
% Text Node
\draw (347.75,137.19) node [anchor=north west][inner sep=0.75pt]  [rotate=-317.44]  {$\dotsc $};
% Text Node
\draw (192.17,136.54) node [anchor=north west][inner sep=0.75pt]  [rotate=-43.58]  {$\dotsc $};
% Text Node
\draw (331.06,67.37) node [anchor=north west][inner sep=0.75pt]  [rotate=-288.15]  {$\dotsc $};
% Text Node
\draw (242.84,95.36) node [anchor=north west][inner sep=0.75pt]  [rotate=-66.3]  {$\dotsc $};
% Text Node
\draw (423.28,207.6) node [anchor=north west][inner sep=0.75pt]  [rotate=-44.58]  {$\dotsc $};
% Text Node
\draw (431.49,156.78) node [anchor=north west][inner sep=0.75pt]  [rotate=-356.95]  {$\dotsc $};
% Text Node
\draw (384.16,118.94) node [anchor=north west][inner sep=0.75pt]  [rotate=-260.8]  {$\dotsc $};
% Text Node
\draw (202.16,101.63) node [anchor=north west][inner sep=0.75pt]  [rotate=-51.92]  {$\dotsc $};

\end{tikzpicture}

    \caption{  $\omega$ is the state on $\mathcal{A}$. The green dots denote actual events/actualities, the blue line denotes the real history and the collection of red lines denote the tree-like structure that indicate all the possibilities for how to branch.   }
    \label{Figure 1}
\end{figure}

 Figure \ref{Figure 1} is visually reminiscent of the branching tree often used to illustrate Everett's many‑worlds interpretation (MWI) of quantum mechanics \cite{everett1957relative}.  
However, the physical mechanisms underlying the ETH approach and MWI are fundamentally distinct. In Everett's `` relative state'' picture, all branches are equally real and the state never collapses; the tree  illustrates a superposition of decohered observer‑memory states and 
probabilities  are recovered only as a measure‑theoretic typicality. Importantly, in Everett's formulation, 
the mathematical decomposition of the total state into classical branches is not unique; the theory itself does not specify which basis of observables should define those branches. In contrast, in the ETH approach, the tree depicts a stochastic branching of all possible histories but 
not a coexistence of parallel worlds. As shown in Figure \ref{Figure 1}, the blue path denotes the single true history; the green dots mark actual, objective events which are triggered exactly via the mechanism discussed before.  The state is then updated via the Collapse Postulate (CP) with Born's rule probabilities as a fundamental law, and the basis of the event is not chosen by hand but is dynamically determined by the algebra of potentialities and the state at that moment. This update is  an ongoing and  irreversible process.
The usual deterministic Schr\"{o}dinger evolution of state arises when one averages over all possible histories. In fact, this ETH formalism formulated by Fr\"ohlich et al.  can be viewed as a concrete realization of the vision of the late Rudolf Haag for a completion of quantum mechanics founded on the concept of events\cite{haag1990fundamental}.

\par A limitation of the ETH formalism is that the stochastic branching process takes place continuously in time, making it challenging to handle such process in a well-controlled manner.  The relatively simple example/model i.e. the model of a very heavy atom coupled to the radiation field in the limit where the speed of light $c$ tends to $\infty$ as shown in \cite{frohlich2022time} that is used to illustrate ETH formalism makes use of the approximation that the time $t$ is discretized i.e. an $UV$ cutoff is introduced. In this paper, I also do not have any new insights on how to realistically model the ETH formalism when the stochastic branching process strictly takes place continuously in time. Nevertheless, I find one tentative explanation provided by Jürg Fröhlich in a lecture  \cite{AuthorYear} particularly insightful: let us consider a scenario that the system is in a state $\omega$  and the time $t$ is discretized with a very small time step $\delta  \ll 1 $, see Figure 2. 

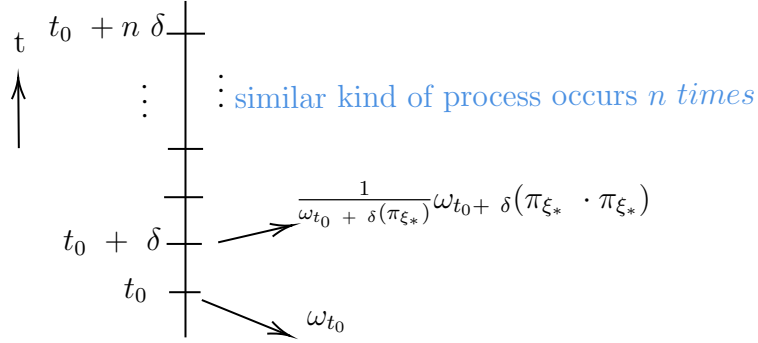
\begin{figure}[h!]
    \centering

\tikzset{every picture/.style={line width=0.75pt}} %set default line width to 0.75pt        

\begin{tikzpicture}[x=0.75pt,y=0.75pt,yscale=-1,xscale=1]
%uncomment if require: \path (0,237); %set diagram left start at 0, and has height of 237

%Straight Lines [id:da7525807534808368] 
\draw    (274.71,26.33) -- (274.71,195.18) ;
%Straight Lines [id:da07695099999170751] 
\draw    (191.07,100.21) -- (190.77,66.64) ;
\draw [shift={(190.75,64.64)}, rotate = 89.49] [color={rgb, 255:red, 0; green, 0; blue, 0 }  ][line width=0.75]    (10.93,-3.29) .. controls (6.95,-1.4) and (3.31,-0.3) .. (0,0) .. controls (3.31,0.3) and (6.95,1.4) .. (10.93,3.29)   ;
%Straight Lines [id:da00801095065720736] 
\draw    (266.5,172.51) -- (280.39,172.51) -- (282.28,172.51) ;
%Straight Lines [id:da42288574243288546] 
\draw    (265.24,148.28) -- (282.91,148.28) ;
%Straight Lines [id:da12107433320145533] 
\draw    (263.98,124.83) -- (283.54,124.83) ;
%Straight Lines [id:da5313701378292598] 
\draw    (265.87,100.6) -- (283.54,100.6) ;
%Straight Lines [id:da9198478994467086] 
\draw    (265.24,42.75) -- (284.49,42.75) ;
%Straight Lines [id:da39119033393217995] 
\draw    (282.91,176.42) -- (325.25,193.65) ;
\draw [shift={(327.1,194.4)}, rotate = 202.14] [color={rgb, 255:red, 0; green, 0; blue, 0 }  ][line width=0.75]    (10.93,-3.29) .. controls (6.95,-1.4) and (3.31,-0.3) .. (0,0) .. controls (3.31,0.3) and (6.95,1.4) .. (10.93,3.29)   ;
%Straight Lines [id:da8418591755184861] 
\draw    (291.12,147.5) -- (326.41,139.35) ;
\draw [shift={(328.36,138.9)}, rotate = 167] [color={rgb, 255:red, 0; green, 0; blue, 0 }  ][line width=0.75]    (10.93,-3.29) .. controls (6.95,-1.4) and (3.31,-0.3) .. (0,0) .. controls (3.31,0.3) and (6.95,1.4) .. (10.93,3.29)   ;

% Text Node
\draw (187.01,40.61) node [anchor=north west][inner sep=0.75pt]   [align=left] {t};
% Text Node
\draw (249.68,87.46) node [anchor=north west][inner sep=0.75pt]  [rotate=-269.74]  {$\cdots $};
% Text Node
\draw (334.35,180.42) node [anchor=north west][inner sep=0.75pt]    {$\omega _{t_{0}}$};
% Text Node
\draw (242.35,163.83) node [anchor=north west][inner sep=0.75pt]    {$t_{0} \ $};
% Text Node
\draw (211.86,138.81) node [anchor=north west][inner sep=0.75pt]    {$t_{0} \ +\ \delta $};
% Text Node
\draw (328.43,115.43) node [anchor=north west][inner sep=0.75pt]    {$\frac{1}{\omega _{t_{0} \ +\ \delta }( \pi _{\xi _{*}})} \omega _{t_{0} +\ \delta }( \pi _{\xi _{*}} \ \cdot \pi _{\xi _{*}})$};
% Text Node
\draw (204.65,31.33) node [anchor=north west][inner sep=0.75pt]    {$t_{0} \ +n\ \delta $};
% Text Node
\draw (297.29,60.62) node [anchor=north west][inner sep=0.75pt]  [rotate=-90.09]  {$\cdots $};
% Text Node
\draw (297.92,67.66) node [anchor=north west][inner sep=0.75pt]  [color={rgb, 255:red, 74; green, 144; blue, 226 }  ,opacity=1 ,rotate=-359.47] [align=left] {similar kind of process occurs $\displaystyle n\ times$};

\end{tikzpicture}  
    \caption{$\delta$ is the time step, $\pi_{\xi_*}$ denotes the projector with an overwhelming probability of occurence at the time $t_0 + \delta$}
    \label{Figure 2}
\end{figure}

When we march from time $t_0$ to $t_{0} + \delta$, suppose an actual event  $\{  \pi_{\xi} (t_0 + \delta) \vert \xi \in \mathcal{X}_{\omega_{t_0 + \delta}}\}$ occurs at time $t_0$ with only one of the projector $\pi_{\xi_{*}}$ which  will have overwhelming probability of occurrence , i.e.
\begin{align}
    \omega_{t_{0} + \delta} (\pi_{\xi_{*}}) = 1- \mathcal{O}(\alpha); \quad \omega_{t_0 + \delta} (\pi_{\xi})=\mathcal{O}(\alpha) \: \forall \xi \neq \xi_{*}  \label{overwhelming}
\end{align}
where $\alpha \ll 1$ is some very small number. After applying CP, we will find an overwhelming chance that the state $\omega_{t_{0}}$ evolves to the state 
$\frac{1}{\omega_{t_{0}+\delta}} \omega_{t_0 + \delta} (\pi_{\xi_*} \cdot \pi_{\xi_*})$ at the time $t_{0}+ \delta$ by the stochastic branching process due to (\ref{overwhelming}). Now, the difference between the state $\omega_{t_0 + \delta}$ and $\frac{1}{\omega_{t_{0} + \delta}(\pi_{\xi_{*}})} \omega_{t_0 + \delta} (\pi_{\xi_{*}} \cdot \pi_{\xi_{*}})$ is negligible. Therefore,  the effect due to the occurrence of the actual event at $t_{0} + \delta$ is almost imperceptible and the description using the evolution in the standard quantum mechanics is a perfectly valid approximation. Suppose that similar kind of process occurs $n$ times and for each time step, CP will lead to the choice of the projector will the biggest probability of occurrence. When $n$ is sufficiently large, the accumulative effect will now make the state $\omega_{t_{0}+ n \delta}$ to be significantly different from the state 
$\frac{1}{\omega_{t_{0} + \delta}(\pi_{\xi_{*},n})} \omega_{t_0 + \delta} (\pi_{\xi_{*},n} \cdot \pi_{\xi_{*},n})$. Here, $\pi_{\xi_{*},n} \in \mathcal{A}_{\geq t_{0} + n \delta}$ is the projector associated to the actual event at the time $t+ n \delta$ which has overwhelming probability of occurence. Therefore,  an observer whose reaction time is bigger than $n \delta$ will really see that an event occurs.

\subsection{Huygens' Principle}
The formalism introduced earlier can be generalized to relativistic quantum field theory \cite{frohlich2020relativistic}. This subsection outlines an implementation of the Principle of Diminishing Potentialities (PDP) while preserving Einstein causality in special relativity. Since this topic is tangential to our main focus, we will treat it briefly. Later, we will see that PDP holds in the large $N$ algebra of  $\mathcal{N}=4$ SYM theory through a mechanism different from that presented in \cite{frohlich2020relativistic}. This provides an affirmative answer to the question of whether  alternatives exist to the physical mechanism used in \cite{frohlich2020relativistic} (so called Huygens' Principle) in deriving the Principle of Diminishing Potentialities.

\par In \cite{frohlich2020relativistic}, an isolated system  $S$ described by a relativistic quantum theory with \textbf{massless modes} on Minkowski space $\mathbb{M}^{d}$ with even spacetime dimension $d=2n$ where $n$ can be any integer. For a given  region $\mathcal{O} \subset \mathbb{M}^d$, the algebra $\mathcal{A}(\mathcal{O})$ is generated by all bounded operators localized in $\mathcal{O}$ that represent physical quantities. The primary regions of interest $\mathcal{O}$ are taken to be forward or backward light-cones with apex in an arbitrary spacetime point $P \in \mathbb{M}^{d}$. We denote $\mathcal{F}_{P}$ to be the $*$ algebra generated by all bounded operators localized in the causal future of the spacetime point $P$. We assume that all these algebras are represented on a separable Hilbert space $\mathcal{H}_{S}$. We also define $\mathcal{A}_{P}$ to be the von Neumann algebra obtained by taking the weak closure of $\mathcal{F}_{P}$. The full algebra in our setting is

\begin{align}
    \mathcal{A} = \overline{\bigvee_{P \in \mathbb{M}^4} \mathcal{A}_{P}}
\end{align}

To develop an intuitive understanding of why the Principle of Diminishing Potentialities (PDP) should hold in this setting, we analyze the scenario illustrated in Figure 3.
\begin{figure}
    \centering \includegraphics[width=0.4\linewidth]{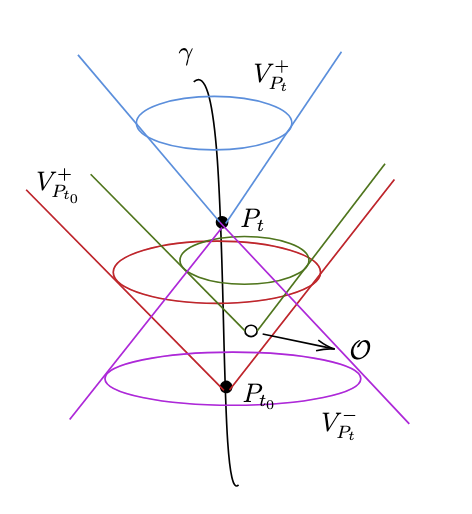}
    \caption{$\gamma$ is a timelike curve. $P_{t} $ is to the future of $P_{t_0}$. The spacetime region $\mathcal{O}$ is located in the future light cone of $P_{t_0}$
      and also in the past light cone of $P_t.$ The blue light cone $V^{+}_{P_t}$ is the future light cone of $P_t$ and the purple light cone $V^{-}_{P_{t}}$ is the past light cone of $P_t$. The red light cone $V^{+}_{P_{t_0}}$ is the future light cone of $P_{t_0}$} 
    \label{Figure 3}
\end{figure}

 We consider a timelike curve $\gamma$ parametrized by its elapsed proper time $\tau$. We denote the spacetime point $P_{t_{0}}$ to be the spacetime point on the curve $\gamma$ where $\tau = t_{0}$ and similarly for $P_{t}$ with $t> t_0$,  It is obvious that $\mathcal{A}_{P_{t}} \subseteq \mathcal{A}_{P_{t_0}}$ since
$P_{t}$ is to the future of $P_{t_0}$. Now, we want to argue that  PDP also holds, i.e. $\mathcal{A}_{P_t} \subsetneqq \mathcal{A}_{P_{t_0}}$. This is due to  that the isolated system $S$ we consider contains massless modes and the spacetime dimension is even, $d=2n$. In even dimensional spacetime, the on-shell  Green function $G (x)$ for a (free) massless field is only non-zero on the light cone $x^2 =0$. This implies that the commutator of two massless fields is only non-zero when the spacetime separation is lightlike, i.e. suppose our theory contains photon, then $[F_{\mu \nu} (x), F_{\alpha \beta}(y)]=0$ unless $(x-y)^2 =0$ where $F_{\mu \nu}$ is the electromagnetic field strngth tensor. This means that the on-shell massless particles do not propogate into the interior of forward light cones but only propagate along the surface of forwards light cones with apices in the point where the source of the massless particles $\mathcal{O}$ is located. Note that in odd dimensional spacetime $d= 2n+1$, the on-shell  Green function $G(x)$ for a (free) massless is non-zero not just on the light cone $x^2=0$ but also inside the light cone $x^2<0$ i.e. the massless particles also propogate into the interior of forwards light cones in odd spacetime dimension.  \par In Figure 3, the spacetime region $\mathcal{O}$  is located in the future light cone of $P_{t_0}$ and also in the past light cone of $P_{t}$. The asymptotic out-fields creating massless particles emanating from $\mathcal{O}$ which escapes to infinity  only propagate along the surface of the light cone and do not propagate into the interior of the light cone. In particular, these massless particles cannot propagate into the future light cone $V^{+}_{P_t}$ of $P_t$. Consequently, such asymptotic out fields will now commute with all operators in the algebra $\mathcal{A}_{P_t}$.  This phenomena is coined with the name \textbf{Huygens' Principle} and it only strictly holds in even spacetime dimension $d=2n$. This gives an intuitive picture why PDP holds: suppose the worldline of an observer $B$ is $\gamma$ , if  $B$ emits a light signal $\lambda$ at the proper time $\tau= t_0$ and $B$ is not paying attention until $\tau=t$, then $B$ will never be able to recover the signal $\lambda$ and the information encoded in $\lambda$ is lost forever to $B$ since $B$ cannot travel faster than the speed of light. Huygens' Principle is also formulated / proved as a mathematical theorem which can be stated in the following way \cite{bucholtz2013} \cite{frohlich2022time} :
 \begin{thm}
     In a relativistic Quantum Field Theory in spacetime dimension $d = 2n, n\geq 1$ with massless modes, the algebra $\mathcal{A}_{P_t}$  which represents physical quantities potentially measured in the causal future of $P_t$ is of Type $\RN{3}_1$, and the relative commutant $\mathcal{A}'_{P_t} \cap \mathcal{A}_{P_{t_0}}$ is also of Type $\RN{3}_1$, for $P_t$ is to the future of $P_{t_0}$.
 \end{thm}
  This theorem immediately implies the Principle of Diminishing Potentialities (PDP)
  \begin{align}
      \mathcal{A}_{P_t} \subsetneqq \mathcal{A}_{P_{t_0}}, \quad P_t \: \textit{is to the future of } P_{t_0}
  \end{align}

The notion of potential event, actual event and Collapse Postulate can be generalized to this relativistic setting in a rather straightforward manner and we do not give a review here. The new feature /axiom in the relativistic case that is not presented in the previous subsection is the so-called 
``
compatibility-locality  '' axiom. Given two spacetime points $P$ and $P'$ which are spacelike separated from each other, if events $\pi^P_{\xi}$ and $\pi^{P'}_{\eta}$ actually happen, then 
\begin{align}
    [\pi^P_{\xi}, \pi_{\eta}^{P'}]=0
\end{align}
This new property will ensure Einstein causality and enable a well-defined 
``history operator'' with no ordering ambiguity. For more details,
see \cite{frohlich2020relativistic}.

 \section{Emergent Large $N$ Algebra in SYM}

\subsection{Generalized Free Field Theory}

The model that we are interested is $\mathcal{N}=4$ supersymmetric Yang-Mills (SYM) theory with gauge group $SU(N)$ in spacetime dimension $d=4$. The full global symmetry group of this theory is $SU(2,2\vert 4)$, whose bosonic subgroup is $SO(2,4)\times SU(4)_R$. Here
$SO(2,4)$ is the conformal extension of the Lorentz group 
and $SU(4)_R$ is the $R$-symmetry group. To simplify the task of explaining the field content of $\mathcal{N}=4$ SYM, we perform a Wick rotation to go to the Euclidean signature(+,+,+,+)for the moment. In this signature, the Lorentz group is $SO(4) \cong (SU(2)_{L} \times SU(2)_{R})/\mathbb{Z}_2$, where the subscript $L/R$ denotes left and right, respectively. Specificially $SU(2)_{L}$ acts on left-handed Weyl spinor while $SU(2)_R$ acts on right-handed Weyl spinor. The field content of $\mathcal{N}=4 $ SYM consists of the following:

\begin{align}
    A_{\mu}, \: \Phi^{I}, \: \psi^{a}_{\alpha}, \:\bar{\psi}_{\dot{\alpha} a}   \label{field content}
\end{align}
where $I=1,..,6$; $a=1,..,4$ and $\alpha=\pm$. Since $SU(4)_R \cong SO(6)_R$, the six scalar fields $\Phi^{I}$ transforms under the vector representation of $SO(6)_{R}$ , with $I$ being the $SO(6)_{R}$ index. The left-handed Weyl spinor $\psi^{a}_{\alpha}$ transforms under the fundamental representation of $SU(4)_R$ and  the fundamental representation of $SU(2)_{L}$. Here, the superscript $a$ denotes the $SU(4)_{R}$ index and the subscript $\alpha$ denotes  $SU(2)_{L}$ index. Conversely,  the right-handed Weyl fermion $\bar{\psi}_{\dot{\alpha} a}$ transforms under the anti-fundamental representation of $SU(4)_R$ and the fundamental representation of $SU(2)_R$, with the subscript $\dot{\alpha}$ representing the $SU(2)_R$ index. Once we analytically continue back to the Lorentzian signature, the Weyl spinor $\psi^{a}_{\alpha} /\bar{\psi}_{\dot{\alpha}a}$ transforms under the fundamental $(\frac{1}{2},0)$ /conjugate  $(0, \frac{1}{2})$ representation of $SL(2, \mathbb{C})$. The field $A_{\mu}$ is the $SU(N)$ gauge field. All of these fields in (\ref{field content}) are in the adjoint representation of the gauge group $SU(N)$ and hence they are all matrix-valued fields i.e. under a general gauge transformation, we have

\begin{align}
    F_{\mu \nu} \rightarrow U F_{\mu \nu} U^{-1}, \quad \Phi^{I} \rightarrow U \Phi^I U^{-1}, \quad  \psi^{a}_{\alpha} \rightarrow U \psi^{a}_{\alpha} U^{-1}, \quad \bar{\psi}_{\dot{\alpha}a} \rightarrow U \bar{\psi}_{\dot{\alpha}a}U^{-1}  \label{observables}
\end{align}
where $F_{\mu \nu}= \partial_{\mu} A_{\nu}- \partial_{\nu} A_{\mu}- [A_{\mu}, A_{\nu}]$ is the field strength tensor and $U \in SU(N)$ is some unitary matrix. The physical observables are gauge invariant observables which are represented by single trace or multi-trace operators. A single trace operator can be constructed by first considering “word” composed by any arbitrary  matrix multiplication of  “letters” consisting of $F_{\mu \nu}, \Phi^I, \psi^{a}_{\alpha}, \bar{\psi}_{\dot{\alpha} a} $ and also derivatives acting  on  
them and then taking the trace over this 
“word”.

\par Our primary focus is on the algebraic structure of this gauge theory in the large $N$ limit. As such, our main analysis is more soft-core, aiming to understand the general structure rather than providing a precise or exact treatment. Therefore, we do not intend to write down the full Lagrangian of $\mathcal{N}=4$ SYM, which is rather complicated. All we need to know for our purpose is that $\mathcal{N}=4$ SYM is a matrix theory and schematically its Lagrangian takes the form

\begin{align}
     \mathcal{L} = \frac{1}{g^2} \Tr \bigg(\frac{1}{2} (\partial A)^{2} + A^{4} + \dots
     \bigg)  \label{lagrangian}
 \end{align}
where $g$ is the Yang-Mills coupling constant, $A$ collectively denotes the matrix valued fields. The trace $\Tr$ here is taken over the gauge group $SU(N)$. Given the matrix nature of the theory, we can systematically apply  't Hooft large $N$ expansion to analyze the $N$- dependence of  correlation functions. We briefly sketch the procedures.  To be able to see the $N$ and $g$ dependence of a particular connected Feynman graph with no external legs, we first view all connected vacuum bubbles (Feynman graphs with no external legs) as partition or triangulations of $2$-surfaces. In fact, for surface with sufficiently high genus $k$, any connected vacuum bubbles can  be drawn on that surface without line crossings, regardless of the complexity of the diagram.  The $N$ dependence of such a connected vacuum bubble is determined by the number of faces $F$ once we put this triangulation on that surface, each vertex will contribute to $g^{-2}$ and each edge (propagator) will contribute $g^{2}$ since the propagator is the inverse of the kinetic term in the Lagrangian.  Therefore, we have
 \begin{align}
     \textit{a connected vacuum diagram} \sim N^{F} (g^2)^{E-V}   \label{t Hooft expansion}
 \end{align}
 where $F$ is the number of faces, $E$ is the number of propagators and $V$ is the number of vertices. We aim to take the large $N$ limit and what we want is that the theory is soluble when $N \rightarrow \infty$. However, (\ref{t Hooft expansion}) does not give a well-defined large $N$ limit in a well-controlled manner, i.e. there is no universal  rule  governing
the $N$ dependence of these vacuum diagrams. To ensure a controlled large $N$ limit, we might take $g \rightarrow 0$ while taking $N \rightarrow \infty$. This reorganizes (\ref{t Hooft expansion}) into

 \begin{align}
     \textit{a connected vacuum diagram} \sim (g^{2}N)^{E-V} N^{F-E+V} \sim \lambda ^{E-V} N^{\chi} \sim \lambda^{E-V} N^{2-2k} \label{t Hooft limit}
 \end{align}

 where $\lambda= g^{2}N$  is the 't Hooft coupling and  $\chi= F-E-V=2-2k$ is the Euler characteristic of the surface (with genus $k$).
Such a manipulation is called the 't Hooft limit, where $\lambda$ remains finite as $g \rightarrow 0$ and $N \rightarrow \infty$. We can see that  the 't Hooft limit gives a sensible large $N$ behaviour. It follows that the logarithm of the partition function $Z$ which is represented by the sum of all these connected vacuum diagrams will also have a sensible $N$-dependence and can be written as genus expansion

\begin{align}
     \log Z = \sum_{k} N^{2-2k} f_{k} (\lambda) \label{genus expansion}
 \end{align}
 where $f_{k} (\lambda)$ is some function which depends on $\lambda$. The diagrams that give the contribution to $k=0$ in the expansion (\ref{genus expansion}) are called planar diagrams and the diagrams that contribute to $k \geq 1$ in (\ref{genus expansion}) are called non-planar diagrams.  Since  the partition function $Z$ has a sensible $N$-dependence, it follows that  generic $n$ point functions will also be under control in the large $N$ limit. To be able to see that, we first note that 
the partition function for any physically sensible Lagrangian that takes the form $\mathcal{L}= N \Tr[...] $  can always be expressed in the form of (\ref{t Hooft limit}). Now, to see how connected $n$ point functions behave, we first introduce another partition function $Z'$ taking into account sources $J_i$

\begin{align}
    Z'[J_{i}]= \int DA_{\mu}\int D \hat{\Phi} \exp{i \int d^{d}x \mathcal{L}[A_{\mu}, \hat{\Phi}] + iN \int d^dx  J_{i} \mathcal{O}_{i}} = \int DA_{\mu} D\Phi \exp{i \int d^{d}x N \Tr[....]} \label{alt lagrangian}
 \end{align}
where $\hat{\Phi}$ denotes a collection of matrix-valued fields other than gauge field $A_{\mu}$ and $\mathcal{O}_i= \Tr[O_i]$ are some gauge invariant single trace local operators. So, we know that  $\log Z'[J_{i}]= \sum_{k} N^{2-2k} g_{k} (\lambda)$ for some functions $g_{k} (\lambda)$. To obtain connected $n$- point functions from  (\ref{alt lagrangian}), we take $n$ functional derivatives of $\log Z'[J_i]$ with respect to sources $J_{i}$ 

\begin{align}
     \langle \mathcal{O}_{1} (x_{1}) \mathcal{O}_{2} (x_{2})\cdot\cdot\cdot  \mathcal{O}_{n} (x_{n}) \rangle^{c} = \frac{\delta}{\delta J_{1}} \cdot\cdot\cdot \frac{\delta}{\delta J_{n}} \log Z' [J_{i}] \Bigr\rvert_{J_{i} = 0} \frac{1}{(iN)^{n}} = \sum_{k} N^{2-2k-n} h_{k} (\lambda)  \label{n correlators}
     \end{align}
where the superscript $c$ denotes the connected correlators.  In the first equality, we need to include a factor $(-N)^{-n}$ because we have manually put in a factor $N$ in (\ref{alt lagrangian}). Therefore,  we see that the $n$-point functions $\sim N^{2-n}$ up to the leading order in $1/N$ expansion i.e. only consider the contribution from planar diagrams. Note that these $n-$point functions are evaluated in the ground state of a prescribed Hamiltonian. Since we want to consider the situation that the theory is formulated on the conformal compactification of Minkowski space i.e.   
$\mathbb{R} \times S^3$ which is the conformal bondary of global $Ads_5$, we take the Hamiltonian $H$ to be the dilatation operator in the radial quantization. The natural preferred time $t$ is then the canonical conjugate of the dilatation operator and the corresponding ground state $\ket{\Omega}$ is the $Ads$ vacuum.

\par We may also consider the thermal state i.e. Gibbs state $\rho_{\beta}=\frac{1}{\Tr e^{e^{- \beta H}}} e^{- \beta H}$ at fome finite temperature $\frac{1}{\beta}$ where the Hamiltonian is still chosen to be the dilatation operator that generate time translation in $\mathbb{R} \times S^3$. The $N$-dependence of the correlation functions with the preferred state $\ket{\Omega}$ will be inherited to thermal correlators where the preferred state is the Gibbs state
\begin{align}
    \langle \mathcal{O}_1 \mathcal{O}_2 \cdots \mathcal{O}_n \rangle^c_{\beta}= \sum_{k} N^{2-2k-n} p_n(k, \beta) \label{thermal correlator}
\end{align}
but now  coefficients $p_n(k, \beta)$ in the genus expansion also depends on the temperature. This is because the way to prepare a thermal system with temperature $\frac{1}{\beta}$ associated to a Hamiltonian in field theory is to first perform a Wick rotation $t= -i \tau$ to go to Euclidean field theory and then we periodically identify the Euclidean time with the inverse temperature i.e. $\tau \sim \tau + \beta$. Note that in the large $N$ limit, i.e. $N \rightarrow \infty$, there will be a first order phase transition. When the temperature is below a threshold called Hawking Page temperature $\frac{1}{\beta_{HP}}$i.e. $\beta > \beta_{HP}$, the free energy $F = \frac{1}{\beta} \log Z\sim \mathcal{O}(1)$ is of order $1$. Above  Hawking Page temperature $\beta < \beta_{HP}$, the free energy $F \sim \mathcal{O}(N^2)$ which is of order $N^2$. The low temperature phase corresponds to the thermal $Ads$ phase while the high temperature phase corresponds to the black hole phase. So, we know that the $N^2$ term for $\log Z$ in (\ref{thermal correlator}), i.e. $p_0 (0, \beta)$ is basically contributed by black hole states (sector with high energy $E \sim N^2$ and high density of states $D(E) \sim e^{N^2}$) in the large $N$ limit because the Boltzmann suppression factor $e^{-\beta E}$ is dominated by the high density of states $ D(E) \sim e^{N^2}$ when $\beta$ is small.  When $\beta > \beta_{HP}$, the contribution to $N^2$ term from the black hole phase  is exponentially suppressed  and the dominant phase will be thermal $Ads$ phase and $\log Z \sim \mathcal{O}(1)$. One immediate explanation from Ads/CFT is that the leading $N^2$ contribution to $\log Z$ is given by the on shell evaluation of the gravitational action $S_{grav}$ given a reference value which is the  pure $Ads$ evaluation of the gravitational action $S_{grav} \vert_{pure \: Ads}$. Since thermal $Ads$ only differs with pure $Ads$ in global structure, we know that $S_{grav}\vert_{thermal  \: Ads}- S_{grav}\vert_{pure \: Ads}=0$ and then the leading contribution will be $p_{0} (1, \beta)$. In particular, at low temperature phase and in  $N \rightarrow \infty$ limit, it is shown in \cite{brigante2006inheritance} by a perturbative analysis that the connected thermal correlators at finite temperature $\frac{1}{\beta}$ can be expressed as \footnote{(\ref{inheritance formula}) assumes that all $\mathcal{O}_i$ are bosonic. If an $\mathcal{O}_i$ is fermionic, then  an additional factor $(-1)^{m_i}$ needs to be included.}
\begin{align}
    G^{c}_{\beta} (\tau_1, ..., \tau_n) = \sum_{m_1, \cdots , m_n = - \infty}^{\infty} G_{0}^{c} (\tau_1-m_1 \beta, ..., \tau_n- m_n \beta)  \label{inheritance formula}
\end{align}
where $G^{c}_{\beta} (\tau_1 \cdots \tau_n)$ denotes Euclidean $n$ point connected correlation function at finite temperature $\beta^{-1}$ while $G_{0}^c (\tau_1, ..., \tau_n)$ denotes Euclidean  $n$ point connected correlation function at zero temperature, i.e. in the vacuum state. This means that at  low temperature phase $\beta > \beta_{HP}$ and in the large $N$ limit, the  thermal correlation function at finite temperature is just inherited  from the zero temperature correlation function by   summing
over images of each operator in the Euclidean time direction. Note that (\ref{inheritance formula}) does not hold at  high temperature phase $\beta < \beta_{HP}$ because $\Tr U \neq 0$ when $\beta < \beta_{HP}$ i.e. deconfined phase. Here, $U$ is the Wilson line of the gauge field wound around the direction of the thermal circle $\tau$ direction. For more details, see \cite{brigante2006inheritance}
.

\par Therefore, we see that in the planar limit (only consider diagrams with  genus $k=0$), the connected $n$-point function in the $Ads$ vacuum $\ket{\Omega}$ or in the thermal state $\rho_{\beta}$ scales as
\begin{align}
  \langle  \mathcal{O}_1 \cdots \mathcal{O}_n \rangle^{c}_{\Omega , \beta} \sim N^{2-n} \label{planar scaling}
\end{align}
In particular, one point function scales with $N$, two point connected function is of order $1$ and $n$-point functions with $n\geq 3$ vanishes in the large $N$ limit

\begin{align}
\langle\mathcal{O} \rangle_{\Omega,\beta}^c \sim N  , \quad \langle\mathcal{O}_1 \mathcal{O}_2 \rangle^{c}_{\Omega, \beta} \sim \mathcal{O}(1), \quad \langle \mathcal{O}_1 \mathcal{O}_2 \cdots \mathcal{O}_{n\geq 3} \rangle_{\Omega, \beta}^c \rightarrow0 \quad \textit{as} \: N \rightarrow \infty
\end{align}
We would like to take the large $N$ limit of the gauge theory. However, naively taking large $N$ limit of the gauge theory is problematic because the expectation value of single trace operators  generally diverge. To address this, Liu and Leutheusser \cite{leutheusser2023emergent}\cite{leutheusser2022subalgebra} introduce a state (background) dependent  `` normal ordering'' prescription for single trace operators. The `` normal ordered '' single trace operators are 
\begin{align}
    \normord{\mathcal{O}} \: := \mathcal{O}- \langle \mathcal{O} \rangle_{\beta}   \label{normal order}
\end{align}
Here, we consider the thermal state $\rho_{\beta}$ to illustrate the idea. Similar procedures work for $Ads$ vacuum $\ket{\Omega}$ exactly in the same way.  By definition, the expectation value of these `` normal-ordered'' single trace operators vanish
\begin{align}
    \langle\normord{ \mathcal{O} } \rangle_{\beta}=0 \label{one point vanish}
\end{align}
Furthermore, since $\langle \mathcal{O} \rangle_{\beta} \sim N$, we can extract out the $N$ dependence by rewriting  the one point function as $\langle\mathcal{O} \rangle_{\beta}= \tilde{\langle \mathcal{O} \rangle}_{\beta} N$ where $\tilde{\langle \mathcal{O} \rangle}_{\beta} \sim \mathcal{O}(1)$. Then, the normal ordered operator can be written as (for $\mathcal{O}= \Tr O$ where $O$ can be any arbitrary ``word'')
\begin{align}
    \normord{\mathcal{O}} \: = \Tr[O -\tilde{\langle \mathcal{O} \rangle}_{\beta} \mathbb{1}] 
\end{align}
where $O- \tilde{\langle \mathcal{O} \rangle}_{\beta} \mathbb{1}$ has no explicit $N$ dependence. Moreover, when we couple a source $J$ to the ``normal-ordered'' operator $\mathcal{O} $, the partition function becomes 
\begin{align}
    \tilde{Z} [J]= \int DA_{\mu} D \hat{\Phi} \exp[-S + \int J(\mathcal{O}- \langle \mathcal{O} \rangle_{\beta})]=Z[J]\exp [-\int J \langle \mathcal{O} \rangle_{\beta} ]
\end{align}
The connected generating functional $W[J]= \log \tilde{Z}[J]$ therefore shifts by a term linear in $J$:
\begin{align}
    W'[J]= W[J] - \int J \langle \mathcal{O} \rangle_{\beta}
\end{align}
Since the connected correlation fucntions are obtained from functional derivatives of $W$, i.e,  $\langle \mathcal{O}_1 \cdots  \mathcal{O}_{n} \rangle_{\beta}^{c}= \frac{\delta^n W}{\delta J_1 \cdots \delta J_n}$, so the linear shifts contributes only to the first derivative which leads to (\ref{one point vanish}). For $n \geq 2$, the shift vanishes upon taking derivatives:
\begin{align}
    \frac{\delta^n W'}{\delta J_1 \cdots \delta J_n}= \frac{\delta ^n W}{\delta J_1 \cdots \delta J_n}  \quad \textit{ for } n \geq 2
\end{align}
Therefore, the rules (\ref{genus expansion}) and  (\ref{planar scaling}) still hold  for these normal ordered operators (\ref{normal order}). It follows that $n$ point functions exhibit Gaussian structure in the sense that Wick theorem holds, i.e.

\begin{align}
    \langle \normord{\mathcal{O}_1} \cdots \normord{\mathcal{O}_n} \rangle_{\beta}=    \left\{ \begin{array}{rcl}
         0 \: \: \: \: & \mbox{for}
         & n \: odd \\  \sum_{\textit{ \{$i_1 \cdots i_n$ \} } \in S_n} \langle  \mathcal{O}_{i_1} \mathcal{O}_{i_2}\rangle_{\beta} \cdots \langle \mathcal{O}_{i_{n-1}} \mathcal{O}_{i_n}  \rangle_{\beta} & \mbox{for} & n  \: even 
                \end{array}\right.
                \end{align}
where $S_n$ is the permutation group for $n$ letters. For an explanation, see Figure 4.

\begin{figure}[ht]
    \centering
    \includegraphics[width=0.6\linewidth]{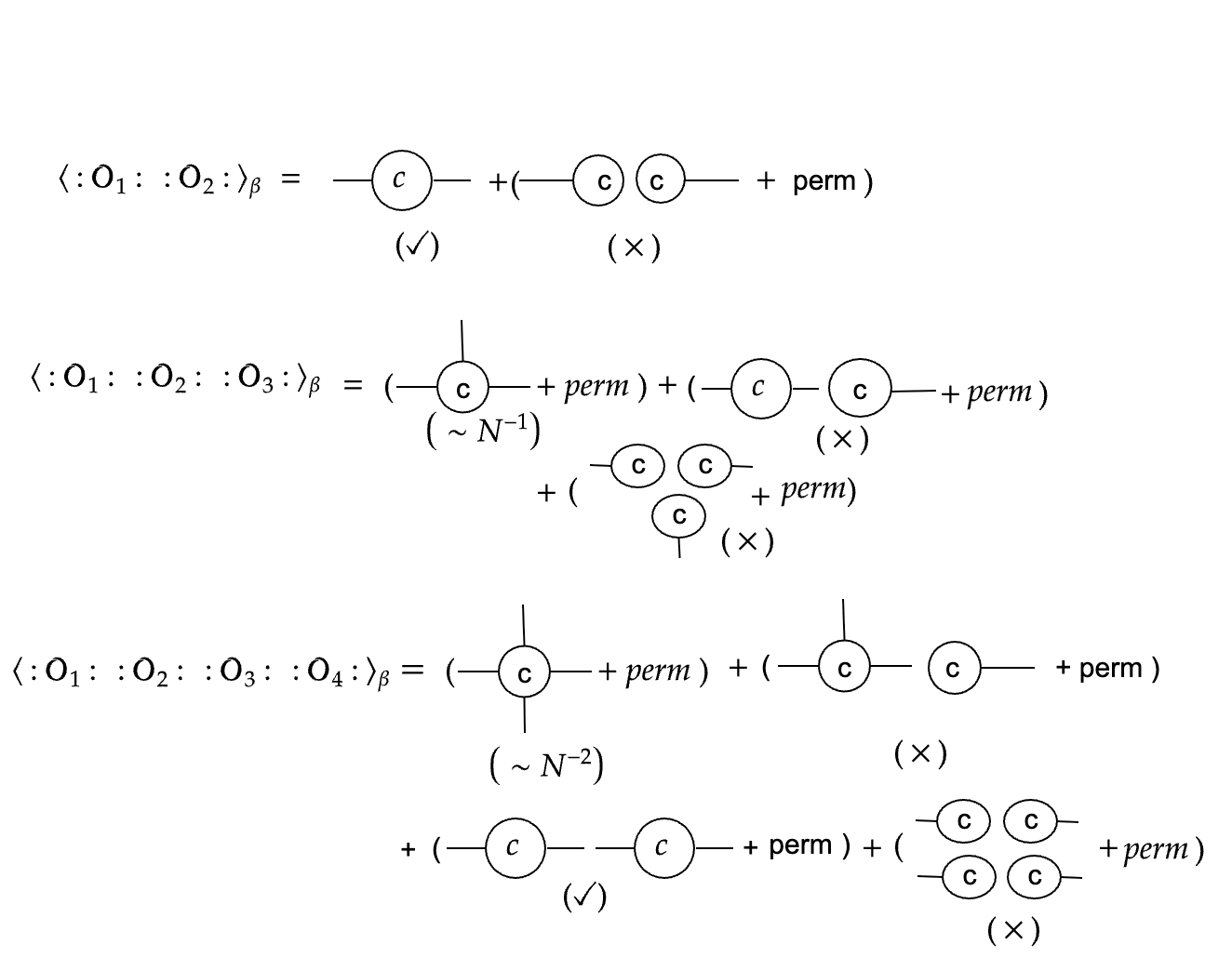}
    \caption{$c$ refers to connected diagram and $perm$ denotes all possible permutations of the labeling of external legs. A tick $\checkmark$ means the class of diagram survives and is well defined. A cross $\times$ means the class of diagram vanishes since it is the product of some diagrams with one point function. All connected $n \geq 3$ function vanishes because they scale with negative power of $N$}
    \label{Figure 4}
\end{figure}

 Hence, we observe that (subtracted) single-trace operators behave as free fields in the large-$N$ limit. Note that the true semiclassical limit corresponds to $N \rightarrow \infty$ and $\lambda \rightarrow \infty$. In fact, this large $N$  theory that we are studying is \textbf{generalized free field theory}.

\subsection{Subregion-Subalgebra Duality}

As demonstrated by Liu and Leutheusser \cite{leutheusser2023emergent}, there is a very rich algebraic structure for the generalized free field theory we consider in the last subsection. Their analysis focus on a system comprising two  entangled $\mathcal{N}=4$ SYM theories. When the gauge group  rank $N$ is  finite, the Hilbert space of this quantum system is simply a tensor product $\mathcal{H}_{L} \otimes \mathcal{H}_{R}$ where $L/R$ denotes  left/right systems, respectively. They furthur consider the canonical purification of Gibbs state to obtain the thermofield double state $\omega_{TFD}$
 \begin{align}
     \omega_{TFD} (\cdot)= \frac{1}{Z_{\beta}} \Tr (e^{- \beta H} \cdot)= \bra{TFD} \cdot \ket{TFD}
 \end{align}
where $Z_{\beta}$ is the thermal partition function. For finite $N$, the TFD state can be explicitly expressed in Dirac bra-ket notation as:
\begin{align}
    \ket{TFD}=\frac{1}{\sqrt{Z_{\beta}}}\sum_{i}e^{- \beta E_i/2} \ket{i}_{R} \ket{i}_L \label{finite N thermofield double}
\end{align}
Here, $\ket{i}$ represents the energy eigenstates and $E_i$ are the corresponding eigenvalues. However, subtleties arise in the large $N$ limit. As previously discussed, state-dependent normal ordering prescription (or subtraction scheme) needs to be applied to single trace operators to obtain a well-defined large $N$ theory which is the generalized free field theory. This serves as a concrete example of how large $N$ theories are inherently state (background) dependent theory. In other words, physical observables are only well defined with respect to a specific state or background and the large $N$ Hilbert space is constructed solely from these regulated observables and the preferred state without any additional inputs. In fact, there is a very insightful discussion in \cite{wittenbackground_independence} which argues that the  background    independent theory for $\mathcal{N}=4$ SYM as a quantum theory is only valid for  finite $N$ theory. If we try to define large $N$ theory which is state independent, we need to divide single trace operators by an extra factor $N$, i.e.
\begin{align}
   \tilde{ \mathcal{O}}= \frac{1}{N} \Tr O
   \label{background independent formula}
\end{align}
ensuring that correlation functions exhibit proper 
$N$ scaling.  When $N = \infty$, the algebra generated by (\ref{background independent formula}) denoted as $\mathcal{A}$ is commutative ( scaling as  $N^{-2}$) and it can be equipped with an antisymmetric bracket satisfying the Jacobi identity, forming a Poisson algebra. The bulk interpretation of this Poisson algebra is that as $G_{N} \rightarrow 0$ the bulk dual of the theory has many classical phase spaces consisting of classical solutions of the relevant gravity. Then, one can deform this Poisson algebra $\mathcal{A}$ order by order in $N^{-2}$ by a certain quantization procedure like deformation quantization to obtain a non-commutative but associative algebra denoted as $\mathcal{A}_{\frac{1}{N^2}}$. It is argued in \cite{wittenbackground_independence}  that $\mathcal{A}_{\frac{1}{N^2}}$ does not have any  distinguished Hilbert space representation within $\frac{1}{N}$ expansion. However, when we expand around a point in the classical phase space, it is possible to have a meaningful Hilbert space representation of $\mathcal{A}_{\frac{1}{N^2}}$. A simple analogous example of deformation quantization of the $2$ sphere is also illustrated there. The  discussion there sheds light on the fundamental nature of large $N$ theories.  For further details, see section $6$ in \cite{wittenbackground_independence}.
\par To construct the large $N$ Hilbert space purely from the operator algebra on the right asymptotic boundary $\mathcal{A}_{R}=\{ \normord{\mathcal{O}}\}$ and thermofield double state $\omega_{TFD}$, we resort to Gelfand-Naimark-Segal (GNS) construction. First, We  check that $\omega_{TFD}(A^{*})= \overline{\omega_{TFD}(A)}$. This can be seen from the following:
\begin{align}
     0 \leq \omega_{TFD} ((A + \mathds{1}) (A^{*}+\mathds{1})) = \omega_{TFD} (AA^{*}) + 1 + \omega_{TFD} (A) + \omega_{TFD}(A^{*})
 \end{align}
Since $\omega_{TFD} (B B^{*}) \geq 0$ (which is real) for all $B \in \mathcal{A}_{R}$ and $\omega_{TFD}(\mathbb{1})=1$, we know that
$\Im{ \omega_{TFD} (A)}= -\Im{\omega_{TFD}(A^{*})}$. Similar argument also tells that
\begin{align*}
    0 \leq \omega_{TFD} ((iA + \mathds{1}) (-i A^{*}+ \mathds{1})) = \omega_{TFD}(AA^{*}) +1 + i \omega_{TFD}(A) -i \omega_{TFD} (A^{*})
\end{align*}
This implies that $\omega_{TFD}(A)-\omega_{TFD}(A^{*})$ cannot have real part. Overall, we know that $\omega_{TFD}(A^{*})= \overline{\omega_{TFD}(A)}$.

Given the state $\omega_{TFD}$, there exists a canonical construction of a Hilbert space representation $(\pi_{{TFD}}, \mathcal{H}_{{TFD}})$.  The map $\pi_{{TFD}}$ is a linear map from $\mathcal{A}_{R}$ into operators on a dense invariant domain $D \subset \mathcal{H}_{TFD}$ (i.e. $\pi_{TFD}(\mathcal{A}_{R})D$ is dense in $\mathcal{H}_{TFD}$) such that $   
\pi_{TFD} (AB) = \pi_{TFD} (A) \pi_{TFD} (B)$ and $\pi_{TFD} (A^{*})= \pi_{TFD} (A)^{*}$.   The Hilbert space $\mathcal{H}_{TFD}$ contains the thermofield double state $\ket{TFD}$ which is a cyclic vector  (cyclicity means that $\pi_{\omega}(\mathcal{A})\Omega_{\omega}$ is dense in $\mathcal{H}_{\omega}$) such that
\begin{align*}
    \omega_{TFD} (A) = \bra{TFD} \pi_{TFD} (A) \ket{TFD}
\end{align*}

and $\pi_{TFD}(\mathcal{A}_R)\ket{TFD}$ is the dense invariant domain $D$ above.
This construction is unique up to unitary equivalence, i.e. any other cyclic representation $(\mathcal{H}'_{TFD}, \ket{TFD'}, \pi'_{TFD})$ is related to $(\mathcal{H}_{TFD},\ket{TFD}, \pi_{TFD})$ via a unitary  $U : \mathcal{H}_{TFD} \rightarrow \mathcal{H}'_{TFD}$ such that $\ket{TFD'}=U \ket{TFD}$ and $U \pi_{TFD} (A) U^{-1} = \pi'_{TFD} (A)$.

\par The construction of $\mathcal{H}_{TFD}$ starts with the complex vector space structure of our algebra $\mathcal{A}_R$. The state $\omega_{TFD}$ provides a map from $\mathcal{A}_{R}$ into $\mathbb{C}$ and we use it to define the scalar product
\begin{align}
    \langle A \vert B \rangle := \omega_{TFD} (A^{*}B)   \label{inner product}
    \end{align}  
The null space $\mathcal{N}$ of the state $\omega_{TFD}$ is called Gelfand ideal:
\begin{align*}
    \mathcal{N}= \{ A \in \mathcal{A}_R \vert \omega_{TFD} (A^{*}A)=0 \}
\end{align*}
$\mathcal{N}$ is a vector space. For later purpose, we note that the property
\begin{align}
    \langle{A}\vert{B} \rangle=0 \quad \forall B \in \mathcal{A}_R,  
 \quad \forall A \in \mathcal{N} \label{null property}
\end{align}
This follows from
\begin{align}
    0 \leq \vert \omega_{TFD}(A^{*}B) \vert^{2} = \vert \langle{A}\vert{B}\rangle \vert^{2} \leq \vert \vert A \vert \vert^{2} \vert \vert B \vert \vert^{2} = 0
\end{align}
where we have used Cauchy-Schwarz inequality in the third inequality and in the last equality we use that $A \in \mathcal{N}$. In order to obtain a Hilbert space, we quotient out our algebra $\mathcal{A}_R$ with the Gelfand ideal $\mathcal{N}$ and complete it using the norm induced by
$(\ref{inner product})$. So, the Hilbert space $\mathcal{H}_{TFD}$ is
\begin{align}
    \mathcal{H}_{TFD} = \overline{\mathcal{A}_R/\mathcal{N}} = \overline{D}
\end{align}
We denote the equivalence class $A + \mathcal{N}$ of $A \in \mathcal{A}_R$ by $[A]$ in $\mathcal{H}_{TFD}$ and the dense invariant subspace $D$ is
\begin{align}
    D= \{ [A]\in \mathcal{H}_{TFD} \vert A \in \mathcal{A}_R \}
\end{align}
We also want to define the representation of $\mathcal{A}_R$ on $\mathcal{H}_{  TFD}$, we simply use the algebra structure of $\mathcal{A}_R$ and we define
\begin{align}
    \pi_{TFD} (A) [B] := [AB]  \quad \textit{on $D$}
\end{align}
We can check that this is well defined. Note that $\mathcal{N}$ is a left ideal on $\mathcal{A}_R$, i.e. $AB \in \mathcal{N}$ for all $A \in \mathcal{A}_R$ and $B \in \mathcal{N}$. This is due to
\begin{align}
    \vert \vert AB \vert \vert ^{2} = \omega_{TFD} ((AB)^{*} AB) = \omega_{TFD} (B^{*} A^{*} AB) = \langle{B} \ket{A^{*}AB} =0
\end{align}
We have used (\ref{null property}) is the last equality. So, this action $\pi_{TFD}$ defines a representation on $D$ with $\pi_{TFD} (AB)= \pi_{TFD}(A) \pi_{TFD}(B)$ and $\pi_{TFD}(A^{*})= \pi_{TFD}(A)^{*}$. All these properties can be checked easily
\begin{align}
    \pi_{TFD}[A]\pi_{TFD}[B] [C]=[ABC]= \pi_{TFD}(AB)[C]
\end{align}
and
\begin{align}
    \bra{B}\pi_{TFD}(A)^{*} \ket{C}= \braket{AB}{C}=\omega_{TFD}(B^{*}A^{*}C)= \langle {B}  \ket{{A^{*}C}} = \langle{B}\ket{\pi_{TFD}(A^{*})C}
\end{align}
We also define the thermofield double state $\ket{TFD}$ in $\mathcal{H}_{TFD}$ to be $\ket{TFD}:= [\mathds{1}]$. This definition  gives the desired property  $\bra{TFD} \pi_{TFD}(A) \ket{TFD}=\langle{\mathbb{1}}\ket{{A}}= \omega_{TFD}(A)$
and cyclicity of $\ket{TFD}$ follows from $\pi_{TFD}(A) \ket{TFD}=[A]$.
\par To see the uniqueness of the construction up to unitary equivalence, we assume another cyclic representation $(\mathcal{H}'_{TFD}, \ket{TFD'}, \pi'_{TFD})$.  We define a map $U: D \rightarrow \mathcal{H}'_{TFD}$ by $U \pi_{TFD}(A) \ket{TFD}= \pi'_{TFD}(A) \ket{TFD'}$. We can check that the map $U$ is an isometry:
\begin{align*}
    \langle{U \pi_{TFD}(A) TFD} \ket{{U \pi_{TFD}(B) }TFD}_{\mathcal{H}'_{TFD}}&=\langle{\pi'_{TFD} (A)TFD'} \ket{{\pi'_{TFD} (B)TFD'}} \\
    &= \omega_{TFD}(A^{*}B)\\
    &=\langle{\pi_{TFD}(A) TFD}\ket{{\pi_{TFD}(B) TFD}}_{\mathcal{H}_{TFD}}
\end{align*}
So, $U$ is bounded. By bounded linear transformation theorem, $U: D \rightarrow U(D)$ can be extended  to $U: \mathcal{H}_{TFD}\rightarrow U(\mathcal{H}_{TFD}) \subset \mathcal{H}'_{TFD}$. Furthermore, since $U$ is isometry, the range of $U$ is closed. Due to cyclicity of $\ket{TFD'}$, we also know the range of $U$ is dense. So, the range of $U$ is $\mathcal{H}'_{TFD}$. Combining all these facts, we know that $U$ is invertible and therefore $U$ is unitary.

\par Now, we define the physical right algebra $\mathcal{M}_{R}$ to be
\begin{align}
    \mathcal{M}_{R}= \pi_{TFD} (\mathcal{A}_{R})''
\end{align}
where we take the double commutant so that the algebra $\mathcal{M}_{R}$ is a von Neumann algebra. Similar construction can also be used to construct the left algebra which is the commutant of the right algebra i.e. $\mathcal{M}_{R}'= \mathcal{M}_L$. Moreover, $\mathcal{M}_{R} \cap \mathcal{M}_L = \mathbb{C} \mathbb{1}$, so $\mathcal{M}_R$ and $\mathcal{M}_L$ are factor. Below Hawking Page temperature $\beta > \beta_{HP}$,  relations (\ref{inheritance formula}) holds and thus the Euclidean thermal 2 point function for scalar operators can be written as (for positive frequency $\omega$)
\begin{align}
    G_E (\tau, \omega) 
    &= \sum_{m= - \infty}^{\infty}   \rho_{0} (\omega)  e^{- \omega \vert \tau- m \beta \vert}
    = \rho_{0} (\omega) \bigg(  \frac{e^{- \omega \tau}+ e^{\omega (\tau- \beta)}}{1- e^{- \omega \beta}}  \bigg) \label{thermal spectrum alternative}
\end{align}
where $\rho_{0}(\omega)$ is the Lehmann spectral function at zero temperature.  For simplicity, we only keep track of the temporal coordinate and we suppressed all spatial coordinates. On the other hand, using Matsubara frequency representation,  the Euclidean thermal $2$ point function can be expressed as
\begin{align}
    G_{E} (\tau, \omega)=   \rho(\omega) \bigg(  \frac{e^{- \omega \tau}+ e^{\omega (\tau- \beta)}}{1- e^{- \omega \beta}}  \bigg)    
    \label{thermal spectrum pure}
\end{align}
where    $\bigg(  \frac{e^{- \omega \tau}+ e^{\omega (\tau- \beta)}}{1- e^{- \omega \beta}}  \bigg)$ is the heat kernel,  $\rho (\omega)$ is the spectral function at finite temperature $\frac{1}{\beta}$. After we analytically continue back to Lorentzian signature and compare (\ref{thermal spectrum alternative}) with (\ref{thermal spectrum pure}), we know that
\begin{align}
    \rho (\omega)=  \theta (\omega) \rho_{0} (\omega) - \theta (- \omega) \rho_{0} (- \omega)
\end{align}
The second term is included because  $\rho (-\omega)= - \rho (\omega)$ for bosonic operators. This can be seen easily from the following formula
\begin{align}
    \rho (\omega)= (1- e^{- \beta \omega}) \sum_{m.n} \delta (\omega- E_{nm}) e^{- \beta E_{m}} \rho_{mn}
\end{align}
where $E_m$ denotes energy eigenvalues with the assumption that the spectrum is discrete and $\rho_{mn}= \vert \bra{m} \mathcal{O}(0) \ket{n} \vert^2$  with $G_{E} (\tau)= \langle \normord{ \mathcal{O}(\tau)} \normord{ \mathcal{O} (0)} \rangle_{\beta}$. It is also known that at zero temperature, the spectral function is
\begin{align}
    \rho_{0}(\omega)\propto \theta(\omega) \sum_{l=0}^{\infty}  c_{l}\delta ( \omega - \Delta-2 l) -\theta(- \omega) \sum_{l=0}^{\infty} c_l \delta(\omega+ \Delta + 2l)
\end{align}
where $c_l$ are some coefficients,  $\Delta$ is the conformal dimension of $\mathcal{O}$. So, we see that correlation function shows discrete spectrum. This implies that $\mathcal{M}_{R}$ and $\mathcal{M}_{L}$ are of Type $\RN{1}$ algebra and they are Type $\RN{1}_{\infty}$ factor \cite{leutheusser2023emergent}.

\par In contrast, above Hawking Page temperature $\beta< \beta_{HP}$, the relation (\ref{inheritance formula}) fails to hold. It is expected that $\rho (\omega)$ at the high temperature phase will be a smooth function with support on the full real $\omega$ axis. Therefore, due to the expected complete spectrum exhibited by the correlation function, it is conjectured that above Hawking Page transition, $\mathcal{M}_{R}$ and $\mathcal{M}_{L}$ are Type $\RN{3}_1$ factor. So, the GNS Hilbert space $\mathcal{H}_{TFD}$ cannot be factorized into $\mathcal{H}_L \otimes \mathcal{H}_{R}$ and 
 $\mathcal{H}_{TFD}$ describes $\mathcal{O}(1)$ perturbation around $Ads$-- Schwarzschild black hole.  The boundary algebra $\mathcal{M}_R / \mathcal{M}_{L}$ is dual to the local algebra  associated to the right / left  exterior region of $Ads$ Schwarzschild black hole in the bulk which we denote as $\tilde{\mathcal{M}_{r}}/ \tilde{\mathcal{M}_l}$. See Figure 5.

\begin{figure}
    \centering
    \includegraphics[width=0.4\linewidth]{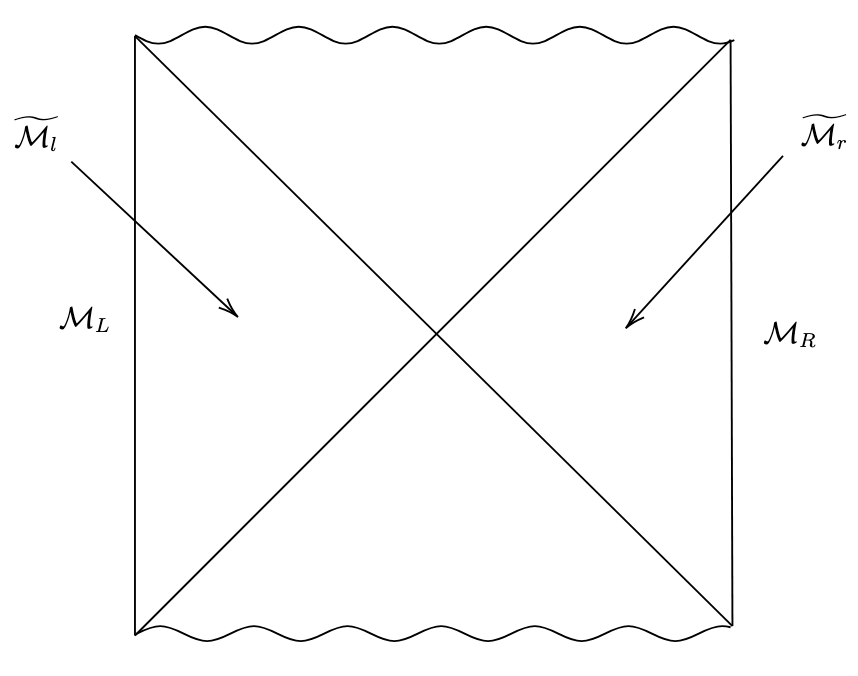}
    \caption{$\mathcal{M}_{R}/ \mathcal{M}_L$} is right/left asymptotic boundary algebra. $\tilde{M}_r/ \tilde{\mathcal{M}_l}$ is the local algebra associated to the right/ left  exterior region of $Ads$ Schwarzschild black hole
    \label{Figure 5}
\end{figure}

\begin{align}
    \mathcal{M}_R = \tilde{\mathcal{M}_r}, \quad \mathcal{M}_L= \tilde{\mathcal{M}_l} \label{algebra duality}
\end{align}
(\ref{algebra duality}) is one of the example of \textbf{subalgebra-subregion duality}. Subalgebra-subregion duality is a statement that any bulk spacetime open region $u$ is associated with an emergent Type $\RN{3}_1$ boundary subalgebra $\mathcal{M}_{\mathcal{U}}$.  More generally, we may start with the finite $N$ boundary algebra $\mathcal{B}^{N}_{\mathcal{U}}$ where $\mathcal{U}$ is some open region on the boundary. In this formalism, it is important to note that the action of restricting the boundary theory to a subregion $\mathcal{U}$ does not commute with the action of taking large $N$ limit. In fact, we have the relation \cite{leutheusser2022subalgebra}

\begin{align}
    \mathcal{X}_{\mathcal{U}}= \pi_{\Psi} (\lim_{N \rightarrow \infty, \Psi} \mathcal{P}_{\mathcal{U}} \mathcal{B}^{N}) \supseteq \mathcal{P}_{\hat{\mathcal{U}}} \pi_{\Psi} (\lim_{N \rightarrow\infty} \mathcal{B}^N)= \mathcal{Y}_{\hat{\mathcal{U}}}
\end{align}
where $\mathcal{P}_{\mathcal{U}}$ is a restriction operation to the region $\mathcal{U}$ and $\hat{\mathcal{U}}$ is the causal completion of $\mathcal{U}$. The algbera $\mathcal{X}_{\mathcal{U}}$ is the entanglement wedge of $\mathcal{U}$ while $\mathcal{Y}_{\hat{\mathcal{U}}}$ is the causal wedge of $\mathcal{U}$. Here, the state $\Psi$ is a general semiclassical state in the sense that there exists a sequence of states $\{  \Psi^{N}\}$ such that correlation functions of (subtracted) single-trace operators have a well defined large $N$ limit. For more discussions, see \cite{leutheusser2022subalgebra}.
 One of the other semiclassical state instead of thermofield double state that we can consider is $Ads$ vacuum state $\omega_{\Omega}$. In this case, we do not need to duplicate the system and we will just have one $\mathcal{N}=4$ SYM. To distinguish the notation with the thermal case, we denote the subtracted single trace operator (subtraction with expectation value in $Ads$ vacuum) as
 
\begin{align}
    \hat{\mathcal{O}}= \mathcal{O}- \langle\mathcal{O}\rangle_{\Omega} \label{vacuum subtraction}
\end{align}

We can also construct $*$-- algebra $\mathcal{A}_{\Omega}$ generated by (\ref{vacuum subtraction}) and perform GNS construction to construct large $N$ Hilbert space $\mathcal{H}_{\Omega}$ and the $GNS$ representation of the algebra
$\pi_{\Omega} (\mathcal{A}_{\Omega})''= \mathcal{M}_{\Omega}$. In this case, the bulk is the pure $Ads$ spacetime.  See \cite{leutheusser2022subalgebra}.

\section{ETH approach in Large $N$ Algebras}

\subsection{Diminishing Potentialities in Large $N$ limit }
In this subsection, we want to present a new argument that the principle of diminishing potentialities (PDP) holds for the generalized free field algebra $\mathcal{M}_{R}$ or $\mathcal{M}_{L}$ in the high temperature phase $\beta< \beta_{HP}$ where the bulk dual is two sided eternal black hole. In contrast,  PDP does not hold below Hawking Page transition $\beta < \beta_{HP}$ and for the vacuum subtracted algebra $\mathcal{M}_{\Omega}$ generated by (\ref{vacuum subtraction}) using our argument. In the former case, the bulk dual is two thermal $Ads$ entangled with each other while for the latter case the bulk dual is pure $Ads$.
As we have discussed in the previous section, the large $N$ boundary theory factorizes into state dependent generalized free field theories. In particular, two semiclassical states $\Psi_{1}, \Psi_{2}$ cannot be in each other $GNS$ Hilbert space and thus there is no correlation / relation
between two GNS Hilbert space $\mathcal{H}^{GNS}_{\Psi_1}$ and $\mathcal{H}^{GNS}_{\Psi_2}$.  In holography, the bulk dual of these large $N$ generalized free field theories are low-energy effective theories when $G_{N} \rightarrow 0$. Moreover, these low energy effective theories in the bulk are just standard free field theories in asymptotically $Ads_{5}$ curved spacetimes \cite{leutheusser2023emergent} \cite{leutheusser2022subalgebra}. This large $N$ dictionary  brings a new perspective on understanding how classical spacetime with its causal structure emerged from the coarse-graining of the microscopic boundary theory. As we have seen, the large $N$ generalized free field theory is primarily constructed by considering the large $N$ limit of correlation functions in semiclassical states.  The (subtracted) single trace operators which survive (have  well defined large $N$ correlation functions) in the large $N$ limit generate the algebra of observables $\mathcal{M}$. However, we also notice that the canonical Hamiltonian of $\mathcal{N}=4$ SYM in $\mathbb{R} \times S^3$ takes the form   \cite{witten2022gravity} \cite{leutheusser2022subalgebra} 
 \begin{align}
     H = \frac{N}{\lambda} \int_{S^{3}} d^3 x \sqrt{g} \Tr [F_{0i}F_{0j} g^{ij}+ \cdots] = N \Tr[\cdots] 
 \end{align}

In particular, the Hamiltonian has an explicit dependence on $N$. In finite $N$ theory, a local operator $A(t, \Vec{x})$ can be expressed as
\begin{align}
    A(t) = e^{i (t-t')H} A(t') e^{-i (t-t') H} \label{Heisenberg evolution}
\end{align}
which is the consequence of Heisenberg equation. However, we note that the Hamiltonian $H$ is not well defined in the large $N$ limit due to the explicit $N$ dependence. This implies that $(\ref{Heisenberg evolution})$ is no longer valid in large $N$ theories and there is no equation of motion for local operators in large $N$ generalized free field theories. This has severe consequence since this will imply that determinism in the sense that the evolution of the  system is uniquely determined by initial conditions/ data is lost in the large $N$ limit. Technically speaking, this means that the time slice axiom does not hold. In algebraic quantum field theory, time slice axiom means that  the algebra of a neighborhood of a Cauchy surface of a given region coincides with the algebra of the full region.

\par To give better insight into how the existence of  on-shell fields (existence of equations of motion) is crucial for time slice axiom to hold, we consider a simple example which is free massive scalar field theory in Minkowski space. We consider an algebra generated by the smeared free massive scalar field $\Phi (f)$ satisfying the following properties (let $f \in C^{\infty}_{0}(\mathbb{M^d})$ be a test function)
\begin{itemize}
    \item $\Phi (f)^* = \Phi (\overline{f})$
    \item $\Phi (Pf)=0$
    \item $[\Phi(f), \Phi(g)]=  \Delta (f, g) \mathbb{1}$
    \item The local algebra $\mathcal{A}(\mathcal{O})$ for an open region $\mathcal{O} \in \mathbb{M}^d$ is given by

    \begin{align*}
        \mathcal{A} (\mathcal{O})= \{  \Phi (f) \vert  supp(f) \in \mathcal{O} \}
    \end{align*}
\end{itemize}
Here, $\Phi (f)= \int \phi (x) f(x) d^{d}x$ where $\phi (x)$ is the free massive scalar field and the symbol $\overline{f}$ means the complex conjugate of $f$. We denote the Klein Gordon operator as $P$, i.e. $P= \partial_{\mu} \partial^{\mu}-m^2$.  We also recall that the commutator of two free massive scalar fields is
\begin{align}
    [\phi (x), \phi (y)] = \Delta_{+} (x,y)- \Delta_{-} (x,y) = \Delta (x,y) \label{commutator free field}
\end{align}
where $\Delta_{+} (x,y)$ is the retarded propagator while $\Delta_{-} (x,y)$ is the advanced propagator. We also recall that $\Delta_{+} (x,y)$ is only non-zero when $x^{0} > y^{0}$ and $\Delta_{-} (x,y)$ is only non-zero when $x^{0} < y^{0}$. The difference between them $\Delta (x,y)$ denotes the usual causal propagator.  Moreover, they satisfy
\begin{align}
    P_{x} \Delta_{+/-} (x,y) = \delta (x-y), \quad P_{x} \Delta (x,y)= 0
\end{align}
We can smear the causal propagator $\Delta$ with one test function to give spatially compact supported solution of Klein Gordon equation (view it as a map $\Delta : C^{\infty}_{0}(\mathbb{M^d}) \rightarrow \mathcal{E}_{sc} (\mathbb{M^d})$) i.e.
\begin{align}
    \Delta_{f} (x) = \int d^{d}y \: \Delta (x,y) f(y) 
\end{align}
Note that the support of $\Delta_{f}$ lies inside  the union of the causal future and the causal past of the support of $f$ , i.e.
\begin{align}
    Supp (\Delta_f) \subset J^{+} (Supp (f)) \bigcup J^{-} (Supp (f))
\end{align}
where $J^{+} $ denotes the causal future and $J^{-}$ denotes the causal past. We note that a function with spatially compacted support means that it vanishes in the causal complement of a compact set. In general, the causal propagator $\Delta$ maps a compactly supported smooth function to a spatially compact smooth function. We denote $\mathcal{E}_{sc}$ to be the space of spatially compacted solution of Klein Gordon equation and $C^{\infty}_{cs}$ to be the space of spatially compacted smooth function.  Furthermore, we can also think of the causal propagator $\Delta$ as a bilinear form , i.e. $\Delta : C^{\infty}_{0} (\mathbb{M}^d) \times C^{\infty}_{0} (\mathbb{M}^d) \rightarrow \mathbb{C}$
\begin{align}
    \Delta (f,g)= \int d^{d}x
    \int d^{d}y f(x) \Delta (x,y) g(y), \quad f,g \in C^{\infty}_{0} (\mathbb{M}^d)
\end{align}
We recall that $\Delta (f,g)=0$ when $Supp (f)$ is spacelike separated from  $Supp (g)$. Now, we consider an exact sequence which will be useful for the proof of time slice axiom for free scalar field theory \cite{fewster2019algebraicquantumfieldtheory}
\begin{align}
    0 \rightarrow C^{\infty}_{0} (\mathbb{M}^d)  \xrightarrow{P} C^{\infty}_{0} (\mathbb{M}^d) \xrightarrow{\Delta} C^{\infty}_{sc} (\mathbb{M}^d) \xrightarrow{P} C^{\infty}_{sc} (\mathbb{M}^d) \rightarrow 0 \label{exact sequence}
\end{align}
In particular,  we note that
\begin{align}
    C^{\infty}_{0} (\mathbb{M}^d)/ \Im P =C^{\infty}_{0} (\mathbb{M}^d)/\ker \Delta  = \Im \Delta = \ker P = \{ g \in C^{\infty}_{sc} (\mathbb{M}^4) \vert P g=0  \}= \mathcal{E}_{sc} (\mathbb{M}^d) =  Sol(\mathbb{M}^d)
 \label{explanation of exact sequence}
\end{align}
where  $ Sol(\mathbb{M}^d)$ denotes the space of solution of Klein Gordon equation. The first equality in (\ref{explanation of exact sequence}) holds because
\begin{align}
    \Delta_{Pf} (x) = \int d^{d}x \int d^{d}y \: \Delta (x,y)\: (P_yf)(y)= \int d^{d}x \int  d^{d}y  \: (P_{y}\Delta(x,y)) f(y)=0
\end{align}
where we have performed integration by part twice to bring the Klein Gordon operator $P$ from the test function $f$ to the causal propagator $\Delta$. 

\par Now, we give a proof of the time slice axiom for free scalar field theory. First, we note that for any solution of Klein Gordon equation $g \in \mathcal{E}_{sc} (\mathbb{M}^d)$, we can use (\ref{exact sequence}) to write $g = \Delta h$ for some $h \in C^{\infty}_{0} (\mathbb{M}^d)$. Now for a Cauchy surface $\Sigma$, we consider an open neighborhood $U$ with $\Sigma \in U$. We also consider another two Cauchy surfaces $\Sigma_1$ and $\Sigma_2$ such that $\Sigma_2$ is to the future of $\Sigma$ and $\Sigma_1$ is to the past of $\Sigma$. Moreover, we take $\Sigma_1 \subset U$ and $\Sigma_2 \subset U$. Next, we construct a test function $f_{U}$ with $Supp(f_{U}) \subset U$ by considering the following
\begin{align}
    f_{U}= P \chi g
\end{align}
where $\chi$ is a smooth cutoff function   such that $\chi=1$ in the future of $\Sigma_2$ and $\chi=0$ in the past of $\Sigma_1$. In the intermediate region between $\Sigma_2$ and $\Sigma_1$, the cutoff function $\chi$ smoothly interpolated between $0$ and $1$. It is important to note that $\chi g$ is  not a compactly supported function and thus $\Delta f_{U} \neq 0$. Moreover, we can write $h= P\Delta_{+}h$ since $P \Delta_{+} = \mathbb{1}$. Therefore, we obtain
\begin{align}
    h- f_{U}= P (\Delta_{+} h- \chi \Delta h)= P \big((1- \chi)\Delta_{+} h + \chi \Delta_{-} h  \big)
\end{align}
Since the supports $Supp (\Delta_{+}h)\subset J^{+}(Supp(h))$,   $Supp (\Delta_{-}h)\subset J^{-}(Supp(h))$, $Supp (\chi)\subset J^{+} (\Sigma_1)$ and $Supp (1-\chi)\subset J^{-} (\Sigma_2)$, we know that
\begin{align}
    Supp ((1- \chi) \Delta_{+}h) \subset J^{-} (\Sigma_2) \bigcap J^{+} (Supp(h)), \quad Supp (\chi \Delta_{-}h) \subset  J^{+} (\Sigma_1) \bigcap J^{-} (Supp(h))
    \end{align}
Therefore, we know that the function $ \Delta_{+} h- \chi \Delta h$ is compactly supported since $\Delta_{+/-}h$ is spatially compacted function. So, we know that 
\begin{align}
    \Phi (h)= \Phi (f_U) + \Phi (P(\Delta_{+} h- \chi \Delta h)) = \Phi(f_{U})
\end{align}
where we use that $\Phi (P k)=0$ for any $k \in C^{\infty}_{0}(\mathbb{M}^d)$. Hence, we know that
\begin{align}
\mathcal{A}(\mathbb{M}^d)= \mathcal{A}(U)   
\end{align}
 This completes the proof. The proof can be extended to free field theory in globally hyperbolic spacetimes and it is expected to hold for any physically reasonable interacting theories. The key step of the proof is the existence of an open neighborhood $U$ such that it can be foliated into Cauchy surfaces. In fact, time slice axiom can be refined to a more exact following statement by following  the same strategies above \cite{fewster2019algebraicquantumfieldtheory} \cite{Haag1962THEPO}:

 \begin{thm}
     Let $ U$ and $V$ be two open sets in $\mathbb{M}^d$ such that $U \subset V$ and $U$ contains a Cauchy surface of $V$, then
     \begin{align}
         \mathcal{A}(U)= \mathcal{A}(V)
     \end{align}
 \end{thm}

Therefore, we see one concrete example how equation of motions / on-shell fields is crucial for time slice axiom to hold. The loss of equation of motion for quantum fields is also equivalent to that the dynamics of quantum fields is no longer governed by Heisenberg equation (\ref{Heisenberg evolution}).   This is manifested in large $N$ generalized free field theory since the Hamiltonian is not well-defined in the large $N$ limit. Moreover, we note that the evolution of local operators in large $N$ generalized free field theory can still be given by modular flow in Tomita-Takesaki theory. For example, since $\omega_{TFD}$ is a normal faithful (equivalently $\ket{TFD}$ in $GNS$ representation is cyclic and separating) state  for the algebra $\mathcal{M}_{R}$ or $\mathcal{M}_L$, there exists a one-parameter modular automorphism of the algebra, i.e. let $A  \subset \mathcal{M}_{R} $
\begin{align}
   A(u)= \Delta^{-iu}_{TFD} A \Delta^{iu}_{TFD},  \quad \Delta^{-iu}_{TFD} \mathcal{M}_{R} \Delta^{iu}_{TFD} =  \mathcal{M}_{R} \: \: \: \forall u \in \mathbb{R}   \label{modular evolution}
\end{align}
Similar relations also hold for $\mathcal{M}_L$. However, we cannot identify (\ref{modular evolution}) as equations of motion since $\Delta^{iu}_{TFD}= e^{-iu h_{TFD}}$ and $h_{TFD}$ depends on the entire asymptotic boundary instead of a single time slice (in contrast with $H$ which depends on a single time slice) \cite{leutheusser2022subalgebra}.

\par Another typical axiom in algebraic quantum field theory that will be useful for our later purpose is the generating property of local algebra. In a given spacetime $M$ with the full algebra $\mathcal{A}$, We call a double cone to any region $\mathcal{O} \subset M$ defined by the intersection of the future open null cone of some point $x \in M$ with the past open null cone of other point $y \in M$.  We assign to each  double cone $\mathcal{O} \subset M$ a local subalgebra $\mathcal{A}(\mathcal{O}) \subset \mathcal{A}$. The collection of all local algebras give the full algebra $\mathcal{A}$:
\begin{align}
    \mathcal{A}= \overline{ \cup_{\mathcal{O}}\mathcal{A}(\mathcal{O})}^{\vert \vert \cdot \vert \vert}  \label{generating property axiom}
\end{align}
The union runs over the sets of all double cones. We choose double cones as our fundamental regions so that it is compatible with time slice axiom.
\par The failure of time slice axiom for large $N$ generalized free field algebras $\mathcal{M}$ (here $\mathcal{M}$ can be $\mathcal{M}_{R}, \mathcal{M}_{L}$ or $\mathcal{M}_{\Omega}$) allows us to consider time band algebra $\mathcal{M}_{I \times S^3}$  where 
$I$ denotes a time interval $I=(t_{1}, t_2)$ with $t_2 > t_1$ which can be a proper subalgebra of $\mathcal{M}$. The time band algebra $\mathcal{M}_{I \times S^3} $ is generated by smeared (subtracted) single trace operators with support in $I \times S^3$ 
\begin{align}
    \mathcal{M}_{I \times S^3}= \mathcal{M} \vert_{I \times S^3} \equiv \mathcal{M}_{I}
\end{align}
 where for notational simplicity we denote it as $\mathcal{M}_{I}$. In this context,  we formulate a version of Principle of diminishing potentialities (PDP) for large $N$ generalized free field algebra
 \begin{equation}\label{PDP large N algebra}
  \mathcal{M}_{(t, \infty) \times S^3} := \mathcal{M}_{\geq t} \subsetneqq \mathcal{M}_{\geq t'} := \mathcal{M}_{(t', \infty) \times S^3}, \quad \forall  t > t'  % Fixed: replaced curly quote with straight prime
\end{equation}
Here, we consider the time band algebra $\mathcal{M}_{(t, \infty) \times S^3}$ with its support in the time interval that starts at the time $t$ and extend all its way to $t \rightarrow \infty$. We also denote it as $\mathcal{M}_{\geq t}$.  We want to argue that (\ref{PDP large N algebra}) holds for $\mathcal{M}_R/ \mathcal{M}_{L}$ above Hawking Page temperature $\beta < \beta_{HP}$ while for the case when $\beta > \beta_{HP}$ and for $\mathcal{M}_{\Omega}$,  (\ref{PDP large N algebra}) does not hold.

\par We start with the case  $\beta < \beta_{HP}$ where the bulk dual is an eternal black hole. We consider two time band  algebras $\mathcal{M}_{R, \geq t}$ and $\mathcal{M}_{R, \geq t'}$ with $t > t'$. The algebra $\mathcal{M}_{R, \geq t}$ is defined by first considering the right algebra $\mathcal{M}_{R}$ generated by (subtracted) single trace operators and then only we restrict it to the region $(t, \infty) \times S^3$, i.e. $\mathcal{M}_{R} \vert_{(t, \infty) \times S^3}$. Therefore,  using the prescription of Liu and Leutheusser, we know that the corresponding  bual dual  algebra $\tilde{\mathcal{M}}_{_{r, \geq t}}$ is the causal wedge of the boundary region $(t, \infty) \times S^3$ or equivalently the timelike envelope  of the region $(t, \infty) \times S^3$  \cite{strohmaier2023timeliketubetheoremcurved}. The causal wedge of an open boundary region $\mathcal{U}$ is defined to be

\begin{align}
    W_{C} (\mathcal{U})= \bigg\{  p \in \textit{bulk} \cup \textit{boundary} \bigg|  p \: \textit{can be reached with causal curves that start and end on }  \hat{\mathcal{U}}\bigg \}
\end{align}
where $\hat{ \mathcal{U}}$ denotes the causal completion of $\mathcal{U}$ with respect to the boundary's causal structure. We denote $W_{C} ((t, \infty)\times S^3])$ as $W_2$ and  
$W_{C} ((t', \infty)\times S^3])$ as $W_{1}$. The region $W_1$ and $W_2$
are both boundary enchored wedge with $W_{2} \subsetneqq W_1$. See Figure $6a$.

\begin{figure}
    \centering
    \includegraphics[width=0.75\linewidth]{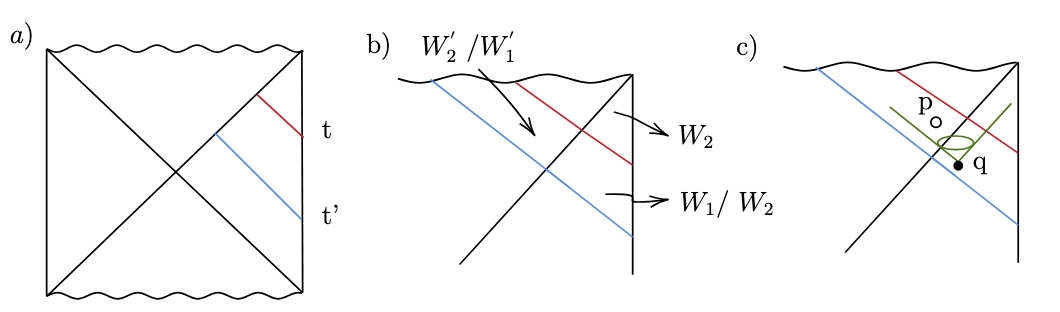}
    \caption{6a) The region $W_2$ bounded by the red line, horizon and the boundary is a boundary anchored wedge while the region $W_1$ bounded by the blue line, horizon and boundary is another boundary anchored wedge with $W_{2} \subsetneqq W_1$. The algebra associated to these regions are $\mathcal{M}(W_{2})= \mathcal{M}_{\geq t}$ and $\mathcal{M}(W_{1})= \mathcal{M}_{\geq t'}$. b) The region spacelike separated from $W_1  (W_2)$ is denoted as $W_{1}^{'}$( $W_{2}^{'})$. The region $W^{'}_{2}/ W_{1}^{'}$ refers to the subregion of $W_{2}^{'}$ with $W^{'}_1$ excluded. Similarly, the region $W_1 / W_2$ is the subregion of $W_1$ with $W_{2}$ excluded. c) Given a spacetime point $p \in W^{'}_2 / W^{'}_1$, it is always possible to find a point $q$ in $W_{1} / W_2$ such that $p \in J^{+} (q)$. }
    \label{fig:enter-label}
\end{figure}

The corresponding bulk dual algebra of these time band algebras are
\begin{align}
    \tilde{\mathcal{M}} _{r, \geq t} = \tilde{\mathcal{M}} (W_2), \quad \tilde{\mathcal{M}}_{r, \geq t'}=\tilde{\mathcal{M}} (W_1)
\end{align}
Since  $W_{2} \subsetneqq W_1$, we know that $\tilde{\mathcal{M}} (W_2) \subset \tilde{\mathcal{M}} (W_1)$. Now, we want to argue that  $\tilde{\mathcal{M}} (W_2)$ is a proper subalgebra of   $\tilde{\mathcal{M}} (W_1)$. First, we note that that bulk dual of generalized free field theory is just standard free field theory in asymptotically $Ads_{5}$ spacetime. Therefore, any local algebra of $\tilde{\mathcal{M}}_{r}$ must satisfy the standard axioms of algebraic quantum field theory including time slice axiom. 
We also note that since $W_{2} \subsetneqq W_1$, it follows that $W^{'}_{1} \subsetneqq W^{'}_{2}$ where $W'$ denotes the causal complement of $W$. The region $W_{1}'$ and $W_2'$ contains regions which are in the interior of the black hole. A bulk local operator in the interior region of the black hole can be  prepared by the unitary evolution of operators in $\mathcal{M}_{R}$ guaranteed by the theorem of half-sided modular inclusion \cite{leutheusser2023emergent}. Moreover, we also know that $\tilde{\mathcal{M}}(W') \subset \tilde{\mathcal{M}} (W)'$.  We denote the subregion of $W_{2}'$ with $W_{1}'$ excluded as $W'_2 / W'_1$. The region $W_{2}' / W'_{1}$ is not spacelike separated from the subregion $W_1$ with $W_{2}$ excluded which we denote as $W_{1}/ W_2$. See Figure $6b$. Therefore, for a given point $p \in W'_{2}/W'_{1}$ it is always possible to find a point $q \in W_1 / W_2$ such that $p \in J^{+} (q)$. See Figure 6c. Using generating property axiom (\ref{generating property axiom}), we know that the algebra $ \tilde{\mathcal{M}} (W'_{2}/ W'_{1})$ is not empty and any element of 
$\tilde{\mathcal{M}} (W'_{2}/ W'_{1})$ commutes with $\mathcal{M} (W_{2})$.  Now, we suppose that $\tilde{\mathcal{M}}(W_{2})= \tilde{\mathcal{M}} (W_{1})$. Then, the algebra $\tilde{\mathcal{M}}(W_{1}/W_{2})$ can be reconstructed from $\tilde{\mathcal{M}}(W_{2})$ i.e. any element in $\tilde{\mathcal{M}}(W_{1}/W_{2})$ can be expressed as an element in  $\tilde{\mathcal{M}}(W_2)$. Therefore, 
$\tilde{\mathcal{M}}(W_{2}'/W_{1}')$ must commute with  $\tilde{\mathcal{M}}(W_1)/\tilde{\mathcal{M} } (W_2)$. However, for a bulk field  $\phi (p)$ localized at $p \in W'_2 /W'_1$, we can always find some other bulk field which we denote as $B$ localized in $W_{1}/W_{2}$ such that $[\phi(p), B]\neq 0$. This is guaranteed because the point  $p$ is not spacelike separated from the region $W_{1}/W_2$ taking into account that the bulk dual of large $N$ generalized free field theory is free field theory in curved spacetime and the commutator of two bulk fields behaves like (\ref{commutator free field}). This contradicts the claim that $[\tilde{\mathcal{M}}(W_{1}/W_{2}), \tilde{\mathcal{M}}(W'_{2}/W'_{1})]=0$. Therefore, our assumption that $\tilde{\mathcal{M}}(W_{1})= \tilde{\mathcal{M}} (W_{2})$ must be wrong. So, it must be that
\begin{align}
    \tilde{\mathcal{M}}(W_{2}) \subsetneqq \tilde{\mathcal{M} }(W_{1}) \label{end of the proof}
\end{align}

Translating (\ref{end of the proof}) back into the boundary language, we obtain that 
\begin{align}
    \mathcal{M}_{R, \geq t} \subsetneqq \mathcal{M}_{R, \geq t'}, \quad \textit{for} \: \: t> t'
\end{align}
which is  PDP that we formulated before for time band algebras. This completes the proof that PDP holds for large $N$ generalized free field theory constructed by using thermofield double state when $\beta < \beta_{HP}$.

\par Next, we want to argue that PDP does not hold for $\mathcal{M}_{R/L}$ when $\beta > \beta_{HP}$ and for $\mathcal{M}_{\Omega}$. In fact, this has been shown by Gesteau and Liu in a different context using the method of exponential type problem \cite{gesteau2024toward}. In \cite{gesteau2024toward}, they want to probe the existence of horizon by checking whether the relative commutant $\mathcal{M}_{I= (- \frac{\mathcal{T}}{2}, \frac{\mathcal{T}}{2})}' \cap \mathcal{M} $ is non-trivial for all $\mathcal{T} \in \mathbb{R}$. Here, $\mathcal{M} $ denotes the algebra associated to a complete asymptotic boundary (it can be $\mathcal{M}_{R}/ \mathcal{M}_{L}$ or $\mathcal{M}_{\Omega}$) since if $\mathcal{M}'_{I} \cap M \neq \mathbb{C}\mathbb{1}$, then $\mathcal{M}_{I} \subsetneqq \mathcal{M}$. If the relative commutant $\mathcal{M}'_{I= (- \frac{\mathcal{T}}{2}, \frac{\mathcal{T}}{2})} \cap \mathcal{M}$ is non trivial for all $\mathcal{T} \in \mathbb{R}$, then horizon exists in the bulk dual while if $\mathcal{M}'_{I} \cap \mathcal{M}$ becomes trivial for a threshold $\mathcal{T}_{0}$, i.e. $\mathcal{M}'_{I=(- \frac{\mathcal{T}}{2}, \frac{\mathcal{T}}{2})} \cap \mathcal{M}= \mathbb{C} \mathbb{1}$ for all $\mathcal{T}> \mathcal{T}_0$, then there is no horizon in the bulk dual.  They showed that above Hawking Page transition,  $\mathcal{M}'_{I=(- \frac{\tau}{2}, \frac{\tau}{2})}\cap \mathcal{M}_{R}$ is non -trivial for all $\mathcal{T} \in \mathbb{R}$ while below Hawking Page transition or for $\mathcal{M}= \mathcal{M}_{\Omega}$, the largest value of $\mathcal{T}$ for which $\mathcal{M}'_{I= (- \frac{\mathcal{T}}{2}, \frac{\mathcal{T}}{2})} \cap \mathcal{M}$ is non-trivial is $\mathcal{T} = \pi$. Their proof for the later case is completely sufficient for our purpose to show the failure of PDP for $\mathcal{M}_{R/L}$ when $\beta > \beta_{HP}$ and for $\mathcal{M}_{\Omega}$. We will briefly sketch their arguments here.

\par The commutator of two smeared single-trace in a semiclassical state $\Psi$ is encoded in the corresponding spectral function $\rho (t, t')$
\begin{align}
    \omega_{\Psi} ([\mathcal{O}(g), \mathcal{O}(f)])= \int dt  \: dt'  \: g(t) f(t')\langle{\Psi} \vert [\mathcal{O}(t), \mathcal{O} (t')] \vert \Psi \rangle = \int dt \: dt' g(t) \rho (t, t') f(t')
\end{align}
where for simplicity we neglect all spatial dependent part. We know that the time band algebra $\mathcal{M}_{I}$ is generated by (subtracted) single trace operators in the form of
\begin{align}
    \mathcal{O} (f) = \int dt \: \mathcal{O}(t) f(t), \quad Supp(f) \in I
\end{align}
Moreover, the relative commutant $\mathcal{M}'_{I} \cap \mathcal{M}$ is generated by $\mathcal{O}(g)= \int dt \: \mathcal{O} (t) g(t) $ where $g$ is a compactly support function with its support in the entire asymptotic boundary such that
\begin{align}
    \int dt \: dt' g(t) \rho (t, t') f(t)=0, \quad \textit{ for all}  \: f \: \textit{with} \: Supp (f) \in I  \label{relative commutant condition}
\end{align}
Put it differently, the convolution $(g* \rho) (t')= \int dt \: g(t) \rho (t-t')$ needs to satisfy
\begin{align}
    Supp(g* \rho) \in I^{c}
\end{align}
where $I^{c}$ denotes the complement of $I$. We can also work in momentum space. After performing Fourier transform, the convolution $(g* \rho)(\omega)$ in momentum space is
\begin{align}
    (g* \rho) (\omega)= -g(\omega) \rho (\omega)
\end{align}
and (\ref{relative commutant condition}) in momentum space is
\begin{align}
    \int d \omega  \: g(\omega) \rho (\omega) f(- \omega)=0
\end{align}
Now, if $\Psi= \Omega$, then we know the bulk dual is pure $Ads$ and the spectral density takes the form 
\begin{align}
    \rho (\omega)= \rho_{0} (\omega)=\sum_{n= -\infty}^{\infty} a_{n} \delta (\omega-2n- \Delta)  \label{vacuum spectral density in last section}
\end{align}
where $\Delta$ is the conformal dimension of $\mathcal{O}$. Given the spectral density (\ref{vacuum spectral density in last section}), we now show that if the convolution $(g* \rho) (t)=0$ within an interval $I$ with width $\vert I \vert > \pi$, then $(g* \rho)(t)$ is identically zero. First, we can express the function $g(\omega)$ as a span of a set of complete basis and write $g(\omega) = \sum_{n}g_{n}(\omega)$. Then we can express $(g* \rho)(\omega)$ as
\begin{align}
    (g* \rho) (\omega)= \sum_{n= - \infty}^{\infty} c_{n} \delta (\omega-2n- \Delta)  \label{convolution in momentum space}
\end{align}
where $c_n= g_{n}a_{n}$. After transforming (\ref{convolution in momentum space}) back to position space, we know that $(g* \rho)(t)$ is periodic in $\pi$ up to a constant phase. Therefore, if $(g* \rho)(t)$ vanishes in an interval $I$ with width $\vert I \vert > \pi$, then $(g* \rho) (t)=0$ for all $t \in \mathbb{R}$. This completes the proof and the results holds for all single trace operator $\mathcal{O}$. Similar result will also hold if $\Psi= \ket{TFD}$ when $\beta> \beta _{HP}$ since in this case the spectral function $\rho(\omega)= \rho_{0} (\omega)$ as we have discussed in the previous section. 

\par The above result is also consistent with  subalgebra-subregion duality of Liu and Leutheusser. 
In the case that $\mathcal{M}= \mathcal{M}_{\Omega}$, the bulk dual is pure $Ads$ while for $\mathcal{M}= \mathcal{M}_{R}$ when $\beta > \beta_{HP}$, the bulk dual is thermal $Ads$. We know that the bulk dual $\tilde{ \mathcal{M}}_{I}$ associated to the time band algebra $\mathcal{M}_{I}$ is the local algebra associated to Rindler wedge $\tilde{\mathcal{M}}(W)$ with its boundary $\partial W= I \times S^3$. When the width $\vert I \vert < \pi$, the bulk region $W' \cap Ads$ is non-empty. When $\vert I \vert \geq \pi$, the bulk region  $W' \cap Ads= \emptyset$ and $W$ contains a complete Cauchy slice. See Figure 7. Therefore, we know that $\tilde{\mathcal{M}}_{\vert I \vert = \pi}= \mathcal{M}$, i.e. the time band algebra $\mathcal{M}_{I}$ with $\vert I \vert = \pi$ can already generate the full algebra $\mathcal{M}$.

\begin{figure}
    \centering
    \includegraphics[width=0.75\linewidth]{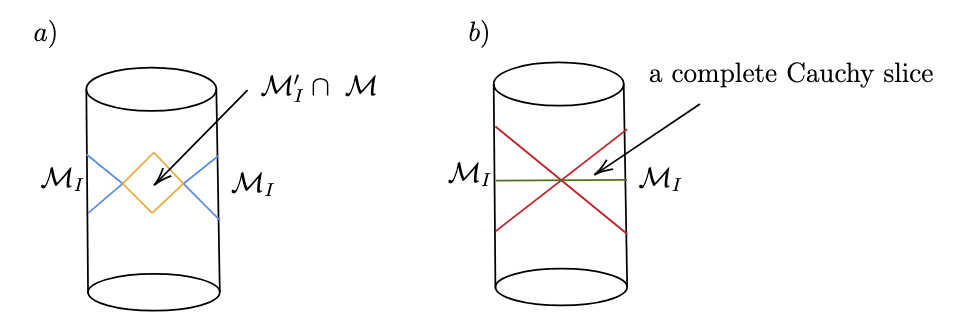}
    \caption{7a) The case when $\vert I \vert < \pi$ where the relative commutant $\mathcal{M}'_{I} \cap \mathcal{M}$ is non-trivial.  7b) The case $\vert I \vert = \pi$ where 
    the relative commutant $\mathcal{M}'_{I} \cap \mathcal{M}$ is trivial and the Rindler Wedge $W$ with its boundary $\partial W= I \times S^3$ contains a complete Cauchy surface.     
    }
    \label{Figure 7}
\end{figure}

This leads to the failure of diminishing potentialities because
\begin{align}
    \mathcal{M}_{\geq t} = \mathcal{M}_{\geq t'}, \quad \forall t,t' \in \mathbb{R}
\end{align}
since $\vert (t, \infty) \vert > \pi$ for all $t \in \mathbb{R}$.

\subsection{No Event Occurs in Thermal Equilibrium}
Now, we would like to understand whether an event occurs in the above setting. Without further considerations, we know immediately that the algebra generated by generalized free fields is not the appropriate framework for the implementation of ETH dynamics because the commutator of generalized free fields are  $c$ numbers and it is completely determined by two point functions of the generalized free fields. This means that the notion of centralizer of a state is completely redundant because the commutator of  generalized free fields in a particular state does not contain any new information about the structure of the algebra that we cannot know by just considering the commutator of  generalized free fields since the commutator of generalized free fields itself is given by $c$ number. In other words, the centralizer coincides with the center of the algebra.  This is consistent with the duality between the large $N$ generalized free fields theory in the boundary and the free fields theory in the bulk. We know that the commutator of free fields is completely determined by the causal structure and the behaviour of commutator in any state is the same. Since the algebra $\mathcal{M}_{R}/ \mathcal{M}_{L}$ is factor, the centralizer of any states in this algebra is always trivial. Therefore, event does not occur at all. However, we can try to extend the algebra of observables by incorporating the additional global charges /symmetries so that the commutator of a large portion of elements in the algebra is no longer given by $c$ numbers.   In fact, this extension can be achieved in large $N$ generalized free field algebra via crossed product and it has been discussed in literatures \cite{witten2022gravity} \cite{chandrasekaran2023large}. Therefore, in this extended framework, it is still possible to implement $ETH$ dynamics and we will discuss this in more details in the next subsection. In this subsection, we want to give an alternative (more rigorous) way to see that event does not occur at all in thermal equilibrium without appealing to the fact that the centralizer of any state coincides with the center of the algebra  which will be useful for the next subsection.

 \par Since the Principle of Diminishing Potentialities (PDP) holds in the algebra $\mathcal{M}_{R}/ \mathcal{M}_{L}$ when $\beta < \beta_{HP}$ where the preferred state  is $\omega_{TFD}$ (in the sene that we formulate for time band algebra),  we will only focus on this case for the subsequent section. We want to check whether the centralizer of the thermofield double state on the algebra $\mathcal{C}_{TFD} (\mathcal{M}_{R})$ is trivial or not. Recall that the  definition of  $\mathcal{C}_{TFD} (\mathcal{M}_{R})$ is
\begin{align}
    \mathcal{C}_{TFD} (\mathcal{M}_{R})= \{  A \in \mathcal{M}_{R} \vert \: \omega_{TFD} ([A,B])=0, \: \: \: \forall B \in \mathcal{M}_R   \}
\end{align}
Now, we also introduce another notion of subalgebra which is the fixed point algebra under the modular automorphism $\mathcal{C}_{fixed} (\mathcal{M}_R, \omega_{TFD})$ which will be useful for later purpose

\begin{align}
    \mathcal{C}_{fixed} (\mathcal{M}_{R})= \{  A \in \mathcal{M}_{R} \vert \sigma_{u}^{TFD} (A)=A , \: \: \forall u \in \mathbb{R} \}
\end{align}
where   $\sigma^{TFD}_{u} (A)= \Delta^{-iu}_{TFD} A \Delta^{iu}_{TFD}$ which is the one-parameter modular flow with respect to the thermofield double state. We also note that a state $\omega$ which acts on an algebra $\mathcal{A}$ such that $\mathcal{C}_{fixed} (\mathcal{M}_{R}, \omega)= \mathbb{C}\mathbb{1}$ is called \textbf{ergodic}. 

\par First, we recall that the correct way to characterize a thermal system with temperature $\frac{1}{\beta}$ in thermodynamic limit is via \textbf{KMS condition} \cite{haag1967equilibrium} \cite{haag2012local}. We consider a state $\omega$ acting on an algebra $\mathcal{A}$. For any two arbitrary elements $A, B \in \mathcal{A}$, we consider a function $F_{AB} : \mathbb{R} \rightarrow \mathbb{C}$
\begin{align}
    F_{AB} (t)= \omega (A \alpha_{t}(B)) \label{FAB Function}
\end{align}
where $\alpha_{t}(B)= e^{i Ht} B e^{-iHt}$ is a one-parameter $*$ automorphism generated by some Hamiltonian $H$. Now, we can also analytically continue the function $F_{AB}(t)$ to a larger domain of the complex plane, i.e. a strip in $\mathbb{C}$ so that (\ref{FAB Function}) is the boundary value of the extended function $F_{AB}(z)$. If $F_{AB}(z)$ is  \textbf{bounded} and \textbf{analytic} in the complex strip $0< \Im{z} < \beta$ with the boundary condition
\begin{align}
    F_{AB} (t + i \beta)= \omega (\alpha_{t}(B)A)
\end{align}
then we say that $\omega$ is a KMS state on the algebra $\mathcal{A}$ with respect to the time evolution $\alpha_t$.

\par In fact, KMS condition is deeply related with Tomita-Takesaki (TT) theory. In TT theory, if a state $\omega$ acting on an algebra $\mathcal{A}$ is normal faithful (equivalently the vector $\ket{\Omega}_{\omega}$ is cyclic and separating in the GNS representation), then Tomita-Takesaki theorem guarantees the existence of a one-parameter modular automorphism group $\sigma^{\omega}_{u}$ on the algebra $\mathcal{A}$ such that
\begin{align}
    \sigma^{\omega}_{u}(\mathcal{A})= \Delta_{\omega}^{-iu}\mathcal{A} \Delta^{iu}_{\omega}= \mathcal{A} \quad \forall u \in \mathbb{R}
\end{align}
Moreover, the state $\omega$ satisfies the KMS condition with respect to modular flow $\sigma^{\omega}_{u}$ with period $1$. More precisely, let $A,B \in \mathcal{A}$ be any two arbitrary elements in the algebra, we define a function $G_{AB}(u)$
\begin{align}
   G_{AB} (u)= \omega (A \sigma^{\omega}_{u} (B)) 
\end{align}
The function $G_{AB}(u)$ is the boundary value of a function $G_{AB}(z)$ which is bounded and analytic in the complex strip $0< \Im{z}< 1$ and whose boundary value for $\Im{z}=1$ is
\begin{align}
    G_{AB} (u+i)= \omega (\sigma^{\omega}_{u}(B)A)
\end{align}

To be able to interpret this result physically, suppose that the state $\omega$ is also a thermal state with temperature $\frac{1}{\beta}$ on the algebra $\mathcal{A}$ with respect to a time translation $\alpha_{t}$ generated by a Hamiltonian $H$, then we have
\begin{align}
    - \log \Delta_{\omega}= h_{\omega}= \beta H
\end{align}
It follows that
\begin{align}
    \sigma^{\omega}_{u} (B)= e^{i h_{\omega}u} Be^{-i h_{\omega}u}=e^{i H (\beta u)}B e^{-{iH (\beta u)}}= \alpha_{t= \beta u} (B)
\end{align}
where $t= \beta u$ is the thermal time. In the case of thermofield double state $\omega_{TFD}$, the modular Hamiltonian $h_{TFD}$ is related to the boundary Hamiltonian in such a way
\begin{align}
    h_{TFD}= \beta (H_{R}- H_{L})
\end{align}
where $H_{R}/ H_{L}$ is the boundary Hamiltonian on the right (left) asymptotic boundary.

\par Now, a useful lemma for us is that the centralizer of thermofield double state is equal to the fixed point algbera with respect to the modular flow
\begin{lemma}
    $\mathcal{C}_{TFD}(\mathcal{M}_R)= \mathcal{C}_{fixed} (\mathcal{M}_{R}, TFD) $
\end{lemma}
We will prove it now. First, suppose that $A \in \mathcal{C}_{TFD}(\mathcal{M}_{R})$, then, we compute
\begin{align*}
    \omega_{TFD}(\sigma^{TFD}_{u}(A)B) &= \omega_{TFD} (A \sigma^{TFD}_{-u} (B))\\
    &= \omega_{TFD}(\sigma^{TFD}_{-u} (B) A)\\
    &= \omega_{TFD} (A \sigma^{TFD}_{-u+i} (B)) \\
    &= \omega_{TFD} (\sigma^{TFD}_{u-i}(A)B)
\end{align*}
In the first and the forth line, we use that $h_{TFD} \ket{TFD}=0$. In the second line, we use the assumption that $A \in \mathcal{C}_{TFD} (\mathcal{M}_{R})$ and the fact that $\sigma^{TFD}_{-u} (B)\in \mathcal{M}_{R}$. In the third line, we use the KMS boundary condition.  Since the function $G(z)= \omega_{TFD} (\sigma^{TFD}_{z} (A)B)$ is bounded and analytic in the lower half complex strip $-1< \Im{z} < 0$ and also $G(u)= G(u-i)$, we can extend $G(z)$ to an entire function $\tilde{G}(z)$ on the whole complex plane. By Liouville theorem, we know that $\tilde{G} (z)= \tilde{G}(0)= G(0)$ for all $z \in \mathbb{C} $. So, it follows that $\sigma^{TFD}_{u}(A)=A$. Hence, we know that $\mathcal{C}_{TFD} (\mathcal{M}_{R}) \subset \mathcal{C}_{fixed} (\mathcal{M}_{R}, TFD)$.
\par Now, we prove the converse $\mathcal{C}_{fixed} (\mathcal{M}_{R}, TFD) \subset \mathcal{C}_{TFD}(\mathcal{M}_{R})$. Suppose that $A \in \mathcal{C}_{fixed} (\mathcal{M}_{R}, TFD)$, we compute
\begin{align*}
    \omega_{TFD} (AB)= \omega_{TFD} (\sigma^{TFD}_{u}(A)B)
    = \omega _{TFD} (B \sigma^{TFD}_{u+i} (A))
\end{align*}
In the first equality we use the assumption that $A \in \mathcal{C}_{fixed}(\mathcal{M}_R, TFD)$. In the second equality we use the KMS boundary condition. Moreover, we also know that the function $H(z)= \omega(B \sigma^{TFD}_{z} (A))$ is analytic and bounded in the upper half complex strip $0< \Im{z} < 1$. Furthermore, the boundary value of $H(z)$ on the real line parametrized by $u$ is a constant, i.e. $H(u)= H(0)$ for all $u \in \mathbb{R}$ because $\sigma^{TFD}_{u}(A)=A$. By the identity theorem in complex analysis, we know that $H(z) = H(0)$ for all $z$ in the upper half complex strip $0 < \Im{z} <1$. Hence, we know that 
\begin{align}
    \omega_{TFD}(AB)= \omega_{TFD} (B \sigma^{TFD}_{u+i} (A))= \omega_{TFD}(B \sigma^{TFD}_{0} (A))= \omega_{TFD}(BA)
\end{align}
So, $ \mathcal{C}_{fixed} (\mathcal{M}_{R}, TFD) \subset\mathcal{C}_{TFD} (\mathcal{M}_{R}) $.
\par Overall, we can conclude that $\mathcal{C}_{TFD}(\mathcal{M}_R)= \mathcal{C}_{fixed} (\mathcal{M}_{R}, TFD) $. Furthermore, since we know the boundary algebra $\mathcal{M}_{R}$ is dual to the bulk algebra $\tilde{\mathcal{M}}_r$ which is the local free field algebra associated to the right exterior region of the eternal black hole. By the theorem of Kay and Wald \cite{kay1991theorems}:
\begin{thm}
    On a spacetime containing a bifurcate Killing horizon,  Hadamard state which is invariant under the corresponding symmetry is unique, In the wedge in which the Killing field is time-like future directed, the state, if it exists, is the KMS state of temperature $T= \frac{1}{2 \pi }\kappa$.
\end{thm}
where $\kappa$ is the surface gravity. We will not go into the rather technical definition of Hadamard state. Roughly speaking, a Hadamard state is a physically reasonable quantum state in curved spacetime whose two point function captures the universal short distance singularity structure. What is important from the  implication of the theorem of Kay and Wald for us is that since the thermofield double state on boundaries is exactly dual to the Hartle-Hawking state of the eternal black hole, we know that the modular flow is geometric, i.e. $- \log \Delta_{TFD}= \frac{2 \pi}{ \kappa} (H_{R}- H_{L})= h_{r}- h_{l}= \hat{h}$ where \cite{witten2022gravity}
\begin{align}
    h_{r}= \int _{\Sigma_r} d \Sigma^{\mu} V^{\nu} T_{\mu \nu}, \: \: \: h_{l}=- \int_{\Sigma_l} d \Sigma^{\mu} V^{\nu} T_{\mu \nu}
\end{align}
Here, $\Sigma_{r} \cup \Sigma_l = \Sigma$ is a complete Cauchy surface of the eternal black hole and $\Sigma_r/\Sigma_l$ is the restriction of the Cauchy surface to the right/left side of the eternal black hole. The notation $V^{\mu}$ denotes the Killing vector field associated to the bifurcate Killing horizon and $T_{\mu \nu}$ is the stress tensor. So, we see that modular flow $\sigma^{TFD}_{u}= \alpha_{t= \beta u}$ that generate the isometric flow which is a one-parameter integral flow of the Killing vector $V^{\mu}$. See Figure 8.

\begin{figure}
    \centering
    \includegraphics[width=0.3\linewidth]{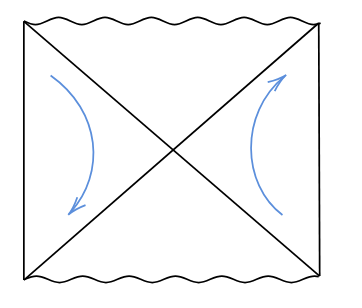}
    \caption{The blue curve is the isometric flow associated to the timelike Killing vector $V^{\mu}$}
\end{figure}

Hence, we know that $\sigma^{TFD}_{u} (A)= e^{i  \hat{h}t} (A) e^{-i \hat{h}t}$ for all $A\in \mathcal{M}_{R}$. Since $\hat{h}$ generates the isometric flow associated to the Killing vector $V^{\mu}$, it follows that the fixed point algebra is trivial, i.e.
\begin{align}
    \mathcal{C}_{fixed} (\mathcal{M}_{R}, TFD)= \mathbb{C} \mathbb{1}
\end{align}
So, the centralizer of thermofield double state $\mathcal{M}_{TFD} (\mathcal{M}_{R})$ is trivial. This implies that no potential events can be actualized in thermal equilibrium.  This is consistent with the other argument we give that the centralizer of any state coincides with the center of the algebra $\mathcal{M}_{R}$. 

\subsection{The First Actual Event in the Extended Algebra}
We see that in exact thermal equilibrium, event does not occur because the thermofield double state (TFD) is ergodic. In order to implement ETH dynamics of state, we  extend the algebra $\mathcal{M}_{R} / \mathcal{M}_{L}$ so that the extended algebra has non-trivial centralizer. One might be tempted to ask whether the modular crossed product improves the situation. Based on Liu and Leutheusser construction, Witten et al. carried out the crossed product construction to include the `` renormalized Hamiltonian'' into the algebra of observables \cite{witten2022gravity} \cite{chandrasekaran2023large}. More precisely, the  modular crossed product algebra $\mathcal{N}_{R}$ is constructed by
\begin{align}
    \mathcal{N}_{R}= \mathcal{M}_{R} \rtimes \mathcal{\mathcal{A}}_{X + h_{TFD}}= \{ \mathcal{M}_{R}, X+ h_{TFD}\}''  \label{cross product right algebra}
\end{align}
where $X$ is a real variable. The explicit form of $X$ depends on the method / physical argument that one uses to carry out the crossed product construction. In \cite{witten2022gravity}, the modular crossed product construction is carried out by taking into account $1/N$ correction to the right Hamiltonian $H_{R}$ and $X= \frac{H_{L}- \langle H_{L}\rangle_{TFD}}{N}$. In contrast, the authors in \cite{chandrasekaran2023large} consider a microcanonical energy window centered at some energy $E_{0} \sim N^2$ on top of the canonical ensemble or thermofield double state and it turns out that the subtracted Hamiltonian $h_{R}= H_{R}- E_{0}$ is then automatically a well-defined operator in the $GNS$ Hilbert space $\mathcal{H}_{TFD}$ and in that case $X= H_{L}- E_{0}$. The algebra given by  (\ref{cross product right algebra}) is generated by operators $a \otimes 1$ with $a \in \mathcal{M}_{R}$ and bounded function of $h_{TFD} + X$, i.e. $e^{is h_{TFD}} \otimes e^{is X}$.   The modular crossed product right algebra $\mathcal{N}_{R}$ acts on the Hilbert space $\mathcal{H}_{TFD} \otimes L^{2} (\mathbb{R})$ where $L^{2} (\mathbb{R})$ is the  space of square integrable  functions of real variable $X$. It also follows that the commutant of (\ref{cross product right algebra}) is given by
\begin{align}
    \mathcal{N}'_{R}= \mathcal{N}_{L} = e^{i P h_{TFD}}\mathcal{M}_{L} e^{-i P h_{TFD}} \rtimes \mathcal{A}_{X}= \{ e^{iP h_{TFD}}\mathcal{M}_{L} e^{-i P h_{TFD}}, X\}''
\end{align}
where $P$ is the canonical conjugate of $X$, i.e. $P= -i \frac{\partial}{\partial X}$. The commutant algebra $\mathcal{N}_L$ is generated by $e^{i P h_{TFD}} \mathcal{M}_{L} e^{-i P h_{TFD}}$ and bounded function of $X$. It might seem that there is an  apparent asymmetry between the algebra $\mathcal{N}_{R}$ and $\mathcal{N}_{L}$. However, as explained in \cite{witten2022gravity}, this asymmetry can be removed by conjugating $\mathcal{N}_{R}$ and $\mathcal{N}_{L}$ with $e^{-i P h_{TFD}/2 }$, i.e.
\begin{align}
    \mathcal{N}_{R}=e^{-i P h_{TFD}/2} \mathcal{M}_{R} e^{i P h_{TFD}/2} \rtimes \mathcal{A}_{X+ \frac{h_{TFD}}{2}}, \quad \mathcal{N}_L= e^{i P h_{TFD}/2} \mathcal{M}_{L}e^{-i P h_{TFD}/2} \rtimes \mathcal{A}_{X- \frac{h_{TFD}}{2}}  \label{symmetry expression}
\end{align}
Since the modular flow $\sigma^{TFD}_{u= \frac{t}{\beta}}$ is geometric and the direction of the flow  in the right exterior region opposite to the left exterior region , the modular Hamiltonian $h_{TFD}$ is odd under the exchange of $\mathcal{M}_{R}$ and $\mathcal{M}_{L}$. Therefore, secretly $\mathcal{N}_{R}$ and $\mathcal{N}_{R}$ are symmetrical up to an unitary transformation and can be treated at the same footing. So, in the subsequent discussion, we will focus on $\mathcal{N}_{R}$.  Moreover, The extended preferred state is
\begin{align}
    \ket{TFD, g}= \ket{TFD} \otimes g(X)
\end{align}
where $g(X)$ can be any  Gaussian function of the real variable $X$ with the condition that 

\begin{align}
    \int dX \: \vert g(X) \vert^2=1
\end{align}
For example, we can take $g$ to be
\begin{align}
    g(X)= \frac{1}{(2 \pi \sigma^2)^{\frac{1}{4}}} e^{-X^2/4 \sigma^2}
\end{align}
where $\sigma^2$ is the variance of the Gaussian distribution. Now, it is obvious that the centralizer of the modular crossed product algebra $\mathcal{C}_{TFD,g} (\mathcal{N}_{R})$ is non trivial and it consists of the operator $e^{i s (h_{TFD} + X)}$. To see this, it suffices to verify it for operators taking the form $a e^{i u (h_{TFD}+X)}$ with $a \in \mathcal{M}_{R}$ because the operators of the form  $a e^{i u (h_{TFD}+X)}$ forms an additive basis of the algebra $\mathcal{N}_{R}$. Let $\hat{\omega} (\cdot)= \langle TFD,g \vert \cdot \vert TFD,g \rangle$, we check that
\begin{align*}
    \hat{\omega} (e^{is (h_{TFD}+X)} (a e^{iu (h_{TFD} +X)})) &= \omega_{TFD}(e^{ish_{TFD}} a e^{iuh_{TFD}}) \int dX \: \vert g(X) \vert^2 e^{i (u+s)X}\\
    &= \omega_{TFD} (ae^{i (u+s)h_{TFD}}) \int dX \vert g(X) \vert^2 e^{i (u+s)X}\\
    &= \hat{\omega}((a e^{iu (h_{TFD} + X)}) e^{is (h_{TFD}+X)})
\end{align*}
In the second line, we use the fact that $h_{TFD} \ket{TFD}=0$. Therefore, we see that the centralizer $\mathcal{C}_{TFD,g} (\mathcal{N}_{R})$ is non-trivial

\begin{align}
\mathcal{C}_{TFD,g} (\mathcal{N}_{R}) \supseteq \{ e^{is (h_{TFD}+X)} , \forall s \in \mathbb{R} \}    
\end{align}

 As discussed in \cite{witten2022gravity} and \cite{chandrasekaran2023large}, the resulting modular crossed product algebra $\mathcal{N}_{R}$ is of Type $\RN{2}_{\infty}$ factor.  In fact, a Type $\RN{2}_{\infty}$ factor admits a renormalized trace $\tau$ up to a multiplicative constant. This leads to that density matrices and entropy difference can be consistently defined for the crossed product algebra $\mathcal{N}_{R}$. Therefore, the bulk dual associated to the algebra $\mathcal{N}_{R}/ \mathcal{N}_{L}$ should be considered as the emergence of `` quantum spacetime'' rather than classical spacetime. This also does not align completely with the subregion-subalgebra duality since now $\mathcal{N}_{R}$ is not just purely generated by large $N$ generalized free fields. Moreover, if we look at (\ref{symmetry expression}), we see that the canonical conjugate  of $X$ which is denoted as $P$ also enters  this new description. The operator $P$ is often interpreted as timeshift operator $\Delta t _{R}+ \Delta t_{L}$\cite{chandrasekaran2023large}. So, the standard deviation of the operator $P$ which we denote as $\Delta P$ captures the fluctuation of time slice in the bulk. It is further shown in \cite{chandrasekaran2023large} that the von Neumann entropy of a semiclassical state $\Psi \in \mathcal{H}_{TFD} \otimes L^{2} (\mathbb{R})$ with $\Delta P (\Psi)  \ll 1$ matches the generalized entropy of the eternal black hole up to an additive constant. Therefore, we see that the extended Hilbert space $\mathcal{H}= \mathcal{H}_{TFD} \otimes L^{2} (\mathbb{R})$ describes both $\mathcal{O}(1)$ perturbation around the eternal black hole and also the fluctuation of time slice. 

\par However, the Principle of Diminishing Potentialities (PDP) does not hold for the modular crossed product algebra $\mathcal{N}_{R}$. In this context, we can also define a time band algebra $\mathcal{N}_{{R},I}$ 

\begin{align}
    \mathcal{N}_{R,I}= \mathcal{M}_{R,I} \rtimes \mathcal{A}_{h_{TFD}+ X} 
\end{align}
In particular, we can conjugate the algebra $\mathcal{N}_{R, \geq t}$ with $e^{-is (h_{TFD}+X)}$ 
where $s >0$

\begin{align}
e^{-is (h_{TFD}+X)} \mathcal{N}_{R, \geq t} e^{is (h_{TFD}+X)}= \mathcal{N}_{R, \geq t-s}= \mathcal{M}_{R, \geq t-s} \rtimes \mathcal{A}_{h_{TFD}+X}
\end{align}
Since $e^{is (h_{TFD}+X)}\in \mathcal{N}_{R, \geq t}$, it follows that
\begin{align}
    \mathcal{N}_{R,\geq t}= \mathcal{N}_{R, \geq t-s}= \mathcal{N}_{R}, \quad \forall s>0
\end{align}
Therefore, PDP does not hold for the modular crossed product algebra.  So, we know that the modular crossed product construction does not really improve the situation in initiating the ETH dynamics although its centralizer is non-trivial. Alternatively, we can consider  the cross product of $\mathcal{M}_{R}$ by a compact group automorphisms $G$. In fact, in this case, the situation is indeed improved and ETH dynamics can be implemented with some further considerations. We will outline the procedures below.

\par In appendices of \cite{witten2022gravity} and \cite{chandrasekaran2023large}, there are very illuminating discussions on how to incorporate the additional symmetries of two sided eternal black hole into the algebra of observables. We will briefly sketch their arguments now. We will consider the eternal black hole with vanishing angular momentum and charges so that the symmetry group is maximized. In this case, the symmetry group $G$ is
\begin{align}
    G=  (Spin (4) \times SU(4)_{R})/\mathbb{Z}_2
\end{align}
The group $SU(4)_{R}$ corresponds to the $R$ symmetry of the boundary theory. A quotient by $\mathbb{Z}_{2}$ is necessary because the group $Spin(4)$ and $SU(4)_{R}$ share a common factor which is $-I$ where $I$ is the action of identity. Since the two sided eternal black hole has two asymptotic boundaries, the full symmetry group is $G_{L} \times G_{R}$ where $G_{R} (G_{L})$ acts on right (left) asymptotic boundary. These symmetry groups are generated by the charge operators $Q^{a}_{R}$ and $Q^{a}_{L}$ respectively with the index $a$ runs over a basis of the Lie algebra $\mathfrak{g}$ of the group $G$. Each $Q^{a}_{R}$ and $Q^{a}_{L}$ does not have large $N$ limit due to divergence fluctuations. However, the difference
\begin{align}
    \hat{Q}^a = Q^{a}_{R}-Q^{*a}_{L}   \label{charge difference}
\end{align}
has a well defined large $N$ limit and it annihilates the thermofield double state
\begin{align}
    \hat{Q}^{a} \ket{TFD}=0
\end{align}
Therefore, we see that the two sided eternal black hole is invariant under the diagonal group $G_{D} \in G_{L} \times G_{R}$. The charge operators $\hat{Q}^{a}$ generates an action of the group $G$ on the Hilbert space $\mathcal{H}_{TFD}$. The Hilbert space representation of an element $g \in G $ is denoted as $W(g)$. The solution of two sided eternal black hole described by $\mathcal{N}=4$ SYM theory is parametrized by the moduli group $G_{\mathcal{M}}$ where $\mathcal{M}$ denotes the moduli space of solutions
\begin{align}
    G_{\mathcal{M}}= \frac{G_{L} \times G_{R}}{G_{D}} \cong G
\end{align}
As explained in \cite{witten2022gravity}, physically, $G_{\mathcal{M}}$ represents the space of Wilson line between the right asymptotic boundary and the left asymptotic boundary. In particular, one can start with a solution denoted by $\mathbb{1} \in G_{\mathcal{M}}$. The action of $G_{L} \times G_{R}$ on the moduli space $G_{\mathcal{M}}$ is given by (let $g \in G_{\mathcal{M}}$)
\begin{align}
    g \rightarrow g_{L}^{-1} g g_{R}
\end{align}
So, we can prepare the solution $g$ by acting an element $\mathbb{1} \times g \in G_{L} \times G_{R}$ on the chosen solution $\mathbb{1}$. Moreover, the stabilizer group $G_{g}$ of the solution $g$ is given by
\begin{align}
    G_{g}= \{  g_{L} \times g_{R} \in G_{L} \times G_{R} \vert g_{L}^{-1} g g_{R}=g \: \textit{or} \: g_{R}=g g_{L}g^{-1}   \} \cong G_{D}
\end{align}
Furthermore, for each solution $g \in G_{\mathcal{M}}$, $\mathcal{O}(1)$ perturbation around the two sided eternal black hole gives a thermofield double Hilbert space denoted as $\mathcal{H}_{TFD,g}$. We can embed the Hilbert space $\mathcal{H}_{TFD,g}$ into a Hilbert space bundle $\mathcal{V}$ with the base space $G_{\mathcal{M}}$ as a fibre at the base point $g$, i.e. $\mathcal{H}_{TFD,g} \hookrightarrow \mathcal{V} \xrightarrow{\pi} G_{\mathcal{M}}$. As we vary the solution $g \in G_{\mathcal{M}}$, we are moving along the base space of the Hilbert space bundle $\mathcal{V}$. Then, we can take the improved Hilbert space $\hat{\mathcal{H}}$ as the space of $L^2$ section of $\mathcal{V}$
\begin{align}
    \hat{\mathcal{H}}= \Gamma_{L^2}(\mathcal{V})
\end{align}
However, as explained by Witten in \cite{witten2022gravity}, there exists a simpler description of the improved Hilbert space taking into account the additional symmetries of the eternal black hole by trivializing the bundle $\mathcal{V}$. Since the base space $G_{\mathcal{M}}$ of the bundle $\mathcal{V}$ is the quotient of the symmetry group $G_{L} \times G_{R}$ by the stabilizer group of any $g \in G_{\mathcal{M}}$, it follows that the group $G_{L} \times G_{R}$ acts transitively on $G_{\mathcal{M}}$, i.e. $G_{\mathcal{M}}$ is a homogeneous space under the action of $G_{L} \times G_{R}$. Therefore, we can apply either $G_{L}$ invariant or $G_{R}$ invariant trivialization of the bundle $\mathcal{V}$. In particular, this can be achieved by picking a trivialization around the solution $\mathbb{1}$ and extend it in $G_{L} (G_{R})$ invariant way by acting the trivialization  with $G_{L} ( G_{R})$. By trivializing the Hilbert space bundle $\mathcal{V}$, the improved Hilbert space $\hat{\mathcal{H}}$ can be written as a tensor product
\begin{align}
    \hat{\mathcal{H}}= \mathcal{H}_{TFD} \otimes L^{2} (G_{\mathcal{M}}) \label{improve tensor product}
\end{align}
If we choose $G_{L}$ invariant trivialization, the action of the group $G_{L} \times G_{R}$  on a state $\Psi_{L}(g) \in \hat{\mathcal{H}}$ defined for $G_{L}$ invariant trivialization is given by
\begin{align}
    L_{g_{L}, g_{R}} (\Psi_{L}(g))= W(g_{R}) \Psi_{L} (g^{-1}_{L}gg_{R})
\end{align}
where the action of $W(g_{L})$ is omitted because we are working in $G_{L}$-invaraint formalism. In contrast, if we choose $G_{R}$ invariant trivialization, the action of $G_{L} \times G_{R}$ on $\Psi_{R} (g)$ defined for $G_{R}$ trivialization is
\begin{align}
    R_{g_{L}, g_{R}}(\Psi_{R} (g))= W(g_{L}) \Psi_{R}(g^{-1}_{L} g g_{R})
\end{align}
In fact, these two formalisms are equivalent and the relation between these two formalisms is determined by a map $\Lambda : \hat{\mathcal{H}} \rightarrow \hat{\mathcal{H}}$ given by

\begin{align}
    \Psi_{R} (g)= (\Lambda \Psi_{L})(g)= W(g) \Psi_{L}(g), \quad R_{g_{L}, g_{R}}= \Lambda L_{g_{L}, g_{R}} \Lambda^{-1}   \label{relation between two approach}
\end{align}
since $W(g)$ is generated by  (\ref{charge difference}). Now, we will mainly continue our discussion using $G_{L}$ invariant trivialization of fibration. In this formalism , $W(g_{L})$ does not act on the improved Hilbert space $\hat{\mathcal{H}}$ at all. Therefore, the action of $G_{R}$ on the improved Hilbert space (\ref{improve tensor product}) is generated by charged operators $\hat{Q}^a + q^a_{R}$ where $q^a_{R} (q^{a}_{L})$ is the generator of the action of $G_{R} (G_{L})$ on $L^{2} (G_{\mathcal{M}})$ while the action of $G_{L}$ on the Hilbert space (\ref{improve tensor product}) is just generated by $q^{a}_{L}$. To incorporate these collective coordinates associated to the symmetry group $G$ into the algebra of observables, we extended the bulk local algebra $\tilde{\mathcal{M}}_r$ associated to the right exterior region of the black hole into a bigger algebra $\mathcal{Y}_{R}$. The algebra $\mathcal{Y}_{R}$ is generated by any operator $a\in \tilde{\mathcal{M}}_r$ and bounded function of $\hat{Q}^{a} + q^{a}_{R}$. It follows that its commutant denoted as $\mathcal{Y}_{L}$ is generated by any operator $b \in \Lambda^{-1}\tilde{\mathcal{M}}_l \Lambda$ and bounded function of $q^{a}_{L}$. In fact, this is exactly the crossed product of the algebra $\tilde{\mathcal{M}}_r$ by the group automorphisms $G$
\begin{align}
    \mathcal{Y}_{R}= \tilde{\mathcal{M}}_r \rtimes G = \mathcal{M}_{R} \rtimes G \label{compact group cross product}
\end{align}
which is completely analogous to the modular crossed product. In the second equality we use the large $N$ holographic duality $\tilde{\mathcal{M}}_r= \mathcal{M}_{R}$.  It turns out that $\mathcal{Y}_{L}$ described before is just the standard commutant of (\ref{compact group cross product}) and the asymmetry between $\mathcal{Y}_{R} $ and $\mathcal{Y}_{L}$ can be eliminated by conjugating everything by $\Lambda^{1/2}$ which is similar to the case of modular crossed product. Furthermore, given the specific setup of this construction (including the compactness of 
$G$), it is expected that the crossed product does not alter the type of the algebra \cite{witten2022gravity} \cite{chandrasekaran2023large}. Hence, $\mathcal{Y}_{R}$ and $\mathcal{Y}_{L}$ are still of Type $\RN{3}_{1}$ factor. In fact, there is an alternative method to obtain the same structure by switching to an ensemble over finite charge states. To accomplish this, one can start by first going to finite $N$  and modify the standard thermofield double state to 
\begin{align}
    \tilde {\ket{TFD}}= \frac{1}{\sqrt{Z_{\beta}}} \sum_{i} e^{- \beta E_{i}/2} g(R_{i}) \ket{i}_{R} \ket{i}_L
\end{align}
where $R_i$ is the irreducible representation of $G$ that contains the state $\ket{i}$and $g$ is a map from the space of irreducible representation of $G$  to $\mathbb{C}$ such that $\sum_{R} \vert g(R) \vert^2 =1$. After taking large $N$ limit and follow some procedures, one can arrive at the same structure as before. We will not review this approach. See appendix B of \cite{chandrasekaran2023large} for more details.

\par In this case, the Principle of Diminishing Potentialities (PDP) holds for $\mathcal{Y}_{R}$ and $\mathcal{Y}_{L}$. Since $\mathcal{Y}_{R}$ and $\mathcal{Y}_{L}$ are symmetric up to a conjugacy by $\Lambda^{1/2}$, we focus on the right algebra $\mathcal{Y}_{R}$. We similarly define the time band algebra $\mathcal{Y}_{R, I}$ 
\begin{align}
    \mathcal{Y}_{R,I}= \mathcal{M}_{R,I} \rtimes G
\end{align}
Since $\mathcal{M}_{R, \geq t} \subsetneqq \mathcal{M}_{R, \geq t'}$ for all $t > t'$ and the automorphism that generate time translation is outer with respect to the algebra $\mathcal{Y}_{R}$, it follows that
\begin{align}
    \mathcal{Y}_{R, \geq t}= \mathcal{M}_{R, \geq t} \rtimes G \subsetneqq \mathcal{M}_{R, \geq t'} \rtimes G= \mathcal{Y}_{R, \geq t'} \quad \textit{for}\: \: t > t'
\end{align}
This is PDP for the algebra $\mathcal{Y}_{R, \geq t}$. Next, we can choose a preferred/initial state $\Psi_{\hat{\omega}} \in \mathcal{H}_{TFD} \otimes L^{2} (G_{\mathcal{M}}) $ given by
\begin{align}
    \ket{\Psi_{\hat\omega}}= \ket{TFD} \otimes f(g) \equiv \ket{TFD,f}
\end{align}
where $f(g) \in L^{2} (G_{\mathcal{M}})$ with the condition 
\begin{align}
    \int_{G} d \mu (g) \:  \vert f(g) \vert^2=1
\end{align}
where $d \mu (g)$ is the Haar measure of $G$. We would like to check whether the centralizer $\mathcal{C}_{TFD,f}(\mathcal{Y}_{R})$ is trivial or not. We first note that the additive basis for the algebra $\mathcal{Y}_{R}$ is given by operators of the form $\{a W(g) w(g) \vert a \in \mathcal{M}_{R}, g \in G   \}$ where $w(R)$ denotes the  action of $G$ on $L^{2} (G_{\mathcal{M}})$ by right multiplication, i.e. 
\begin{align}
    w(g_1) f(g)= f(gg_1)
\end{align}
 The operators of the form $a W(g)w(g)$ form a set of additive basis because
\begin{align}
    \bigg( a W(g_1)w(g_1) \bigg) \circ \bigg( b W(g_2) w(g_2)   \bigg)= a W(g_1)bW(g_2) w(g_1) w(g_2)= a \sigma_{g}(b)W(g_1 g_2)w(g_1 g_2)
\end{align}
where $\sigma_{g}(b)= W(g) b W(g^{-1})$ and it is automorphisms of the group $G$ on $\mathcal{M}_{R}$. So, $\sigma_{g} (b)\in \mathcal{M}_{R}$ and thus $a \sigma_{g} (b) \in \mathcal{M}_{R}$.  Now, we check that whether the  operators of the form $\{ W(g)w(g) \vert g \in G  \}$ lie inside $\mathcal{C}_{TFD,f} (\mathcal{Y}_{R})$. To check it, we compute (let $a$ be any non-trivial element in the algebra $\mathcal{M}_{R}$)
\begin{align*}
    \hat{\omega} \bigg( (W(g_1) w(g_1)) (a W(g_2) w(g_2))    \bigg)&=  \omega_{TFD} (W(g_1) a W(g_2)) \int_{G} d \mu(g) \bar{f}(g) f(gg_1 g_2) \\
    &= \omega_{TFD} (a W(g_2) W(g_1)) \int_{G} d \mu(g) \bar{f} (g) f(g g_1 g_2)\\
&\overset{\textit{if}\: [g_{1},g_2]=0}{=} \hat{\omega} \bigg( (a W(g_2) w(g_2)) ( W(g_1) w(g_1)  )  \bigg) \quad \forall g_{2} \in G 
    \end{align*}

In the second line, we use that $\hat{Q}^{a} \ket{TFD}=0$ so that $W(g) \ket{TFD}=\ket{TFD}$. In the third line, we note that if $[g_{1}, g_{2}]=0$ for all $g_{2} \in G$, then $f(gg_1 g_2)= f(g g_2 g_1)$. This will lead to that $W(g_{1})w(g_1) \in \mathcal{C}_{TFD,f} (\mathcal{Y}_{R})$. However, things do not go as we wish because the symmetry group $G$ we consider is a non-abelian group so in general there does not exist a group element except identity that can commute with all elements of the group $G$. In fact, the operators of the form $aW(g)w(g)$ with $a \in \mathcal{M}_{R}$ a non-trivial element also does not lie in $\mathcal{M}_{TFD, f} (\mathcal{Y}_{R})$ because similar calculation as above show that these operators do not commute with operators of the form $W(g)w(g)$ in the state $\ket{TFD, f}$. Therefore,
\begin{align}
    \mathcal{C}_{TFD, f} (\mathcal{Y}_{R})= \mathbb{C} \mathbb{1}
\end{align}
    This is pretty disappointing after all the effort. However, there is a rescue if we only consider the maximal \textbf{abelian} subgroup $H \in G$.  The maximal abelian subgroup $H \in G$ corresponds to the exponential of the Cartan subalgebra (maximal abelian subalgebra)$\mathfrak{h} \in \mathfrak{g}$.  In our case, the Lie algebra $\mathfrak{g}= so(4) \times su(4)_{R}$. Its Cartan subalgebra $\mathfrak{h} \in \mathfrak{g}$ consists of 
    \begin{align}
        \mathfrak{h}= \{ J_1, J_2, R_1, R_2, R_3   \}
    \end{align}
    where $J_{1}, J_{2}$ belong to the Cartan subalgebra of $so(4) \cong su(2)_{L} \times su(2)_R$ and $R_1, R_2, R_3$ belong to the Cartan subalgebra of $su(4)_{R}$. In this case, the the maximal abeliam subgroup $H= U(1)^5$. Now, if we only perform the crossed product of $\mathcal{M}_{R}$ by the maximal abelian  subgroup $H$ instead of the full group $G$, then the centralizer of an appropriate chosen state will be non-trivial. Therefore, we consider the right algebra $\mathcal{Y}^{H}_{R}$
    \begin{align}
        \mathcal{Y}^{H}_{R} = \mathcal{M}_{R} \rtimes H
    \end{align}
    We can repeat the above procedures by just restricting to a smaller group. Since $H \in G$ is also compact,  thus in this particular setup,  $\mathcal{Y}^{H}_{R}$ is also of Type $\RN{3}_1$. We denote 
    \begin{align}
\hat{J}^{i}= J^{i}_{R}- J^{i}_{L}, \quad \hat{R}^{\alpha}  = R^{\alpha}_{R}- R^{\alpha}_{L}
    \end{align}
   where $i=1,2$ and $\alpha=1,2,3$. The right algebra $\mathcal{Y}^{H}_{R}$ is obtained by conjugating $\mathcal{M}_{R}$ by $\tilde{J}^{i}= \hat{J}^{i}+j_{R}^{i}$ and also $\tilde{R}^{\alpha}= \hat{R}^{\alpha}+ r^{\alpha}_{R}$. The operators $j^{i}_{R}$ and $r^{a}_R$ are generators of the right action of $H$ on $L^{2} (H)$. In this case, the extended Hilbert space is
   \begin{align}
       \hat{\mathcal{H}}^{H} = \mathcal{H}_{TFD} \otimes L^{2} (H)
   \end{align}
   We can also choose a preferred/initial state $\ket{TFD, k} \in \hat{\mathcal{H}^H}$ where $k(h) \in L^{2} (H)$ with the condition that
   \begin{align}
       \int_{H} d \mu(h) \vert k(h) \vert^2 =1
   \end{align}
   where $d \mu (h)$ is the Haar measure of the group $H$.  Repeating the same procedures as above, we know that PDP still holds for $\mathcal{Y}^{H}_{R}$, i.e. $\mathcal{Y}^{H}_{R,\geq t } \subsetneqq \mathcal{Y}^{H}_{R, \geq t'} $ for all $t> t'$. Moreover, we know that the operators of the from $\{ W(h) w(h) \vert h \in H  \}$ lies within the centralizer $\mathcal{C}_{TFD,k} (\mathcal{Y}^{H}_{R})$. To see that these are the only operators in the centralizer $\mathcal{C}_{TFD,k} (\mathcal{Y}^{H}_{R})$, for each operator of the form $a W(h)w(h)$ with $a \in \mathcal{M}_{R}$ a non trivial element, we need to find  one element in $\mathcal{Y}^{H}_{R}$  that does not commute with it in the state $\ket{TFD, k}$. A straightforward calculation leads to
   \begin{align}
       [aW(h_1) w(h_1), bW(h_2) w(h_2)]= \bigg( a \sigma_{h_1}(b)- b \sigma_{h_2} (a)  \bigg) W(h_1 h_2) w(h_1 h_2),    \quad \forall b \in \mathcal{M}_{R},\: \: \forall h_{2} \in H \label{final commutator}
   \end{align}
   Evaluating (\ref{final commutator}) in the state $\ket{TFD,k}$, we obtain
   \begin{align}
       \omega_{TFD,k} ([aW(h_1) w(h_1), b W(h_2) w(h_2)])= \omega_{TFD} (a \sigma_{h_1}(b)-\sigma_{h_{2}^{-1}}(b) a) \int_{H} d \mu(h) \bar{k} (h) k (h h_1 h_2)  \label{final very important calculation}
   \end{align}
   for all $b \in \mathbb{\mathcal{M}}_{R}$ and for all $h_{2}\in H$.
   Now, from previous subsection, we know there always exist one element $c$ in $\mathcal{M}_{R}$ such that $\omega_{TFD} (ac) \neq \omega_{TFD} (ca)$. Now, We have the freedom to choose $h_{2}= h_{1}^{-1}$. Therefore, the first factor in  (\ref{final very important calculation}) becomes $\omega_{TFD} (a \sigma_{h_1} (b) - \sigma_{h_1}(b)a)$. We can also choose $b$ such that $\sigma_{h_1}(b)=c$. Therefore, $\omega_{TFD} (a \sigma_{h_1}(b)- \sigma_{h_1}(b)a) \neq 0$. Moreover, upon choosing $h_2 = h^{-1}_{1}$, the second factor  in (\ref{final very important calculation}) becomes $ \vert \vert k      \vert \vert^2_{L^2 (H)}$ which is non-zero and positive. Therefore, for each element $a W(h_1)w(h_1)$, we find one element $b W(h^{-1}_{1})w(h^{-1}_1)$ such that they do not commute with each other in the state $\ket{TFD, k}$.

   \par Hence, we know that
   \begin{align}
       \mathcal{C}_{TFD,k} (\mathcal{Y}^{H}_{R})= \{  W(h) w(h) \vert h \in H \}
   \end{align}
   Furthermore, we know that $\mathcal{C}_{TFD,k} (\mathcal{Y}^{H}_{R})$ is by itself commutative because the group $H$ is abelian. Therefore, the center of the centralizer of the state $\ket{TFD,k}$ is equal to itself

   \begin{align}
       Z_{TFD, k} (\mathcal{Y}^{H}_{R})= \mathcal{C}_{TFD,k} (\mathcal{Y}^{H}_{R})
   \end{align}
   The center of the centralizer $Z_{TFD,k} (\mathcal{Y}^{H}_{R})$ is generated by the generators of $\{   W(h) w(h) \vert h \in H   \}$. Therefore, the \textbf{first actual event} is  given by the spectral projections of $\{ \tilde{J}^{i}, \tilde{R}^{\alpha}   \}$ with $i=1,2$ and $\alpha=1,2,3$. In general, via spectral theorem these charge operators can be written as

   \begin{align}
       \tilde{J}^{i} = \int_{\sigma (\tilde{j}^{1}, \tilde{j}^{2}, \tilde{r}^{1}, \tilde{r}^{2}, \tilde{r}^{3})} \tilde{j}^{i}  dP(\tilde{j}^{1}, \tilde{j}^{2}, \tilde{r}^{1}, \tilde{r}^{2}, \tilde{r}^{3}),  \quad \tilde{R}^{\alpha}= \int_{\sigma (\tilde{j}^{1}, \tilde{j}^{2}, \tilde{r}^{1}, \tilde{r}^{2}, \tilde{r}^{3})} \tilde{r}^{\alpha} dP(\tilde{j}^{1}, \tilde{j}^{2}, \tilde{r}^{1}, \tilde{r}^{2}, \tilde{r}^{3}) 
   \end{align}
   where $\tilde{j}^{i}$ denotes eigenvalues of $\tilde{J}^{i}$ and $\tilde{r}^{\alpha}$ denotes eigenvalues of $\tilde{R}^{\alpha}$. The symbol $\sigma (\tilde{j}^{1}, \tilde{j}^{2}, \tilde{r}^{1}, \tilde{r}^{2}, \tilde{r}^{3}) \subset \mathbb{R}^{5}$ denotes the joint spectrum while $dP(\tilde{j}^{1}, \tilde{j}^{2}, \tilde{r}^{1}, \tilde{r}^{2}, \tilde{r}^{3})$ denotes the spectral measure. Now, we can partition the joint spectrum  $\sigma (\tilde{j}^{1}, \tilde{j}^{2}, \tilde{r}^{1}, \tilde{r}^{2}, \tilde{r}^{3}) \subset \mathbb{R}^{5}$ into countably many disjoint set $I_{\mathcal{\xi}}$ with $\mathcal{\xi} \in \mathcal{X}$ where $\mathcal{X}$ is a countable set to form a partition unity of orthogonal projectors $\{ P_{\mathcal{\xi}} \vert \mathcal{\xi} \in \mathcal{X} \}$
\begin{align}
    P_{\mathcal{\xi}}= \int_{\sigma (\tilde{j}^{1}, \tilde{j}^{2}, \tilde{r}^{1}, \tilde{r}^{2}, \tilde{r}^{3})} \chi_{\mathcal{\xi}}(\tilde{j}^{1}, \tilde{j}^{2}, \tilde{r}^{1}, \tilde{r}^{2}, \tilde{r}^{3}) dP(\tilde{j}^{1}, \tilde{j}^{2}, \tilde{r}^{1}, \tilde{r}^{2}, \tilde{r}^{3})
\end{align}
where $\chi_{\mathcal{\xi}}(\lambda)$ is the characteristic function 
\begin{align}
    \chi_{\mathcal{\xi}}=    \left\{ \begin{array}{rcl}
         1 & \mbox{if} & (\tilde{j}^{1}, \tilde{j}^{2}, \tilde{r}^{1}, \tilde{r}^{2}, \tilde{r}^{3}) \in I_{\mathcal{\xi}} \\  0 & \mbox{if} & (\tilde{j}^{1}, \tilde{j}^{2}, \tilde{r}^{1}, \tilde{r}^{2}, \tilde{r}^{3}) \not\in I_{\mathcal{\xi}}
                \end{array}\right.
                \end{align}
    Certainly, the choice of partitioning the joint spectrum $\sigma (\tilde{j}^{1}, \tilde{j}^{2}, \tilde{r}^{1}, \tilde{r}^{2}, \tilde{r}^{3})$ into countable many disjoint sets is not unique at all. It would be nice if there exists a rule or principle to tell which choice is preferred/ canonical. We will not go deep into this issue in this paper. However, we note that even if there is no canonical way to partition the joint spectrum, this does not contradict the spirit of the stochastic nature of ETH dynamics. After all, what is certain is only the probability outcome in the Collapse Postulate (CP) which is given by Born's rule $P(\xi)= \langle TFD, k \vert P_{\xi} \vert TFD,k \rangle$.  
    
    \par Therefore, the actual event $\{ P_{\mathcal{\xi}} \vert \mathcal{\xi} \in \mathcal{X}  \}$ serves as the \textbf{first actual event initiating the whole ETH dynamics}. See Figure 9.  After the first actual event occurs, the initial state $\ket{TFD,k}$ is replaced by
    \begin{align}
        \ket{\Psi(\xi_*)}= \frac{1}{\langle TFD ,k \vert P_{\xi_*} \vert TFD,k \rangle} P_{\xi_*} \ket{TFD,k}
    \end{align}
    where $\xi_{*}$ is some point in $\mathcal{X}$ with $\omega_{TFD,k} (P_{\xi_*}) >0$. So far, we only restrict our exploration to the right asymptotic boundary. In fact,  it is more natural to consider ETH formalism taking into account both right and left asymptotic boundaries algebras because the description of the full evolution of the system depends on both right side and left side. We can consider the time band algebra $\mathcal{Y}^{H}_{R, \geq t} \cup \mathcal{Y}^{H}_{L, \leq -t}$. See Figure 10. Note that for the left side we reverse the direction of time because the time translation generated by modular flow $- \log \Delta_{TFD}$ on the left side is opposite to the time direction on the right side.  See the last point in Section 5 for more descriptions.

\begin{figure}
    \centering
    \includegraphics[width=0.5\linewidth]{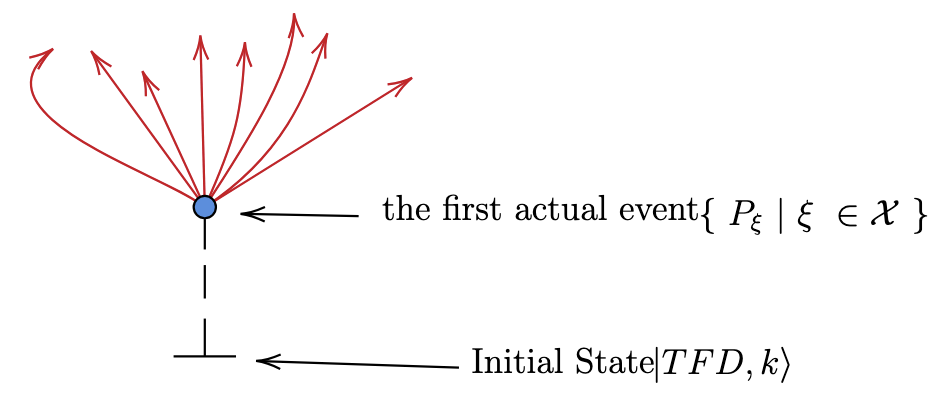}
    \caption{The first actual event}
    \label{Figure 9}
\end{figure}
 \par Admittedly, the construction presented in this subsection is rather ad hoc. Nevertheless, together with the earlier insight that no event can occur in exact thermal equilibrium, the consideration here can be viewed as an example of a broader physical intuition: in equilibrium, any significant intervention perturbs the system and drives it away from equilibrium. To measure the temperature of a thermal system, for instance, one must bring a thermometer into contact with it  i.e, an action that is not itself in equilibrium with the system and therefore constitutes a genuine event. It is precisely this kind of external disturbance that the extended algebraic framework is meant to capture. In our construction, performing the crossed product of $\mathcal{M}_R$ by the maximal abelian subgroup $H$ 
 amounts to coupling the system to an external “device” (the collective coordinates of $H$). Moreover by a suitable choice of $k(H) \in L^2 (H) $ which can mimic a genuine `` external measuring device'' (e.g, $\vert k(h) \vert^2 \neq 1$ ) typically leads to that the extended state $\omega_{TFD,k}$  no longer satisfies the KMS condition with respect to the original modular flow $\sigma^{TFD}_{t}$ on $\mathcal{Y}^{H}_{R}$ in line with the idea that the ``device'' is not in thermal equilibrium with the black hole. At the same time, the center of the centralizer  $Z_{TFD,k} (\mathcal{Y}^{H}_{R})$ is non-trivial and serves as the trigger for the first actual event.  Thus, although the construction is not fully satisfying (natural), it is not entirely ad hoc from a physical perspective, as it reflects the very intuition that a ``measurement apparatus'' disturbs the thermal equilibrium of the system.

\section{Discussion}
We give a brief summary of what we have done in this paper. In section $2$, we summarized the \textbf{Events-Trees-Histories Approach} (ETH) to quantum mechanics-- both in the setting of non-relativistic and relativistic  quantum theory. The ETH approach is a proposal for the completion of quantum mechanics featuring an isolated-open system characterized by the \textbf{Principle of Diminishing Potentialities} (PDP). In this framework, the evolution of state is given by a stochastic branching process and the usual unitary evolution of state arises by averaging over all histories. In section $3$, we review the algebraic approach to $\mathcal{N}=4$ SYM theory with gauge group $SU(N)$. This approach is particularly useful to elucidate the structure of the  large $N$ limit / asymptotic limit of the gauge  theory as generalized free fields theory. We also discuss its implications  for holographic duality. In section 4, it is shown that  PDP holds for the large $N$ algebra , i.e. $\mathcal{M}_{\geq t} \subsetneqq \mathcal{M}_{\geq t'}$ for $t> t'$ in thermal equilibrium when the temperature is above Hawking Page temperature $\beta < \beta_{HP}$ using techniques in algebraic quantum field theory and large $N$ holographic duality. In contrast, we also give arguments that PDP does not hold when the temperature is below Hawking Page temperature $\beta >\beta_{HP}$ and also for the case when the temperature is zero $\beta \rightarrow \infty$. In this framework, we provide a mathematical proof that event does not occur at all in thermal equilibrium. This is compatible with the observation that the the centralizer of the thermofield double state $\mathcal{C}_{TFD} (\mathcal{M})$ coincides with the center of the algebra $\mathcal{M}$ because the commutators of generalized free fields are given by $c$ numbers. Finally, we extend the large $N$ generalized free field algebra $\mathcal{M}$ in the case above Hawking Page transition by performing crossed product of the algebra $\mathcal{M}$ by the maximal abelian subgroup $H$ of a compact symmetry group $G$. The group $G$ corresponds to additional symmetries of two sided eternal black hole other than time translation. The resulting algebra  $\mathcal{Y}^{H}$ still satisfies PDP. When restricting to the right boundary, we show that the centralizer of a natural preferred state (an extension of thermofield double state) is given by operators of the form $\{  W(h) w(h) \vert h \in H\}$ where $W(h)$ is the action of $H$ on the GNS Hilbert space $\mathcal{H}_{TFD}$ and $w(h)$ is the right action of $H$ on $L^{2} (H)$. The \textbf{first actual event that initiates the entire ETH dynamics} is given by the generators of the center of this centralizer which is the spectral projectors associated to the representation of  Cartan subalgebra $\mathfrak{h}$ of $H$   on the extended Hilbert space $\mathcal{H}_{TFD} \otimes L^{2} (H)$.

\par We also list down some natural questions that arise from our study:
\begin{itemize}
    \item In this paper, I strictly follow ETH formalism formulated by
Fr\"ohlich et.al. A natural inquiry arises: Does there exist an alternative formulation of the ETH approach that preserves its foundational motivation and spirit while offering greater flexibility in application?

\item How to really characterize a continuous sequence of events in the evolution of the system? This is certainly related to the previous item. In the final paragraph of Section $2.1$, we also write down a tentative hint/ direction provided by J.Fr\"ohlich in a seminar \cite{AuthorYear}. Is it necessary to discretize the time or introduce a cutoff? 

\par In fact,  there is  a fairly satisfactory and beautiful method developed by Fr\"ohlich et.al (see \cite{Frohlich:2024kyo} \cite{Frohlich:2026epd} )  by combining Principle of Diminishing Potentialities (PDP), 
Gorini-Kossakowski-Sudarshan-Lindblad (GKSL) master equation and  infinitesimal perturbation theory (first order perturbation theory) to give a self-consistent stochastic  mechanism to describe the updating of state in some simple \textbf{non-relativistic quantum mechanical experiments} (where actual events appear as `` quantum Poisson jumps'') and the continuous limit is fully under control, yielding a well-defined stochastic law of dynamics.  The main open challenge is to determine whether a method of the same spirit can be developed for \textbf{relativistic quantum field theory}?

\item Is there other mechanisms for the Principle of Diminishing Potentialities (PDP)? Certainly a natural physical origin of PDP is the Huygens Principle \cite{frohlich2020relativistic} \cite{bucholtz2013}. In Section 3, we provide unconventional arguments that PDP holds for  time band algebras in the black hole phase. Our simple arguments only depend on the causal structure of the causal wedge associated to the time band in the large $N$ limit. The price to pay is that no events can occur because the centralizer of the thermal state coincides with the center of the algebra which is trivial. Therefore,  we need to extend the large $N$ generalized free field algebra by incorporating additional symmetries of two sided eternal black hole so that the centralizer of the extended thermal state is no longer trivial.

\item In a series of works(see \cite{Satishchandran:2025cfk} and references therein),  Wald, Satishchandran and Danielson have argued that the black hole itself, specifically the presence of a Killing horizon can decohere spatial superpositions, as shown through a Stern–Gerlach type Gedanken experiment. Their conclusion is very much in the spirit of the discussion in this paper. A natural question, then, is whether the ETH approach to quantum mechanics can be systematically integrated with their discovery, potentially offering a unified framework in which black hole horizons act as universal sources of event‑inducing decoherence.

\item It is notable that in all known mechanisms giving rise to the Principle of Diminishing Potentialities, notions of locality and causality play crucial role. This raises an intriguing question: could the completion of quantum mechanics in the sense of a dynamics that intrinsically incorporates measurement and the occurrence of events, itself be only an emergent concept? After all, in a more ultraviolet‑complete theory, locality and causality typically lose their fundamental status and may themselves emerge only in suitable limits.

\item Is it possible that ETH formalism can shed light on the black hole information paradox? In this paper, we always restrict ourselves to one asymptotic boundary when discussing PDP and whether event occurs or not. We might also consider time band algebra associated to two boundaries i.e. $
\mathcal{Y}^{H}_{R, \geq t}\cup \mathcal{Y}^{H}_{L, \leq -t}$. See Figure 10. In the black hole phase, PDP still holds for this algebra, i.e. $\mathcal{Y}^{H}_{R, \geq t}\cup \mathcal{Y}^{H}_{L, \leq -t}\subsetneqq\mathcal{Y}^{H}_{R, \geq t'}\cup \mathcal{Y}^{H}_{L, \leq -t'}$ for $t > t'$. The initial algebra corresponds to $t \rightarrow -\infty$ which is  the union of the left and right algebra $\mathcal{Y}^{H}_{R} \cup \mathcal{Y}^{H}_{L}$. We also note that after conjugating $\mathcal{Y}^{H}_{R}$ with $\Lambda^{-1}$ and then replacing all right action with left action, we will obtain $\mathcal{Y}^{H}_{L}$. Taking this into account, we know that the centralizer    $\mathcal{C}_{TFD,k}  (\mathcal{Y}^{H}_{R} \cup \mathcal{Y}^{H}_{L} ) $ is given by
\begin{align}
\mathcal{C}_{TFD,k} ( \mathcal{Y}^{H}_{R} \cup \mathcal{Y}^{H}_{L} )= \{ W(h_{R} ) w(h_{R}), w(h_{L}) \vert h_{R} \in H_{R}, \: \: h_{L} \in H_{L}  \}    
\end{align}
where $w(h_L)$ is the left action of $H$ on $L^{2} (H)$
\begin{align}
   w(h_L) f(h)= f(h^{-1}_{L} h) , \quad f(h) \in L^2 (H)
\end{align}
Moreover, since $W(h_R)w(h_{R})$ commutes with $w(h_{L})$, we know that the center of the centralizer $Z_{TFD,k} (\mathcal{Y}^{H}_{R} \cup \mathcal{Y}^{H}_{L})$  coincides with the centralizer itself. Hence, the first actual event is given by the spectral projectors that generates $Z_{TFD,k} (   \mathcal{Y}^{H}_{R} \cup \mathcal{Y}^{H}_{L})$. A new feature is that while the state $\omega_{TFD,k}  \vert_{\overline{\mathcal{Y}^{H}_{R} \cup \mathcal{Y}^{H}_{L}}^{\vert \vert \cdot \vert \vert_{\hat{\mathcal{H}}^H}}} $ (with respect to the full algebra where the closure with respect to the norm of the Hilbert space $\hat{\mathcal{H}}^{H}$ is taken) might be pure, the initial state $\omega_{TFD,k}\vert_{\mathcal{Y}^{H}_{R} \cup \mathcal{Y}^{H}_{L}}$ (i.e. $t \rightarrow- \infty$) already appears to be a mixed state; and  it will continue to evolve to another mixed state $  \omega_{\Psi} \vert_{\mathcal{Y}^{H}_{R, \geq t} \cup \mathcal{Y}^{H}_{L, \leq -t}}$ according to the evolution given by the ETH formalism.
\end{itemize}

\begin{figure}
    \centering
    \includegraphics[width=0.4\linewidth]{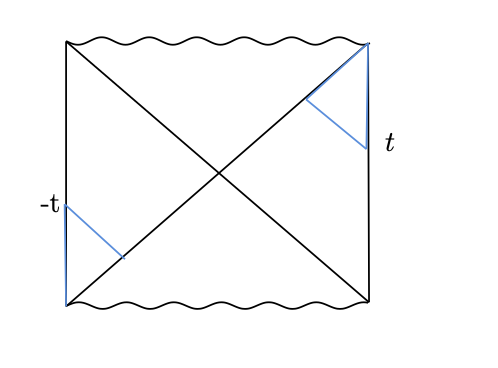}
    \caption{The time band algebra $\mathcal{Y}^{H}_{R, \geq t} \cup \mathcal{Y}^{H}_{R, \leq t}$ associated to two boundaries}
    \label{fig:placeholder}
\end{figure}

\section*{Acknowledgements}
I thank Daniel Galviz for suggestions on how to make figures. The author is supported by Chinese Government Scholarship (CSC).

\bibliography{references}

\bibliographystyle{utphys}

\end{document}